%% file: main.tex
\documentclass[floatfix, superscriptaddress,reprint,amsmath,amssymb, pra,showkeys]{revtex4-2}

\usepackage{graphics}
\usepackage{dcolumn}
\usepackage{bm}

\usepackage{array}
\usepackage{booktabs}
\usepackage{multirow}
\usepackage{esint}
\usepackage{color}
\usepackage{graphicx}
\usepackage{epstopdf}
\usepackage[version=4]{mhchem}
\usepackage{verbatim}
\usepackage{float}
\usepackage{fancyvrb}
\usepackage[export]{adjustbox}
\usepackage{svg} 
\usepackage{comment} 

\usepackage[section]{placeins} 
\usepackage[labelformat=simple]{subcaption}

\usepackage{dblfloatfix} 
\usepackage{float}
\usepackage[ruled, lined]{algorithm2e} 
\usepackage{algpseudocode} 

\usepackage{tikz}
\usepackage{pgfplots}
\usepackage{pgfplotstable}
\pgfplotsset{compat = newest}
\usetikzlibrary{shapes.geometric, arrows}

\usepackage[colorlinks=true, linkcolor=blue, citecolor=blue, urlcolor=blue]{hyperref}

\makeatletter

\let\revtex@sect@ltx\@sect@ltx
\def\@sect@ltx#1#2#3#4#5#6[#7]#8{%
  \GetTitleString{#7}%
  \let\@currentlabelname\GetTitleStringResult
  \revtex@sect@ltx{#1}{#2}{#3}{#4}{#5}{#6}[#7]{#8}%
}

\let\revtex@ssect@ltx\@ssect@ltx
\def\@ssect@ltx#1#2#3#4#5#6[#7]#8{%
  \GetTitleString{#7}%
  \let\@currentlabelname\GetTitleStringResult
  \revtex@ssect@ltx{#1}{#2}{#3}{#4}{#5}{#6}[#7]{#8}%
}

\makeatother

\usepackage{xr-hyper}

\usepackage{soul,xcolor}
\setstcolor{red}

\newcommand{\RNum}[1]{\uppercase\expandafter{\romannumeral #1\relax}}
\usepackage{textcomp, gensymb}
\DeclareUnicodeCharacter{0308}{HERE!HERE!}

\usepackage{scalerel}
\usepackage{tikz}
\usetikzlibrary{svg.path}
\definecolor{orcidlogocol}{HTML}{A6CE39}
\tikzset{
  orcidlogo/.pic={
    \fill[orcidlogocol] svg{M256,128c0,70.7-57.3,128-128,128C57.3,256,0,198.7,0,128C0,57.3,57.3,0,128,0C198.7,0,256,57.3,256,128z};
    \fill[white] svg{M86.3,186.2H70.9V79.1h15.4v48.4V186.2z}
                 svg{M108.9,79.1h41.6c39.6,0,57,28.3,57,53.6c0,27.5-21.5,53.6-56.8,53.6h-41.8V79.1z M124.3,172.4h24.5c34.9,0,42.9-26.5,42.9-39.7c0-21.5-13.7-39.7-43.7-39.7h-23.7V172.4z}
                 svg{M88.7,56.8c0,5.5-4.5,10.1-10.1,10.1c-5.6,0-10.1-4.6-10.1-10.1c0-5.6,4.5-10.1,10.1-10.1C84.2,46.7,88.7,51.3,88.7,56.8z};
  }
}
\newcommand\orcidicon[1]{\href{https://orcid.org/#1}{\mbox{\scalerel*{
\begin{tikzpicture}[yscale=-1,transform shape]
\pic{orcidlogo};
\end{tikzpicture}
}{|}}}}

\begin{document}
\include{revtex_preamble}
\title{Surprisingly High Redundancy in Electronic Structure Data Across Materials Explained by Low Intrinsic Dimensionality}

\author{Sazzad Hossain \orcidicon{0009-0005-4064-5668}}
\affiliation{Department of Mechanical and Aerospace Engineering, Michigan Technological University, Houghton, MI, USA.}
\author{Ponkrshnan Thiagarajan \orcidicon{0000-0003-3946-3902}}
\affiliation{Hopkins Extreme Materials Institute, Johns Hopkins University, Baltimore, MD, USA.}
\author{Shashank Pathrudkar \orcidicon{0000-0001-8546-8056}}
\affiliation{Department of Mechanical and Aerospace Engineering, Michigan Technological University, Houghton, MI, USA.}
\author{Stephanie Taylor \orcidicon{0000-0000-0000-0000}}
\affiliation{Department of Materials Science and Engineering, University of California, Los Angeles, CA, USA.}
\author{Abhijeet Sadashiv Gangan \orcidicon{0000-0002-8937-7984}}
\affiliation{Department of Materials Science and Engineering, University of California, Los Angeles, CA, USA.}
\author{Amartya S. Banerjee \orcidicon{0000-0001-5916-9167}}
\email{asbanerjee@ucla.edu}
\affiliation{Department of Materials Science and Engineering, University of California, Los Angeles, CA, USA.}
\author{Susanta Ghosh \orcidicon{0000-0002-6262-4121}}
\email{susantag@mtu.edu}
\affiliation{Department of Mechanical and Aerospace Engineering, Michigan Technological University, Houghton, MI, USA.}
\affiliation{Center for Artificial Intelligence, Michigan Technological University, Houghton, MI, USA.}
\date{\today}

\vspace{0.5em}
\begin{abstract}
\begin{center}
    \textbf{\abstractname}
\end{center}
\vspace{-0.5em}
Machine learning (ML) models for electronic structure typically rely on large datasets generated by computationally expensive Kohn–Sham density functional theory calculations, as it is not known a priori which portions of the data are essential for accurate learning. Here, we reveal significant redundancies in electronic structure datasets across diverse material systems and attribute them to the low intrinsic dimensionality of the underlying data. We show that even random pruning can substantially reduce dataset size with minimal degradation in predictive accuracy. Moreover,  a state-of-the-art coverage-based pruning strategy that samples data across all learning difficulties almost always preserves chemical accuracy and maintains model generalizability while using up to two orders of magnitude less data and reducing training time by a factor of three or more.  We further demonstrate that the essential electronic structure information lies on a low-dimensional, non-linear manifold, providing a  potential geometric explanation for the observed prunability. These observations are consistent with the predominance of local atomic environments in determining electronic properties, as suggested by nearsightedness arguments, and indicate that large-scale datasets may contain highly overlapping information. Our findings challenge the prevailing assumption that such extensive datasets are necessary for accurate ML-based electronic structure predictions and open a path toward identifying minimal, representative datasets for each material class. 
\end{abstract}
\maketitle

\section*{Introduction}
The calculation and analysis of the electronic structure of materials underpin much of modern computational materials science. From predicting thermodynamic stability and charge transport to modeling chemical reactivity, the electronic structure serves as a fundamental description of how interacting atoms and electrons give rise to observable properties. Consequently, electronic structure calculations often serve as an important building block of multiscale materials modeling~\citep{tadmor2011modeling},  with a significant share of contemporary high performance computing resources being devoted to such simulations \cite{hafner2008ab, curtarolo2013high, lejaeghere2016reproducibility}.

Kohn-Sham Density Functional Theory (KS-DFT) serves as the primary workhorse of modern electronic structure calculations \citep{kohn1965self, hohenberg1964inhomogeneous, hafner2006toward}. Although formulated in terms of the so-called Kohn-Sham orbitals, the fundamental unknown in KS-DFT is the \emph{electron density}. Indeed, many ground state material properties, e.g.   structural parameters and elastic constants, may be computed from the electron density field, and it also serves as a starting point for \emph{ab initio} descriptions of excited state phenomena. However, practical calculations using KS-DFT, especially for complex systems such as disordered alloys or materials with extended defects, can be limited by the cubic scaling computational cost of the method with respect to system size. As such, a considerable share of research has been devoted to the development of specialized solution techniques and high performance computing algorithms to overcome this bottleneck \citep{gavini2023roadmap, goedecker1999linear, banerjee2016chebyshev, banerjee2018two, motamarri2014subquadratic, lin2014siesta, dogan2023real, banerjee2021ab, banerjee2016cyclic}.

In recent years, Machine Learning (ML) has enabled the development of powerful surrogate models that can enable the prediction of the ground state electron density and related properties at a fraction of the computational cost of conventional first principles simulations \cite{lewis2021learning, jorgensen2022equivariant, zepeda2021deep, chandrasekaran2019solving, fiedler2023predicting, brockherde2017bypassing, del2023deep, schleder2019dft, kulik2022roadmap, pathrudkar2022machine, pathrudkar2024electronic, pathrudkar2024HEA}. These ML models are typically trained on KS-DFT data and use descriptors of the atomic environment to produce the electron density as the output. Recent studies have demonstrated the success of such ML models in predicting electron densities across a wide range of systems, including pure metals, alloys, insulators, organic molecules and nanostructures ~\citep{zepeda2021deep, li2025image, chandrasekaran2019solving, pathrudkar2024electronic, pathrudkar2024HEA, pathrudkar2022machine}. Moreover, these models show promising extrapolation capabilities, accurately predicting electron densities for unseen compositions in complex alloys, atomic configurations with defects and dislocations, and for chemical species beyond training \cite{li2025image, pathrudkar2024HEA, pathrudkar2024electronic}. The ML models have also been continuously improved to address several challenges, including  the enforcement of physical symmetries, incorporation of uncertainty quantification capabilities, and improvements in training and prediction efficiency \citep{jorgensen2022equivariant, teh2021machine, pathrudkar2024electronic}. 

The paradigm shift brought on by the development of ML-based electronic structure techniques has motivated the creation of extensive datasets \cite{ramakrishnan2014quantum, nandi2023multixc, pinheiro2020machine, liang2019qm} --- sometimes comprising millions of structures --- under the implicit assumption that large, exhaustive sampling is essential to achieve high predictive accuracy and generalizability \cite{faber2017prediction, smith2018less, musil2019fast}.  While this assumption has driven progress in materials informatics \cite{jain2013commentary, saal2013materials, jain2020materials}, it also raises questions about the efficiency and necessity of such data collection efforts, particularly given the computational cost and environmental footprint associated with high-throughput quantum mechanical calculations.

In this work, we examine the following fundamental but largely overlooked questions: How much of this electronic structure data is genuinely essential? Furthermore, how can such essential datasets be systematically obtained? Finally, how can the existence of such essential datasets be explained? To answer these questions, we methodically probe the redundancy in electron density data spanning molecular systems, elemental metals, and chemically complex concentrated alloys, and show that a surprisingly small subset is often sufficient to train ML models to within chemical accuracy. In the context of this work, we define redundancy as the presence of data points whose removal does not significantly impact ML model accuracy, generalizability, or derived physical quantities. Furthermore, to probe the origin of such redundancy, we explore the geometric structure of the data, and are able to identify its low intrinsic dimensionality across materials. Our findings challenge the prevailing view that predictive fidelity and generalizability require vast amounts of data and suggest that it is possible to identify minimal datasets that dramatically reduce computational cost without sacrificing accuracy. Thus our exploration of the redundancy of electronic structure data can potentially lead to a small foundational dataset for each material class, which in turn can be used for benchmarking and training foundation models.

The exploration of redundancy in chemical datasets has received prior attention in the literature \cite{li2023exploiting, li2024md, chen2024beyond}. For instance, Li et al.~\citep{li2023exploiting} have demonstrated a high degree of redundancy in datasets of scalar material properties such as the formation energy and band gaps, and have proposed an iterative strategy to reduce the size of training data. However, there has not been a systematic exploration of  redundancies in the electronic structure itself, particularly in the context of ML models. Notable prior efforts include the observation that electronic structure data for various simple systems often lie on low-dimensional manifolds, enabling reconstruction via low-rank tensors \citep{blesgen2012approximation, motamarri2016tucker} or a small number of principal component modes \citep{teh2021machine, pathrudkar2022machine}. 
Somewhat relatedly, recent contributions have also exploited redundancies in the electron density to develop targeted grid-point sampling schemes,  primarily on molecules and simple metals~\cite{focassio2023linear, feng2025efficient}, although the effectiveness of such approaches may depend sensitively on the manner in which core states are modeled.  These approaches use physical intuition based on density or descriptor-space linear independence to guide pruning, but do not incorporate standard ML-based data-pruning techniques, such as importance scoring computation and dataset coverage quantification. Thus, although previous studies have addressed redundancy in various contexts, statistically grounded one-shot pruning methods --- free from heuristics and  having data coverage guarantee --- have yet to be explored for electronic structure data, particularly for chemically complex concentrated alloys. 
The present work adopts systematic ML-based pruning techniques that are applicable to any class of materials without the need for domain knowledge and is in principle applicable to fields beyond the electron density, e.g. the electronic wavefunctions or local density of states.

The observed redundancy in electronic structure data is perhaps closely connected to Kohn’s nearsightedness principle \citep{prodan2005nearsightedness, kohn1996density}, which states that the electronic properties of a system are predominantly determined by its local environment. Indeed, this result has been used to deduce that the single particle density matrix that appears in electronic structure calculations can be very sparse \cite{benzi2007decay, benzi2013decay}, and has led to a large number of electronic structure codes that exploit it \cite{gavini2023roadmap, baer1997sparsity, suryanarayana2017nearsightedness, mohr2015accurate, shimojo2001linear, SURYANARAYANA2018288, nakata2020large, garcia2009linear, goedecker1999linear,  skylaris2005introducing}. In the context of ML-based electronic structure prediction, nearsightedness suggests a closely related but largely unexplored implication: if local atomic environments primarily determine electronic properties, then many distinct global configurations --- differing only in distant or weakly coupled regions  --- should map to nearly identical electronic descriptors. As a result, datasets constructed by exhaustively sampling configurations are expected to contain a high degree of redundancy, particularly for large or chemically complex systems. From a geometric perspective, this locality-induced redundancy implies that electronic structure data across materials occupy a low-dimensional, nonlinear manifold embedded in a much higher-dimensional descriptor space. ML models can therefore learn the essential physics efficiently from a relatively small, well-chosen subset of data. Despite its conceptual alignment with nearsightedness, this perspective has not yet been systematically examined for modern neural-network-based electronic structure models, nor has its implications for dataset prunability and intrinsic dimensionality been systematically examined.

Neural network performance generally improves with data and model size, often following a power law scaling~\citep{kaplan2020scaling, hestness2017deep}. However, beyond a certain level, marginal accuracy gains require  significantly more data~\citep{sorscher2022beyond, zhai2022scaling}. Thus, power law scaling suggests that much of the training data is redundant and can be pruned without notable performance loss~\citep{sorscher2022beyond}. Commonly used approaches for selecting a smaller, representative subset of a large dataset that preserves ML model performance are referred to as \emph{data pruning} techniques or \emph{coreset selection} methods. These techniques are often broadly categorized into the following three major approaches~\citep{tan2024data}, \emph{importance-score} based methods~\citep{paul2021deep, toneva2018empirical, tan2024data, he2024large}, \emph{coverage}-driven methods~\citep{zheng2023_CCS, sener2018coreset, xia2022moderate} and \emph{optimization}-based methods~\citep{yang2022dataset, mirzasoleiman2020coresets, killamsetty2021grad, killamsetty2021glister, killamsetty2021retrieve}.

Most existing pruning methods have been evaluated primarily on classification problems, whereas electronic structure prediction is a regression task. Therefore, identifying the degree of prunability of electronic structure data, as well as the most suitable pruning technique remain unexplored. In the present work, two state-of-the-art methods: GraNd~\citep{paul2021deep} and Coverage-Centric Coreset Selection (CCS)~\citep{zheng2023_CCS} are adopted. 
Optimization-based methods are not considered in this work because, despite their strong theoretical foundation and impressive performance, they are computationally quite expensive, and can be challenging to implement on large-scale datasets~\citep{yang2024mind}. The GraNd technique is adopted because of its intuitive approach, ease of implementation, and ability to identify important examples early in the training. GraNd, an \emph{importance-score} based method, assigns lower scores to easy-to-learn examples, which are deemed redundant and removed while pruning. This can lead to poor data coverage at high pruning factors, as only difficult-to-learn examples are retained, causing a severe performance drop \cite{zheng2023_CCS, guo2022deepcore}. CCS, a \emph{coverage}-driven method, overcomes this limitation by proposing a probability density-based coverage strategy to improve the coverage of the coreset~\citep{zheng2023_CCS}. 
Among other coverage-driven methods, $k$-center greedy~\citep{sener2018coreset} is computationally prohibitively expensive for datasets of the size considered here (e.g. data size is 160 million for the CrFeCoNi system), while Moderate Coreset~\citep{xia2022moderate} is developed for classification problems, as it selects points by their distance to class centers computed from class labels. CCS is therefore adopted in the present work due to its exceptional performance at high pruning factors and its flexibility in incorporating any importance score~\citep{zheng2023_CCS}.

Through the use of these data-pruning techniques, we explore the degree of redundancy in electronic structure data across a wide variety of material systems --- a simple metal, chemically  disordered concentrated complex alloy systems, and molecules.  We show that high degrees of redundancy appear to be a generic feature of electronic structure data: even random pruning can substantially reduce dataset size with minimal loss in predictive accuracy, while state-of-the-art coverage-based methods unveil even more dramatic levels of redundancy. Thus, a remarkably large fraction of the dataset across all the materials studied can be systematically pruned while preserving prediction accuracy and model generalizability. 
We have further quantified the quality of predictions via calculations of the ground-state energy from predicted electron densities and have found that chemically accurate predictions (i.e., energy errors less than 1 kcal/mol or 1.6 mHa/atom) are obtained even after pruning up to  $90-99\%$ of the data for the in-distribution test cases. We found that for several out-of-distribution tests pruned datasets  largely retain the generalization behavior. In particular, our results for complex disordered metallic alloys \emph{hold across composition space}, thus suggesting the potential to identify a minimal, essential dataset representative of each material class.
Additionally, we found that for the materials considered here, ML model training time on the pruned datasets can be reduced by a factor of three or more, thus enabling accelerated hyperparameter tuning and foundation model development. However, we note that the present study is based on datasets generated from independent AIMD/MD trajectories at various temperatures and compositions, with sufficiently large time intervals between consecutive snapshots. This sampling strategy minimizes structural similarity and temporal correlation within the dataset. Nevertheless, other data generation approaches may exhibit different degrees of redundancy depending on the methodology employed. 

To explain the high redundancy, we analyze the data geometry following the \textit{manifold hypothesis }\citep{Kiani_NurIPS2024}, which posits that high-dimensional data lies on or near a low-dimensional nonlinear manifold, and is supported empirically across domains \citep{MaFu2011Manifold,Carlsson2009Topology,pope2021the}. The dimensionality of the nonlinear manifold is referred to as the \emph{intrinsic dimension} (ID). The intrinsic dimension  estimation methods  quantify the manifold dimensionality using local geometric statistics, including maximum likelihood estimator (MLE)~\citep{levina2004maximum}, Two Nearest Neighbors ~\citep{facco2017estimating}, and  geometric MLE ~\citep{gomtsyan2019geometry}. 
In this work, we employ various nearest-neighbor based methods, such as Gride~\cite{denti2022gride}, MiND\textsubscript{ML}\cite{rozza2012mind}, IID~\cite{qiu2023underestimation},  and MID~\cite{qiu2023underestimation}. Our results show that the estimated IDs agree closely across nearest-neighbor estimators, and that even the most conservative estimates --- obtained through the MID method --- remain below the ambient descriptor dimensions across materials, a possible explanation for the high redundancy.

The remainder of this paper is structured as follows: Section ``\nameref{sec:results}'' presents our findings, followed by a discussion in Section ``\nameref{sec:discussion}'', and methodology details in Section ``\nameref{sec:Methods}''.

\section*{Results}
\label{sec:results}
In this section, we quantify the redundancy in electronic structure data, as well as the intrinsic dimensionality, for four different types of materials. We present the error in the ML model for various pruning factors to identify the fraction of data that can be pruned without significantly compromising the model's performance and generalizability, and further quantify this through the accuracy of the energy computed from the ML-predicted electron density. 

The four material systems considered here are pure aluminum, bulk water, and  two different alloy systems across composition space --- SiGeSn and CrFeCoNi. The first two serve as representative examples of simple metals and molecular systems, respectively, while the latter two illustrate technologically important chemically complex concentrated alloys. Notably our results for the alloy systems (discussed below) are particularly striking since the ML predictions are carried out over the entire composition space of each alloy \cite{pathrudkar2024electronic}.

The dataset consists of snapshots of KS-DFT simulation taken at intermittent time steps, with each snapshot capturing the atomic coordinates, species information, and the resulting electron density over the simulation cell. Given a grid point in the simulation domain, the ML model provides a map to the electron density at that point, from the nearby atomic configuration. Details of the KS-DFT simulations and the ML map are provided in Section ``\nameref{sec:Methods}''. 
\begin{figure}[hb!]
    \centering
    \includegraphics[width=\linewidth]{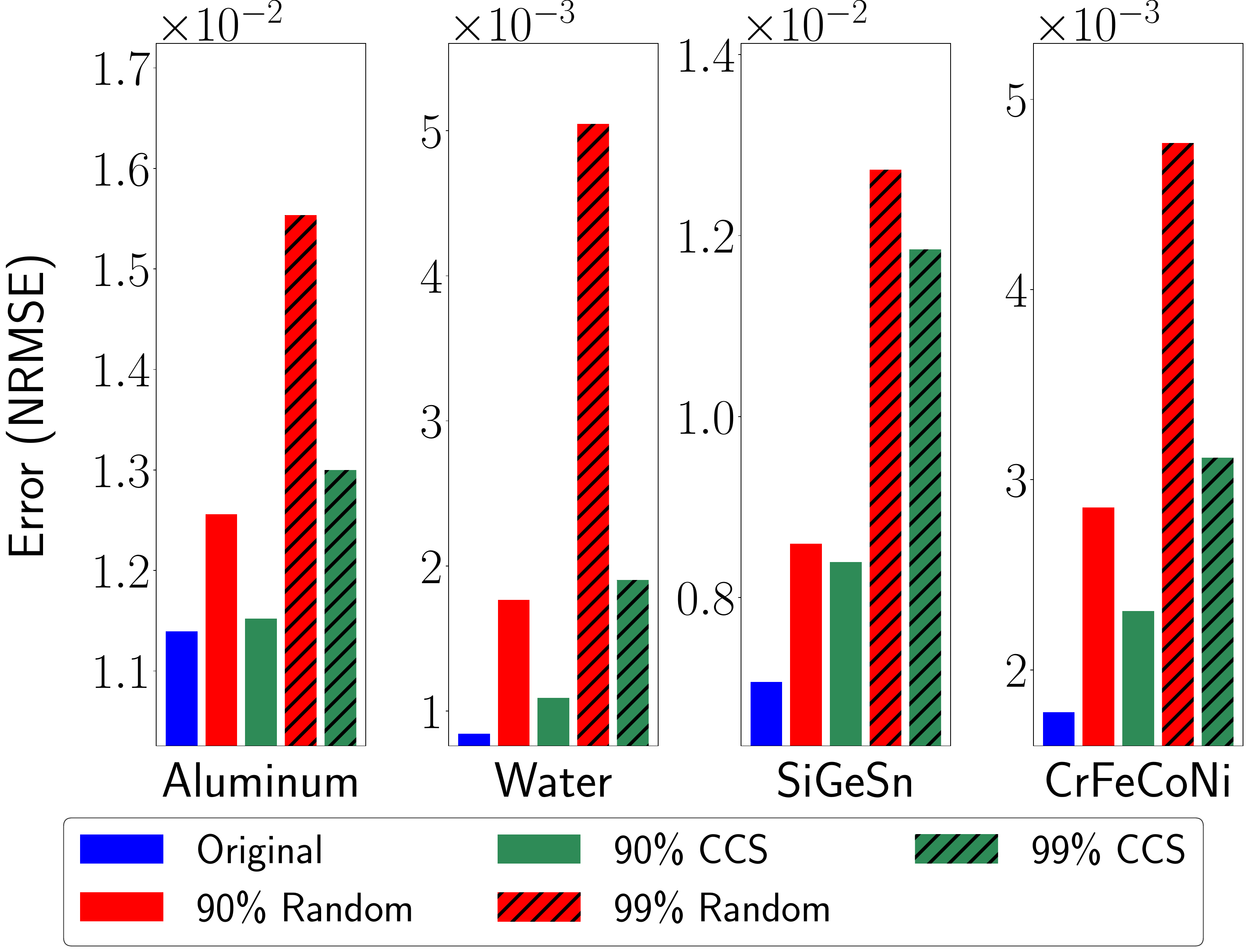} 
    \caption{Error in ML-predicted electron density for the original dataset, 90\% and 99\% randomly, and CCS-based pruned datasets. Each ML model was trained three times and the mean error is reported.}  
    \label{fig:consolidated_nrmse} 
\end{figure}

\begin{figure}[hb!]
    \centering
    \includegraphics[width=\linewidth]{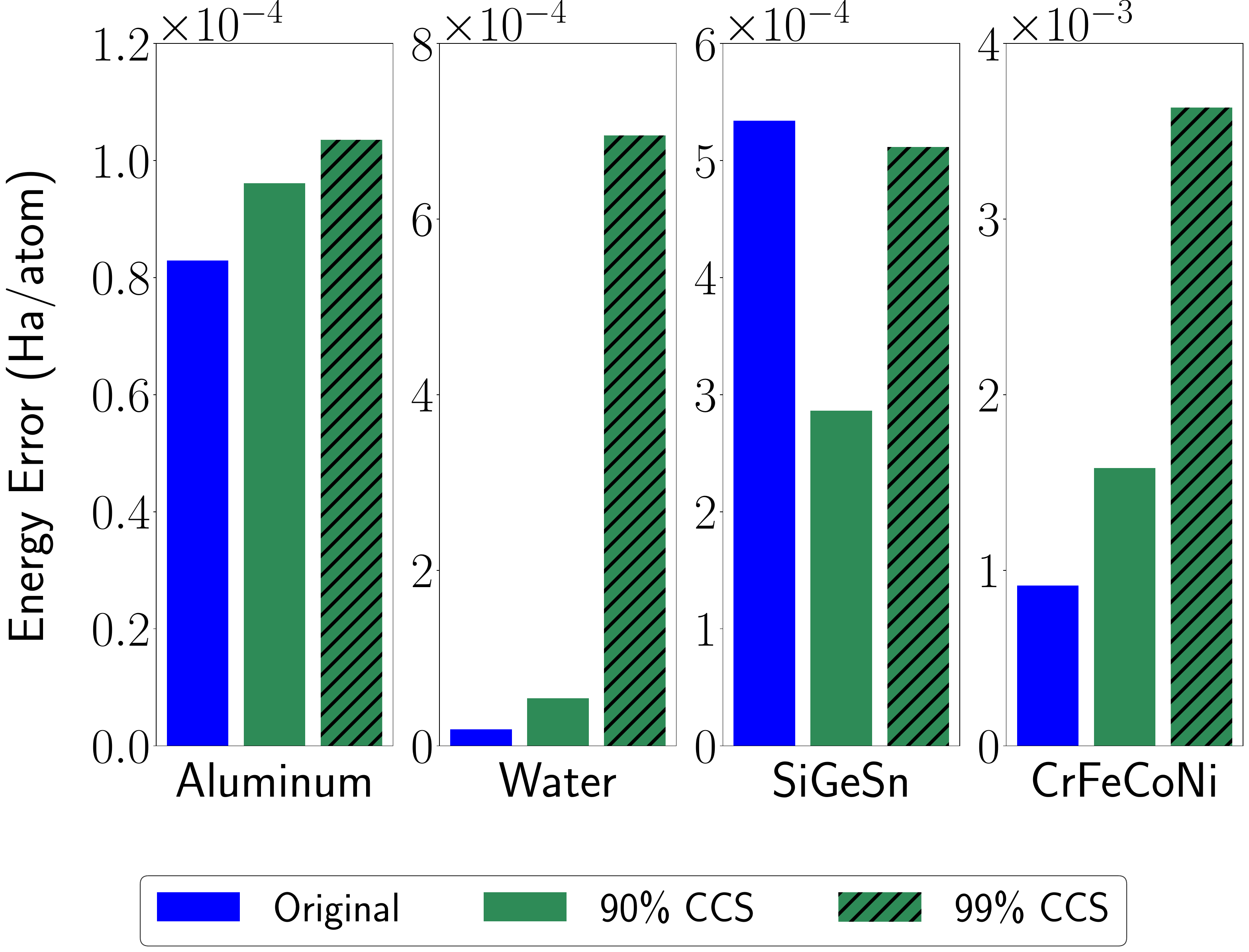} 
    \caption{Error in energy with respect to KS-DFT, as obtained from the ML-predicted electron density for the original dataset, 90\% CCS, and  99\% CCS-based pruned datasets. The electron density prediction from one of the ML models was postprocessed in each case.}  
    \label{fig:consolidated_energy} 
\end{figure}

\begin{figure*}[h]
    \centering
    \includegraphics[width=\linewidth]{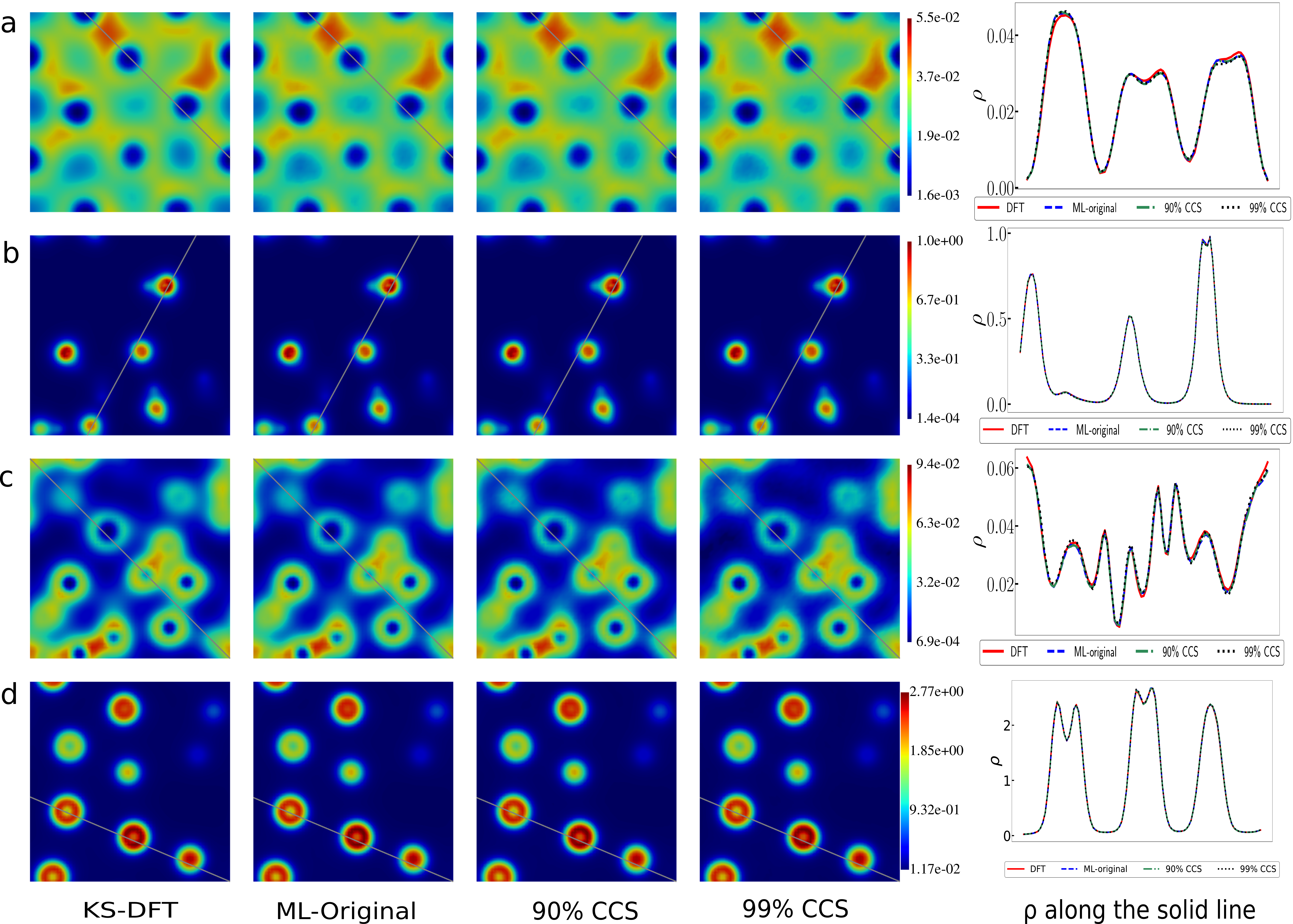}
        \caption{Two-dimensional (2D) slices showing electron density obtained by KS-DFT, ML model trained on the original dataset, 90\% CCS, and 99\% CCS-based pruned dataset for (a) Aluminum at 1500K (b) Water at 600K (c) SiGeSn at 2400K (d) CrFeCoNi for 5000K. The electron density($\rho$) along the solid line is also compared with the KS-DFT and the ML-predicted electron densities show remarkable agreement with the KS-DFT, true for all three ML models across four systems. The unit of electron density is $\text{Bohr}^{-3}$ in atomic units.}
    \label{fig:consolidated-rho-plots}
\end{figure*}

\subsection*{Redundancy in diverse material datasets}

To explore the redundancy, pruning is performed by removing a subset of the grid points of the original dataset. We explore three approaches for data pruning:  random grid-point-wise pruning, GraNd score-based pruning, and CCS-based pruning. In random grid-point-wise pruning, grid points from the original dataset are removed randomly. The importance score (i.e. GraNd score) is computed for each datapoint and used for the GraNd and CCS approaches. Upon training ML models on the pruned datasets, the errors in the predicted electron density on a testing dataset are computed.  We report the commonly used normalized root mean squared error (NRMSE)~\citep{teh2021machine} as an error metric, defined in the Supplementary Section~\ref{sm:metrics}). 

Figure~\ref{fig:consolidated_nrmse} highlights the redundancy in electronic structure data. For all four systems, $90\%$ of the data can be pruned through the CCS-method, without compromising accuracy in the predicted electron density. Even $99\%$ redundancy can be identified by the CCS method: at such a high pruning factor, the errors increase as expected, but the predicted electron densities still result in chemically accurate energies (i.e., errors less than $1$ kcal/mole or $1.6$ mHa/atom) across  all systems except the CrFeCoNi system, as shown in Figure~\ref{fig:consolidated_energy}. For CrFeCoNi, the 99\% pruning slightly exceeds the chemical-accuracy threshold. 
The total energy of the system is calculated from the ML-predicted electron density, via postprocessing \cite{pathrudkar2024HEA}. Notably, while the errors in the electron density and the corresponding energy errors are correlated, the one-to-one relationship between them can break down when the density errors and hence the energy errors are themselves low. This situation can arise for low pruning percentages across all models, or for very high pruning percentages, for some models (e.g. those involving the alloy systems considered in this study). An additional issue to keep in mind is that in general, the CrFeCoNi alloy system is more difficult to predict accurately through ML models, because of the involvement of harder pseudopotentials, the inclusion of semi-core electrons for transition metals in the KS-DFT calculations, and the large amount of compositional variation \cite{pathrudkar2024HEA}. Thus, the overall ML predictions are slightly less accurate  for this system, as opposed to the other systems studied here. However, our key finding, in this regard is that there is still a high degree of redundancy in the CrFeCoNi electronic structure data, and the elimination of this redundancy does not reduce ML model prediction accuracy further.

Pruning also does not introduce many significant locally large pointwise errors with respect to the ML-original model. We assess this through computing the relative $L_\infty$ norm of the error and the full pointwise error distributions before and after pruning. Further details can be found in the Supplementary Section~\ref{sm:robustness}. (see Supplementary Figures~\ref{fig:error-linf} and \ref{fig:error-comparison}).

The ML models not only capture the electron density fields but also smooth variations in them. The errors in smoothness exhibit a trend similar to that of the electron density. Further details are provided in the Supplementary Section~\ref{sm:H1}. (see Supplementary Figure~\ref{fig:consolidated-H1}).

We compare the electron density fields obtained from KS-DFT with those predicted by ML models trained on three datasets --- the original dataset, the $90\%$, and  $99\%$ CCS-based pruned dataset --- for all four material systems, as shown in Figure~\ref{fig:consolidated-rho-plots}.
ML models trained on the pruned datasets show excellent agreement with those trained on the original dataset and with KS-DFT calculations across all systems, confirming the high degree of near-lossless pruning.

\begin{figure}[t]
    \centering
    \includegraphics[width=\linewidth]{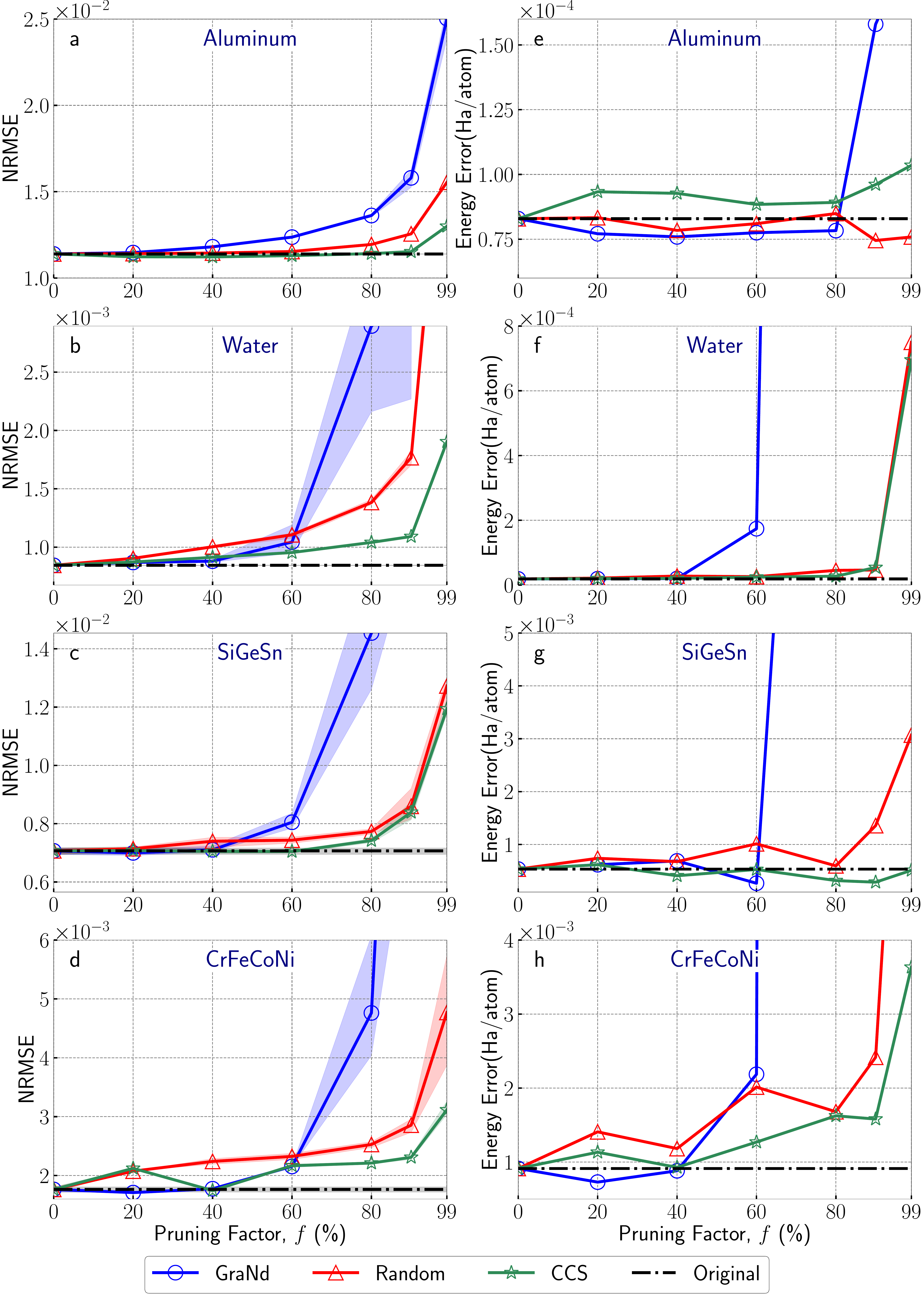}
    \caption{\textbf{Comparison of pruning strategies across material systems.}(a-d) The error in the electron density prediction for various pruning factors for all systems. The shaded region represents the range (maximum to minimum) of three ML models around the mean, shown by a solid line. (e-h)The error in the total energy obtained via postprocessing from the predicted electron density for various pruning factors for all systems}
    \label{fig:consolidated-all-nrmse-energy}
\end{figure}

\subsection*{Comparative assessment of pruning strategies}

In the following, we study the performance of three prominent pruning techniques --- random, GraNd and CCS --- for various pruning factors. ML models for each pruning factor and pruning strategy were trained three times, and the results are reported accordingly. The errors (NRMSE) in electron density predictions for these three pruning approaches are plotted against various pruning factors across all four materials systems in Figure~\ref{fig:consolidated-all-nrmse-energy}(a-d). The ML models exhibit minor random fluctuations in their predictions, which is expected due to inherent randomness in the dataset and the model. All three methods perform well at smaller pruning factors across materials, maintaining the accuracy of the original dataset. However, the GraNd method fails catastrophically, with errors rising sharply beyond 40–60\% pruning across different materials. Random pruning shows a steady increase in error with pruning factor,  preserving the accuracy of the original dataset up to 40–80\% and exhibiting rapid error growth beyond that. In contrast, the CCS method outperforms both GraNd and random pruning, maintaining the accuracy of the original dataset up to 80–90\% pruning and showing a slower increase in error than random beyond 90\%. Remarkably, CCS continues to deliver  reasonably accurate results at 99\% pruning across all materials, with only a modest increase in error compared to the original dataset. 
The corresponding errors in the total energy of the system, as obtained from post-processing the ML-predicted electron density, are plotted against various pruning factors in Figure~\ref{fig:consolidated-all-nrmse-energy}(e-h). For each pruning factor and strategy, the predicted electron density from one of the three trained ML models was post-processed. All three methods predict the energy nearly as accurately as the original dataset for smaller pruning factors, up to 40\%. However, for  aluminum all three methods are chemically accurate until 99\% pruning, despite a sharp jump in error for the GraNd method beyond 80\%. 
The GraNd method retains chemical accuracy up to 40\% pruning for all materials; however, at higher pruning factors, its energy error increases very rapidly for all materials.
Random pruning retains chemical accuracy up to 40\% pruning for all materials. It maintains chemical accuracy for simple materials, such as aluminum and water, up to 99\%, whereas for complex alloys, the limit decreases to 90\% for ternary systems and  40\% for quaternary systems.
The CCS method consistently outperforms the other two approaches, nearly preserving the accuracy of the original dataset up to 90\% pruning across all materials. Remarkably, it continues to yield chemically accurate energy predictions even at 99\% pruning, except CrFeCoNi, where CCS achieves chemically accurate level of predictions until 90\% pruning. Notably, unlike the pure materials considered here --- bulk water and aluminum metal --- the errors in energies computed for complex alloy systems appear to exhibit random fluctuations with respect to pruning factor for high pruning factors. Thus, for such cases, the errors in the energy, although related to the electron density errors, are not necessarily monotonic with respect to pruning factors. This is consistent with our recent observations for ML-based electronic structure prediction of complex quaternary alloys, whereupon the NRMSE in electron density errors alone was found to be not sufficient in completely determining the energy errors \citep{pathrudkar2024HEA}. Nevertheless, {for the in-distribution tests,} we observe that for CCS-based pruning, {the energy predictions for the quaternary CrFeCoNi system remain within chemical accuracy even at very high pruning factors, whereas aluminum, water, and SiGeSn maintains energy errors substantially below the chemical-accuracy threshold. } \\ 
Energy accuracy depends on both the predicted density and the post-processing pipeline. To isolate ML model accuracy from post-processing effects, we evaluated the predicted density directly using mean, maximum, and smoothness errors. All three remain low even at high CCS pruning fractions, confirming that accurate energies stem from accurate density predictions rather than from a robust post-processing pipeline.
Further details on these pruning results are provided in the Supplementary Section~\ref{sm:additional-results}. (see Supplementary Figures ~\ref{fig:water-consolidated-full}, \ref{fig:sigesn-consolidated-full}, and \ref{fig:crfeconi-consolidated-full} for the unscaled version of the Figure~\ref{fig:consolidated-all-nrmse-energy} for Water, SiGeSn and CrFeCoNi respectively. Also see Supplementary Figures~\ref{fig:al-iso-rho} -- ~\ref{fig:crfeconi-iso-error} for the iso-surface plots of the density and error fields through the original dataset, 90\%, and 99\% CCS-based pruned datasets for all systems studied here). 

\subsection*{Generalizability of the ML models}\label{sec:Generalization}
\color{black}

\begin{figure*}[t]
    \centering
    \includegraphics[width=\textwidth]{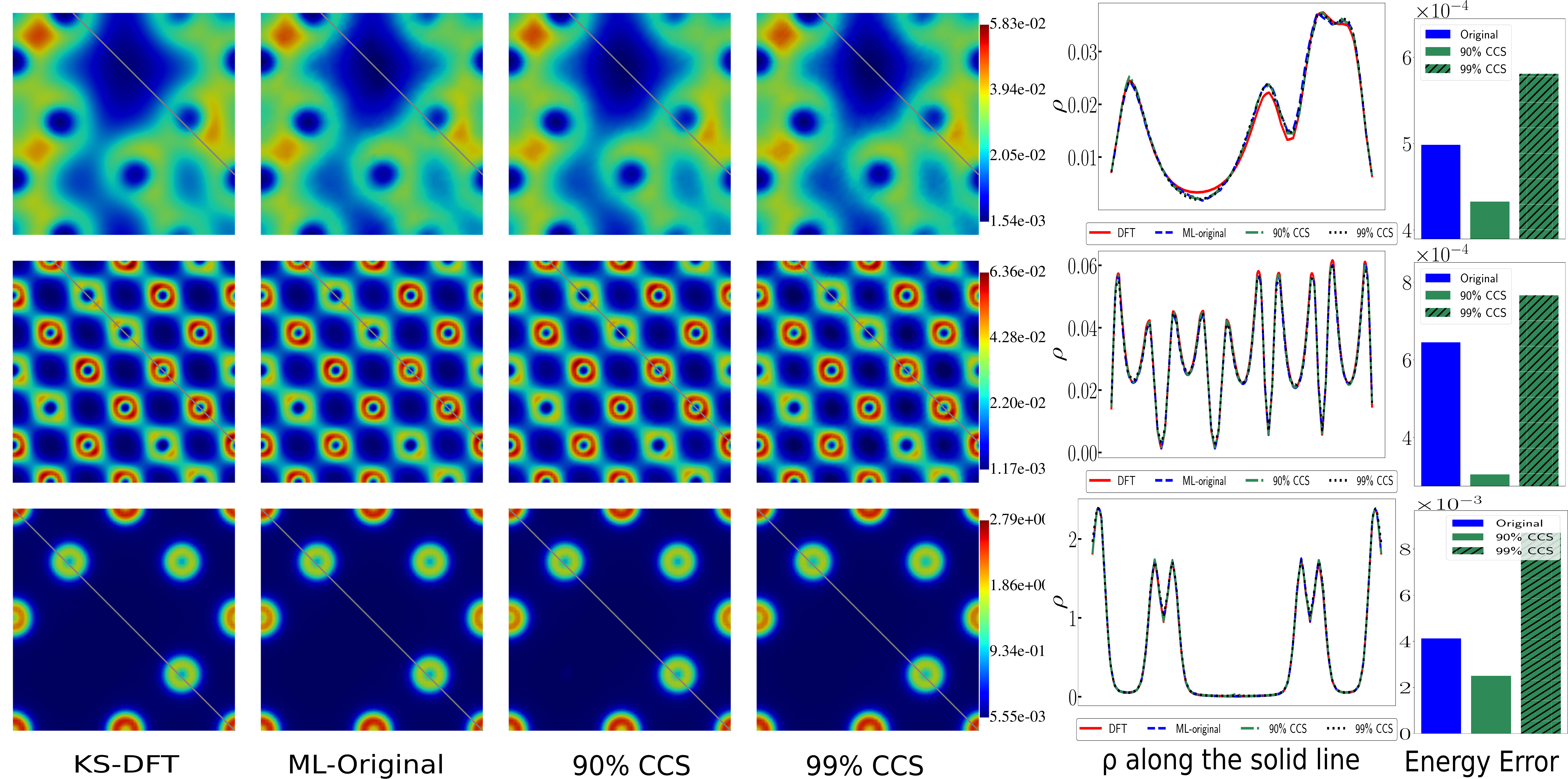}
   \caption{\textbf{Model generalizability to unseen configurations:} 2D slices of electron density obtained by KS-DFT, ML model trained on the  original dataset, 90\% CCS, and 99\% CCS-based pruned datasets. The electron density($\rho$) along the solid line is also compared with KS-DFT, showing remarkable agreement. The energy errors(Ha/atom) are also shown for each case. \textit{(Top)} Mono-vacancy defect for 32-atoms aluminum systems. \textit{(Middle)} 216-atoms SiGeSn checkerboard systems \textit{(Bottom)} Di-vacancy defect for 32-atoms CrFeCoNi systems. The unit of electron density is $\text{Bohr}^{-3}$ in atomic units.}
    \label{fig:generalizability}
\end{figure*}

\begin{table*}[htbp]
\centering
\begin{tabular}{llccc}
\toprule
 {System} &  {Generalization test} &  {Original} &  {90\% CCS} &  {99\% CCS} \\
\midrule
\multirow{2}{*}{Aluminum}
& Structural defects (mono-vacancies) & below & below & below \\
& Temperature extrapolation (2100\,K) & below & below & below \\
\midrule
\multirow{2}{*}{SiGeSn}
& Checkerboard systems (64 and 216 atoms)   & below & below & below \\
& Size extrapolation (216 atoms)      & below & below & near \\
\midrule
\multirow{2}{*}{CrFeCoNi}
& Structural defects (mono-/di-vacancies) & near & near & above \\
& Unseen compositions                 & below & below & above \\
\bottomrule
\end{tabular}
\caption{ {Summary of energy error for generalization tests relative to the chemical-accuracy threshold ($\sim1.6$\,milli Ha/atom or 1 kcal/mole) for models trained on the original, 90\% CCS-, and 99\% CCS-pruned datasets. ``below'', ``near'', and ``above'' indicate that the energy error is below the chemical-accuracy threshold, above but in its immediate vicinity(around $2 - 3$\,milli Ha/atom), and above the threshold, respectively. Detailed results are provided in the Supplementary Section~\ref{sm:generalization}.}}
\label{tab:generalization}
\end{table*}

The limited generalization capability of ML models beyond their training data remains a fundamental issue of the field. In particular, the generalization capability of ML models developed using pruned data is yet unexplored. 
To address this issue, we evaluate the generalization capabilities of the ML models developed here, by assessing their performance on a diverse set of systems that differ significantly from, and were not included in, the training data. These tests span four categories: (i) structural defects, namely mono- and di-vacancies (aluminum, CrFeCoNi) and handcrafted checkerboard systems featuring species segregation (SiGeSn) (ii) unseen compositions (CrFeCoNi) (iii) extrapolation to temperatures beyond the training range (aluminum) and (iv) extrapolation to system sizes larger than those used in training (SiGeSn).\\
The ML-predicted densities via CCS ($90$\% and $99$\% pruning) exhibit remarkable agreement with KS-DFT across all generalization cases, as illustrated in Figure~\ref{fig:generalizability} and in Supplementary Figure~\ref{fig:generalization-extra}.
We further compute each system's energy from the ML-predicted electron densities, for the original, 90\% CCS, and 99\% CCS pruned datasets. The predicted energy is compared against the chemical accuracy threshold in the Table~\ref{tab:generalization} and the detailed results are provided in the Supplementary Section~\ref{sm:generalization}. We observe two trends. First, we find that the generalization capability of  90\% CCS-pruned datasets remains unaffected across all four categories. Second, at 99\% pruning, the result depends on the material: generalization is fully retained for aluminum, largely retained for SiGeSn, and degrades only for CrFeCoNi, where the energy errors exceed the chemical-accuracy threshold but the error is still relatively low ($\mathcal{O}(10^{-3})$  Ha/atom). For the CrFeCoNi defect configurations, however, even the original model lies near the chemical-accuracy threshold --- a reflection of the intrinsic difficulty of learning this system rather than of pruning. \\
Additional results confirming the aforementioned conclusions regarding generalization ability are provided in the Supplementary Section~\ref{sm:generalization}. (see Supplementary Figures~\ref{fig:al-defects} and \ref{fig:aluminum-2100k} for Aluminum, Supplementary Figures~\ref{fig:sigesn-handcrafted} and \ref{fig:sigesn-216atoms} for SiGeSn, Supplementary Figures~\ref{fig:crfeconi-defects} -- \ref{fig:crfeconi-unseen} for CrFeCoNi).

\subsection*{Intrinsic dimension of electronic structure data: a route to explaining high redundancy}\label{sec:Low_id}
\begin{figure}[!h]
    \centering
\includegraphics[width=0.9\linewidth]{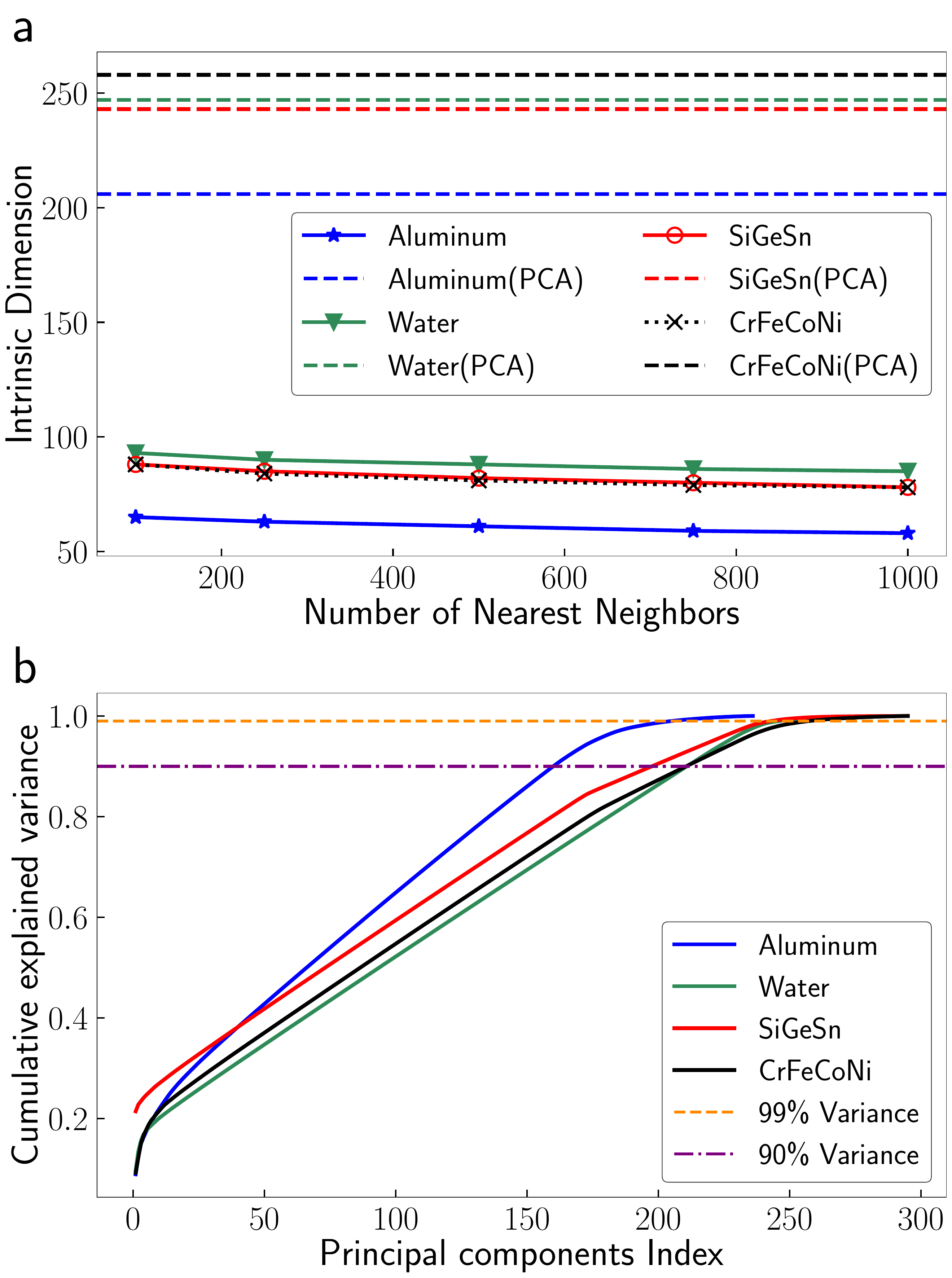}
   \caption{\textbf{Intrinsic-dimensionality of electronic-structure descriptors.} (a) Intrinsic dimension (ID) estimates for four materials using MLE with varying neighbor counts on 2 million randomly sampled points; PCA-based ID for 99\% variance are shown. (b) Cumulative variance for PCA.}
    \label{fig:lowid}
\end{figure}
In the following, we estimate the intrinsic dimension of the geometric structure of the data embedded in the high-dimensional descriptor space. The dimensions of descriptors are $236$ for Aluminum, and $295$ for SiGeSn, CrFeCoNi, and water, in our model. To this end, first, we perform the principal component analysis (PCA) to identify the number of dimensions required to capture $99\%$ variance in the data. The estimated dimensions through PCA for Aluminum, SiGeSn, CrFeCoNi, and water are $206, 243, 259$, and $247$ respectively, as shown in Figure~\ref{fig:lowid}(a,b). Therefore, PCA does not show significant reductions in dimension, indicating that the true geometric structure of the data might be nonlinear or genuinely high-dimensional. In view of this, we next employ the maximum likelihood estimator (MLE) method~\citep{levina2004maximum}, which utilizes nearest neighborhood distances to compute the intrinsic dimension. The estimated intrinsic dimensions using the MLE estimator for various number of nearest neighbors are approximately between $65 - 90$, as shown in Figure~\ref{fig:lowid}(a), for different materials. This shows that the intrinsic dimension of each material class is much smaller than the descriptor dimension. It also suggests that the data has a nonlinear geometric structure, with a much lower intrinsic dimension than what linear methods like PCA identify. Other nearest neighbor based methods, such as Gride~\cite{denti2022gride}, MiND\textsubscript{ML}~\cite{rozza2012mind}, iterative-based intrinsic Dimension(IID) also show similar intrinsic dimensions (see Supplementary Materials). However, as it has been mentioned in the literature, the intrinsic dimension may be underestimated because of the nonlinearity in the manifold~\citep{qiu2023underestimation}. So, we employed modified intrinsic dimension(MID)~\citep{qiu2023underestimation} estimator and have reported the IDs through various estimators in the Supplementary Materials. Among all these, MID provides the largest value(i.e. most conservative estimate) which is still very low than the ambient dimension of the dataset(see Supplementary Section~\ref{sm:id} for further details). Since the data lies on a much lower dimensional nonlinear manifold, it facilitates efficient learning from a smaller subset of data, thus explaining the high degree of data redundancy in each material class. 

\section*{Discussion}\label{sec:discussion}

We have explored the hidden redundancy in electronic structure datasets for diverse material systems. To the best of our knowledge, this marks the first exploration of redundancy in electronic structure datasets through systematic pruning approaches. In addition to random pruning, we have employed two ML-based pruning techniques, namely GraNd \citep{paul2021deep} and CCS \citep{zheng2023_CCS}. 
GraNd based pruning selects a subset of data having the highest importance scores. The CCS approach selects a subset that ensures comprehensive coverage of the dataset based on a probability density-based coverage strategy. 
These two ML-based pruning methods are compared against random grid-point-wise pruning --- with the goal of revealing the aforementioned redundancy, and helping select the minimal essential dataset without compromising accuracy. Overall, we are able to identify a large fraction of the electronic structure data that is redundant and which can be eliminated from ML training, without compromising the accuracy or the smoothness of the ML predicted fields. 
Random grid-point-wise pruning turns out to be an effective strategy, aligning with findings in the computer vision literature. Although its error rate increases steadily, it can outperform several ML-based pruning methods at higher pruning factors ~\citep{guo2022deepcore, zheng2023_CCS}. 

We have also verified that the observed redundancy is not attributed to temporal correlation in the underlying MD trajectories. The training snapshots are sampled at intervals of at least 100\,fs, which allow us to generate uncorrelated (or structurally dissimilar) snapshots along the MD trajectories, and snapshots at different temperatures and compositions originate from entirely independent simulations. 

More details on this aspect are provided in Supplementary Section~\ref{sm:temporal}.
For all four material systems, at lower pruning rates, GraNd captures the redundancy of the dataset effectively, closely matching or outperforming the original dataset. However, at higher pruning rates, GraNd’s performance drops catastrophically. This occurs because, at large pruning factors, most of the easy-to-learn data points are removed, significantly altering the data distribution and leading to poor data coverage. 

We found that pruning via CCS consistently outperforms the other two methods, particularly at higher pruning factors. It can systematically identify an order of magnitude smaller subset of the dataset that maintains the performance of the original dataset (i.e., it is practically  near-\emph{lossless}) for all materials studied here. We have also verified this trend via computing the system's energy from the ML-predicted electron densities. Furthermore, pruned datasets which are two orders of magnitude smaller than the original (i.e., $99\%$ pruning fraction) also demonstrate surprisingly high accuracy, and result in chemically accurate energy predictions  for Aluminum, SiGeSn, and water. However, for CrFeCoNi system the $99\%$ pruning fraction shows slightly higher energy error than chemical accuracy level.

We demonstrate that $90$\% and $99$\% CCS-based pruned datasets retain significant generalization capability of the  ML-original model, through tests on out-of-distribution data including structural defects (mono- and di-vacancies), ordered handcrafted systems, extrapolation to higher temperature or larger system size, and unseen compositions. We also compute the system's energy from the ML-predicted electron densities. Across all generalization tests, the predicted energies for aluminum and SiGeSn maintain chemical accuracy up to 99\% pruning, whereas for CrFeCoNi, chemical or near-chemical accuracy is achieved at 90\% pruning, but at 99\% pruning the energy errors slightly exceed the chemical-accuracy threshold.

CCS retains grid points in regions of high and rapidly varying electron density --- e.g. near atomic cores and bonding regions --- while predominantly pruning the regions having low variation in density. For aluminum the variation of density is less rapid and the retained points appear to be more uniform. In all cases, a small number of scattered interstitial points are also retained, consistent with its stratified sampling across the importance-score. This selection pattern of grid points emerges purely from the ML-based importance scores and data-coverage criteria, without any physics-informed input. Further details are provided in the Supplementary Section~\ref{sm:viz-pruning}.  Importantly, comparable prunability is also observed for water and CrFeCoNi, whose harder pseudopotentials require fine, convergence-tested real-space meshes, indicating that the observed redundancy is unlikely to arise from grid oversampling alone.

The high redundancy across materials suggests an underlying physical origin. We attribute this to the fact that the data lie on a low-dimensional nonlinear manifold embedded in high-dimensional descriptor space, consistent with the predominance of local atomic environments in determining electronic properties, as suggested by the nearsightedness principle \cite{prodan2005nearsightedness}.
Consequently, many grid-points provide highly overlapping information about the same manifold geometry rather than introducing new or independent geometric information.
Thus, the ML models need to learn only the structure of the low-dimensional nonlinear manifold in each material class, rather than the entire high-dimensional descriptor space, which may explain the high degree of near-lossless pruning. We emphasize, however, that we only qualitatively indicate that a low intrinsic dimension facilitates a high degree of pruning; we make no claim of establishing a precise quantitative relationship between intrinsic dimension and the degree of pruning, and a formal investigation of this relationship is beyond the scope of the present work.

While the present results demonstrate substantial redundancy in the datasets considered, establishing the universality of this behavior requires further investigation across a broader range of materials including systems with highly inhomogeneous electron density distributions (such as materials that possess regions of extreme charge concentration adjacent to regions of severe charge depletion), and systems with strong electron localization along with complex crystal structures, such as those found in cuprates (which often feature dynamic charge stripes and fluctuating domains), doped perovskite manganites, and heavy-fermion materials (e.g., cerium or uranium alloys). A systematic investigation of ML-based electron density predictions for the aforementioned systems, along with electronic structure data pruning, remains the scope of future work. At the same time, we believe that the systems chosen here, including simple metals, molecules, and complex alloy systems, all represent the type of systems that are of interest to typical computational materials practitioners, thus underscoring the relevance of the present findings.
In addition to identifying the essential subset of the data, pruning can also accelerate ML model training due to the reduced size of the subset. For the 90\% and the 99\% CCS-pruned datasets we find  speedups of roughly $2\times$ - $5\times$ and $3\times$ - $6\times$ respectively for various materials. Further details are provided in the Supplementary Section~\ref{sm:TimingComparison}.
Thus, such pruned datasets can be used to accelerate hyperparameter tuning. In turn, these advancements can be critical in developing ML models that work across chemical space.   Data pruning might also be utilized for adaptive dataset generation. One possible direction is monitoring the novelty of local atomic environments to guide sampling, continuing AIMD simulations only until sufficient coverage of the configuration space is achieved. Another possible direction is a multi-fidelity workflow, where inexpensive electronic structure calculations enable snapshot-level pruning before performing computationally expensive KS-DFT calculations only on the selected configurations.

\section*{Methods}\label{sec:Methods}
The electron density data needed for training, validation, and testing of the ML models was produced via Kohn-Sham Density Functional Theory (KS-DFT) calculations \cite{kohn1965self, Martin_ES} as implemented in the finite-difference based SPARC code \cite{xu2021sparc,xu2020m, ghosh2017sparc}. Norm conserving (ONCV) pseudopotentials \cite{hamann2013optimized, schlipf2015optimization, van2018pseudodojo} and GGA PBE exchange-correlation \cite{perdew1996generalized} were employed. Prior to using the pseudopotentials for electronic structure data generation, tests were performed to verify that  lattice parameters and other physical quantities predicted were consistent with the literature and that the energies and forces were sufficiently converged with respect to the mesh fineness.  The mesh spacing chosen for the aluminum, water, SiGeSn and CrFeCoNi systems were $0.25$, $0.25$, $0.40$ and $0.20$ Bohr respectively. A tolerance of $10^{-6}$ was chosen for the self-consistent field (SCF) iteration convergence, and the Periodic-Pulay scheme \citep{banerjee2016periodic} was employed for convergence acceleration. The KS-DFT calculations for ML data generation only sampled the gamma point in reciprocal space, as is common in the simulation of large condensed matter systems.

While the aforementioned calculations provided the electron densities for a given atomic configuration, additional techniques were employed to efficiently generate a large sample set of atomic configurations. For the aluminum and SiGeSn alloy systems, \textit{Ab Initio} Molecular Dynamics (AIMD) simulations were employed, and the electron density data needed for the ML models was extracted from AIMD ``snapshots'' at regular simulation intervals \cite{pathrudkar2024electronic, zepeda2021deep}. A standard Nos\'e Hoover thermostat \cite{evans1985nose} with temperatures between $298$ K and $2100$ K was employed for the aluminum systems, while a temperature of $2400$ K was employed for the SiGeSn systems. Fermi-Dirac electronic smearing at a fixed electronic temperature of $631.554$ K was used. As pointed out in our recent contributions, use of thermalized systems to sample the configuration space leads to more generalizable ML models of the electronic structure \cite{pathrudkar2024electronic, pathrudkar2024HEA}, due to better sampling of the input features space.  

Although AIMD simulations are convenient for data generation, they are  computationally costly. In order to reduce the data generation expense for the water and CrFeCoNi alloy systems, atomic configurations were generated through  classical Molecular Dynamics (MD) simulations  employing machine learning based interatomic potentials \cite{pathrudkar2024HEA}. No accuracy is sacrificed by this method choice since the electron densities corresponding to each configuration  are still produced from high-fidelity KS-DFT calculations. At the same time, by running these classical MD simulations at elevated temperatures and over long times, a wide variety of the atomic configurations can be conveniently generated by sampling the resulting MD trajectories. We employed the Materials Graph Library (MatGL) Python package \cite{chen2021learning, chen2022universal}, which includes the Materials 3-body Graph Network (M3GNet) --- a universal machine-learned interatomic potential --- for these purposes. The NVT ensemble with  Nos\'e Hoover thermostat was employed for these MD runs as well. The CrFeCoNi  alloy data was collected from high-temperature simulations between $2500$ K and $7000$ K, while the water data was obtained from simulations with temperatures between $300$ K and $600$ K.

The timestep for both  AIMD and MD simulations was set to $1$ femtosecond.
After a brief simulation period to allow the system to thermally equilibrate, atomic configurations (and also the corresponding electron densities, in case of AIMD) were sampled from the simulation trajectories. Additionally, a number of different initial configurations were used to seed these simulations, in order to allow a larger variety in atomic configurations to be sampled.
\subsection*{Machine learning map for electron density prediction}\label{sub-sec:ml-mapping}
ML approaches for predicting electron densities use descriptors that encode the atomic configuration of a local spatial neighborhood as input. Effective descriptors include interatomic distances and angles, as well as the Superposition of Atomic Densities (SAD), both of which have been successful in capturing local atomic environments \citep{chandrasekaran2019solving, pathrudkar2024electronic, pathrudkar2024HEA, li2025image}. 
The output of these ML models, the electron density, is commonly represented in one of two ways: (1) as sums of atom-centered basis functions \citep{grisafi2018transferable, fabrizio2019electron}, which offer computational efficiency but may struggle with complex densities, and (2) as grid-based predictions \citep{zepeda2021deep, fiedler2023predicting}, which provide higher accuracy but require significant computational resources due to fine-mesh evaluations.

The present ML model predicts the ground state electron density values at a set of grid points  within a computational domain, based on the coordinates and atomic numbers of the atoms. The mapping is computed in two steps: \textit{First}, atomic neighborhood descriptors are derived for each grid point using the atomic coordinates and species information. \textit{Second}, a neural network is employed to map these descriptors into the corresponding electron density at each grid point.

As proposed in our recent contribution ~\citep{pathrudkar2024HEA}, we adopt the position vectors of atoms in body-attached local frames and the atomic species as the descriptors. This formulation ensures that the descriptors are invariant with respect to rotations, translations and species permutations of the system. Moreover, as shown in ~\citep{pathrudkar2024HEA} the descriptors maintain approximately the same descriptor-vector size even if the number of species in the system increases, thus ensuring efficient representation of atomic neighborhoods across chemical space.
The local frame of reference for calculating the descriptors is obtained using Principal Component Analysis (PCA) of an atomic neighborhood consisting of $M$ atoms. 
The components of the position vector ($\textbf{r}_j$) of the $j$-th atom in the local reference frame are denoted by $(x, y, z)$. 
The species information is encoded by using the atomic number of the species. The atomic number of the $j$-th atom is denoted as  $Z_j$. Thus the descriptors for $i$-th grid point ($\mathcal{D}_i \in \mathbb{R}^{5M}$) are given as,
\begin{equation}
    \mathcal{D}_i = \bigg\{ Z_j ,   \| \textbf{r}_j \|,  \frac{x}{\| \textbf{r}_j \|},   \frac{y}{\| \textbf{r}_j \|},  \frac{z}{\| \textbf{r}_j \|} \bigg\}_{j=1,\cdots,M}
    \label{eq:descriptor}
\end{equation}

\subsection*{Pruning strategies}
We briefly recapitulate the pruning strategies used in this work in addition to the random pruning: first, the GraNd score-based pruning \cite{paul2021deep,sorscher2022beyond}, followed by a Coverage-centric Coreset Selection (CCS) method \cite{zheng2023_CCS} applied to the GraNd scores. \\

{In the following, we discuss \textbf{random pruning}.} In the present work, a group of grid points from all KS-DFT snap-shots are randomly selected and pruned away. Note that grid points are randomly pruned instead of individual snap-shots. \\

{In the following, we discuss \textbf{{GraNd} score based pruning.}} We consider a supervised regression setting, where the training set is $S=(x_i, y_i)_{i=1}^N$, drawn i.i.d. from an underlying distribution $P$. Here, $x_i \in \mathbb{R}^d$ and $y_i \in \mathbb{R}^+$ denote the descriptors at $i$-th grid point and the electronic charge density at that grid point, respectively. Furthermore, $d$ is the dimension of the atomic-neighborhood descriptors. For a fixed neural network, let the $\textbf{w}$ be the weights and $l$ be the loss of the neural network at a given step. Let $\textbf{w}_0$, $\textbf{w}_1$, ... , $\textbf{w}_T$ be iteratively  updated weights of the neural network, where, for some sequence of minibatches $S_0$, $S_1$, ..., $S_{T-1}$ $\subseteq S$ of size $M$, we have $\textbf{w}_t = \textbf{w}_{t-1} - \eta \sum_{(x, y) \epsilon S_{t -1}} g_{t-1}(x, y)$. Here $g_t(x, y) = \nabla_{\textbf{w}_t}l(x, y)$, $t = 1,..., T$, and $\eta$ is the learning rate. The iterations can be perceived as time evolution, with the time derivative of the loss denoted by, $\Delta_t((x,y), S_t) = -\frac{dl}{dt}$. 
Due to the random initialization of the neural network, the weight vector at time $t > 0$, $\mathbf{w}_t$, is considered as a random variable. 

The GraNd score of a training example $(x, y)$ is defined at time $t$ as $\chi_t(x,y) = \mathbb{E}_{\mathbf{w}_t} || g_t(x, y) ||_2\,$, \citep{paul2021deep}.  
It represents  the impact of removal of an example $\left(x_j, y_j\right)$ from the training set ($S$) on the change in the loss for other examples. 

The change in the time derivative of the loss of any example $(x^*, y^*)$ resulting from the removal of a training example $(x_j, y_j)$ from the dataset $S$ is bounded by the \emph{GraNd score} associated with the removed example.
Formally,  for all $(x^*, y^*)$, there exists a constant $c$ such that $\big\|\Delta_t((x^*, y^*); S) - \Delta_t((x^*, y^*); S_{-j})\big\| \leq c \, \|g_t(x_j, y_j)\|$. Where $S_{-j} = S \setminus \{(x_j, y_j)\}$ and $\Delta_t((x^*, y^*); S) = -\frac{d\ell}{dt}$.  
The constant $c$ does not depend on the training examples, $(x,y)$s. Therefore, the contribution of an example $(x,y)$ to change in the loss of any other example $(x^*, y^*)$ is bounded by the gradient norm, $\|g_t(x_j, y_j)\|$, at any time $t$, given the weights $\mathbf{w}_t$.  
In practice, the gradient of the loss, $l$, is computed with respect to the model's weights $\mathbf{w}_t$ at epoch $t$ for each training example. The GraNd score is obtained by computing the expectation over several training runs of $L_2$ norm of the gradient. The examples are ranked based on their GraNd scores, with higher scores indicating greater difficulty in learning. \\

{In the following, we discuss \textbf{Coverage-centric Coreset Selection (CCS)} method.} The purpose of coreset selection is to choose a  subset ($\mathcal{S'}$) of the data ($\mathcal{S}$) that minimizes the loss on the test set \cite{zheng2023_CCS}. 
This selection process can be represented through the following optimization problem  \cite{sener2018coreset}:
\begin{equation}
    \min_{\mathcal{S}' \subseteq \mathcal{S} : \frac{|\mathcal{S}'|}{|\mathcal{S}|} \leq 1- 
 f} \mathbb{E}_{x,y \sim {P}} \left[ l(x, y; h_{\mathcal{S}'}) \right],
\end{equation}
where $l$ is the loss function, $h_{\mathcal{S}'}$ is the model trained with the coreset $\mathcal{S}'$, and $f$ is the pruning factor. In other words, $\mathcal{S}'$
is a subset of $\mathcal{S}$ which has a size at most $(1-f)$ of $S$, and which minimizes the expected test loss over the distribution ${P}$. 

In a classical coverage setting, a set $\mathcal{S'}$ is called an r-cover set of $\mathcal{S}$ if a ball of radius $r$ centered at each data point of $\mathcal{S'}$ covers the entire $\mathcal{S}$. Zheng et al. \cite{zheng2023_CCS} extended the classical geometric set cover to cover a probability distribution—referred to as the \textit{density-based distribution cover}. The associated coreset selection method is called  the \textit{Coverage-centric Coreset Selection} (CCS). 
They also introduced the concept of a partial cover percentage $p$, of the probability distribution $P$, instead of the full cover. 

A subset $\mathcal{S}$ of a metric space $(X, d)$ is defined as the 
\textit{p-partial r-cover} of a distribution $P$, on the space $X$ if:
\begin{equation}
        \int_{X} \mathbf{1}_{\cup_{x \in \mathcal{S}} B_d(\textbf{x}, r)}(\textbf{x})\,d\mu(\textbf{x}) = p 
\end{equation}
Here $B_d(\textbf{x}, r) = \{\textbf{x}' \in X : d(\textbf{x}, \textbf{x}') \leq r\}$ denotes a $r$-radius ball centered at $\textbf{x}$, $p$ quantifies coverage (in percentage), and $\mu$ represents the probability measure associated with the probability distribution $P$. The probability density information over the input space is reflected in the measure $\mu$. The percentage of coverage $p$ increases with the increasing radius $r$ for a given set $\mathcal{S}$ and a distribution $P$. Given a coverage percentage $p$ and a set $\mathcal{S}$, the minimum covering radius $r$ is the smallest value such that $\mathcal{S}$ achieves at least $p$ percentage coverage under distribution $P$. In practice, the distribution $P$ is unknown and so an approximate estimation of the coverage is required. Thus, the area under the p-r curve ($\text{AUC}_{pr}$), is proposed as a metric to assess the coverage of a coreset \cite{zheng2023_CCS}. The quantity $\text{AUC}_{pr}$ is defined as the expectation of the minimum distance between the examples that follow the distribution $P$ and the data in $\mathcal{S}$: 
$\text{AUC}_{pr}(\mathcal{S}) = \mathbb{E}_{x \sim P} \left[\min_{x' \in \mathcal{S}} d(\textbf{x}', \textbf{x})\right]$. 
The lower the value of the $\text{AUC}_{pr}$, the better the coverage by the coreset. 

State-of-the-art methods often tend to eliminate low-importance data points from high-probability density regions, which decreases coverage (increases the $\text{AUC}_{pr}$), and in turn, this can degrade model performance under high pruning rates. 
In our implementation of CCS, we used the GraNd score as the importance score. However, the coverage obtained by CCS is distinct from that obtained by GraNd based pruning alone. Traditional GraNd removes only low-scoring examples, leading to poor coverage. In contrast, CCS maintains better data coverage in high-probability density regions, reducing $\text{AUC}_{pr}$ and improving performance, even under significant pruning. 
CCS allocates more sampling budget to high-probability density areas with easy examples, improving coverage of the data distribution. Compared to random sampling, CCS allocates more resources to low-probability density regions, focusing on challenging examples that are critical for training.

{In the following, we describe the steps for the CCS method.}  Consider a dataset $\mathcal{S}$ = $\{(x_i, y_i, s_i)\}_{i=1}^n$ where $s_i$ is the importance score for $\textit{i}$-th example $(x_i, y_i)$. 
First, CCS selects a pruning rate $\beta$, a tunable hyperparameter representing the percentage of hard examples to remove. Pruning $\mathcal{\beta}\,\%$  of hard examples from $\mathcal{S}$ yields $\mathcal{S}_o$. Second, CCS partitions $\mathcal{S}_o$ into $k$ strata based on importance (GraNd) scores, using fixed-width score intervals. While each stratum has the same score range, the number of examples per stratum may vary. Third, given the pruning factor $f$, the desired coreset size is determined as $m = (1 - f)\, n$. This total budget $m$ is distributed evenly across the $k$ strata (another hyperparameter), allocating $m/k$ samples per stratum. If a stratum has fewer examples than its budget, all its examples are included. The final coreset ($\mathcal{S'}$) is formed by combining the sampled examples from all strata.

\section*{Author contributions}
SH, PT and SP worked on developing the framework for pruning and other machine learning aspects. ST and AG worked on the KS-DFT data generation and post-processing calculations. ASB and SG were involved in conceptualization, methodological design, supervision, and securing funding/resources. All authors contributed to writing the manuscript. 

\section*{Competing Interests}
\noindent The authors declare no competing financial or non-financial interests.
Author ASB is an editorial board member of the Journal of Materials Science: Materials Theory, and is currently editing a special collection on ``Machine Learning for Electronic-Structure Theory and Methods''. ASB was not involved in the journal’s review of, or decisions related to, this manuscript.

\section*{Data Availability}\label{subsec:Data_availability}
\noindent The datasets generated and/or analyzed during the current study are not publicly available due to their large size and associated storage and data-transfer constraints. Raw data were generated at Hoffman2 High-Performance Compute Cluster at UCLA's Institute for Digital Research and Education (IDRE) and National Energy Research Scientific Computing Center (NERSC).  Derived data supporting the findings of this study are available from the corresponding author upon request.

\section*{Code Availability}\label{subsec:Code_availability}
Codes supporting the findings of this study are available from the corresponding authors upon reasonable request. The code developed and used for the study is not currently available publicly due to ongoing updates and further developments in the workflow. It is available from the corresponding authors on reasonable request.

\section*{Materials \& Correspondence}
Correspondence and requests for materials should be addressed to:  \\
Amartya S. Banerjee (email: asbanerjee@ucla.edu), and \\
Susanta Ghosh (email: susantag@mtu.edu) 
\vspace{10pt}

\section*{Acknowledgements}

The authors thank UCLA's Institute for Digital Research and Education (IDRE), Michigan Technological University's MRI computing clusters, and the National Energy Research Scientific Computing Center (NERSC) for providing computational resources. The authors thank Abhishek Keripale for assistance with the development of later versions of the CrFeCoNi machine-learning models. The authors used GPT-5 (OpenAI) for language editing and proofreading and subsequently reviewed and edited all content.

\section*{Funding}

This work was primarily supported by the U.S. Department of Energy (DOE), Office of Science, Basic Energy Sciences (BES), under Award DE-SC0023432, which provided primary support for the overall project conception and integrated KS-DFT--ML investigations. The University of California National Laboratory Fees Research Program under Grant L25CR9003 supported data generation and post-processing at UCLA (ASB, ST, AG). The National Science Foundation under Award 2442313 supported data-pruning and dimensionality analyses at Michigan Technological University (SG, SH, PT, SP). This research also used resources of the National Energy Research Scientific Computing Center (NERSC), a U.S. Department of Energy Office of Science User Facility supported under Contract No. DE-AC02-05CH11231, through NERSC awards BES-ERCAP0037205, BES-ERCAP0033206, BES-ERCAP0025205, BES-ERCAP0025168, and BES-ERCAP0028072.

\clearpage
\bibliography{main, pruning, dimension}
\clearpage

\include{SuppMat}
\end{document}

%% file: revtex_preamble.tex
%

\newcommand{\rz}{\mathbb{R}}
\newcommand{\gz}{\mathbb{Z}}
\newcommand{\cz}{\mathbb{C}}
\newcommand{\qz}{\mathbb{Q}}
\newcommand{\nz}{\mathbb{N}}

\newcommand{\bfa}{{\bf a}}
\newcommand{\bfb}{{\bf b}}
\newcommand{\bfc}{{\bf c}}
\newcommand{\bfd}{{\bf d}}
\newcommand{\bfe}{{\bf e}}
\newcommand{\bff}{{\bf f}}
\newcommand{\bfg}{{\bf g}}
\newcommand{\bfh}{{\bf h}}
\newcommand{\bfi}{{\bf i}}
\newcommand{\bfj}{{\bf j}}
\newcommand{\bfk}{{\bf k}}
\newcommand{\bfl}{{\bf l}}
\newcommand{\bfm}{{\bf m}}
\newcommand{\bfn}{{\bf n}}
\newcommand{\bfo}{{\bf o}}
\newcommand{\bfp}{{\bf p}}
\newcommand{\bfq}{{\bf q}}
\newcommand{\bfr}{{\bf r}}
\newcommand{\bfs}{{\bf s}}
\newcommand{\bft}{{\bf t}}
\newcommand{\bfu}{{\bf u}}
\newcommand{\bfv}{{\bf v}}
\newcommand{\bfw}{{\bf w}}
\newcommand{\bfx}{{\bf x}}
\newcommand{\bfy}{{\bf y}}
\newcommand{\tbfy}{{\tilde{\bfy}}}
\newcommand{\bfz}{{\bf z}}
\newcommand{\bfA}{{\bf A}}
\newcommand{\bfB}{{\bf B}}
\newcommand{\bfC}{{\bf C}}
\newcommand{\bfD}{{\bf D}}
\newcommand{\bfE}{{\bf E}}
\newcommand{\bfF}{{\bf F}}
\newcommand{\bfG}{{\bf G}}
\newcommand{\bfH}{{\bf H}}
\newcommand{\bfI}{{\bf I}}
\newcommand{\bfJ}{{\bf J}}
\newcommand{\bfK}{{\bf K}}
\newcommand{\bfL}{{\bf L}}
\newcommand{\bfM}{{\bf M}}
\newcommand{\bfN}{{\bf N}}
\newcommand{\bfO}{{\bf O}}
\newcommand{\bfP}{{\bf P}}
\newcommand{\bfQ}{{\bf Q}}
\newcommand{\bfR}{{\bf R}}
\newcommand{\bfS}{{\bf S}}
\newcommand{\bfT}{{\bf T}}
\newcommand{\bfU}{{\bf U}}
\newcommand{\bfV}{{\bf V}}
\newcommand{\bfW}{{\bf W}}
\newcommand{\bfX}{{\bf X}}
\newcommand{\bfY}{{\bf Y}}
\newcommand{\bfZ}{{\bf Z}}
\newcommand{\vphi}{{\varphi}}
\newcommand{\eps}{{\varepsilon}}
\newcommand{\Nhat}{\hat{\mbox{\tiny {\bf N}}}}
\newcommand{\ehat}{\hat{\bf e}}
\newcommand{\nhat}{\hat{\bf n}}
\newcommand{\uhat}{\hat{\bf u}}
\newcommand{\phihat}{\hat{\varphi}}
\newcommand{\xihat}{\hat{\xi}}
\newcommand{\fhat}{\hat{f}}
\newcommand{\Vhat}{\hat{V}}
\newcommand{\adj}{{\mbox{adj }}}
\newcommand{\beq}{\begin{equation}}
\newcommand{\eeq}{\end{equation}}
\newcommand{\beqs}{\begin{eqnarray}}
\newcommand{\eeqs}{\end{eqnarray}}
\newcommand{\beql}{\begin{equation} \label}

\newcommand{\normp}[1]{\| #1 \|}
\newcommand{\brho}{\boldsymbol{\rho}}
\newcommand{\expchar}[2]{\textrm{e}^{\frac{2\pi \textrm{i} #1}{#2}}}
\newcommand{\expcharconj}[2]{\textrm{e}^{-\frac{2\pi \textrm{i} #1}{#2}}}

\newcommand{\half}{\frac{1}{2}}
\newcommand{\calA}{{\cal A}}
\newcommand{\calB}{{\cal B}}
\newcommand{\calC}{{\cal C}}
\newcommand{\calD}{{\cal D}}
\newcommand{\calE}{{\cal E}}
\newcommand{\calF}{{\cal F}}
\newcommand{\calG}{{\cal G}}
\newcommand{\calH}{{\cal H}}
\newcommand{\calI}{{\cal I}}
\newcommand{\calJ}{{\cal J}}
\newcommand{\calK}{{\cal K}}
\newcommand{\calL}{{\cal L}}
\newcommand{\calM}{{\cal M}}
\newcommand{\calN}{{\cal N}}
\newcommand{\calO}{{\cal O}}
\newcommand{\calP}{{\cal P}}
\newcommand{\calQ}{{\cal Q}}
\newcommand{\calR}{{\cal R}}
\newcommand{\calS}{{\cal S}}
\newcommand{\calT}{{\cal T}}
\newcommand{\calU}{{\cal U}}
\newcommand{\calV}{{\cal V}}
\newcommand{\calW}{{\cal W}}
\newcommand{\calX}{{\cal X}}
\newcommand{\calY}{{\cal Y}}
\newcommand{\calZ}{{\cal Z}}

\newcommand{\hilb}{\mathsf{H}}
\newcommand{\thilb}{\tilde{\mathsf{H}}}
\newcommand{\hamil}{\mathfrak{H}}
\newcommand{\thrbyto}{\frac{3}{2}}
\newcommand{\Fhat}{\hat{F}}
\newcommand{\tlambda}{\tilde{\lambda}}
\newcommand{\fivbyto}{\frac{5}{2}}

%
%
%
%


\newenvironment{myproof}[1][Proof:]{\begin{trivlist}
\item[\hskip \labelsep {\bfseries #1}]}{\qed\end{trivlist}}

\newenvironment{myexample}[1][Example:]{\begin{trivlist}
\item[\hskip \labelsep {\bfseries #1}]}{\end{trivlist}}

\newenvironment{myremarks}[1][Remarks:]{\begin{trivlist}
\item[\hskip \labelsep {\bfseries #1}]}{\end{trivlist}}

\newenvironment{myremark}[1][Remark:]{\begin{trivlist}
\item[\hskip \labelsep {\bfseries #1}]}{\end{trivlist}}


\newcommand{\abs}[1]{\lvert#1\rvert}
\newcommand{\norm}[2]{\lVert#1\rVert_{#2}}
\newcommand{\innprod}[3]{\langle#1,#2\rangle_{#3}}
\newcommand{\biginnprod}[3]{\mathbf{\Big\langle}#1,#2\mathbf{\Big\rangle}_{#3}}
\newcommand{\esssup}[2]{\displaystyle\text{ess sup}_{#1}#2}
\newcommand{\Lpspc}[3]{\textsf{L}^{#1}_{#2}(#3)}
\newcommand{\infm}[2]{\displaystyle\inf_{#1}{#2}}
\newcommand{\supm}[2]{\displaystyle\sup_{#1}{#2}}
\newcommand{\argmin}[2]{\displaystyle\textrm{argmin}_{#1}{#2}}
\newcommand{\argmax}[2]{\displaystyle\textrm{argmax}_{#1}{#2}}
\newcommand{\mspan}[1]{\text{span}\{#1\}}
\newcommand{\pd}[2]{\frac{\partial#1}{\partial#2}}
\newcommand{\hpd}[3]{\frac{\partial^{#3}#1}{\partial#2^{#3}}}
\newcommand{\od}[2]{\frac{d#1}{d#2}}
\newcommand{\hod}[3]{\frac{d^{#3}#1}{d#2^{#3}}}
\newcommand{\isom}[2]{(#1|#2)}
\newcommand{\orb}[2]{\text{Orb}(#1,#2)}
\newcommand{\stab}[2]{\text{Stab}(#1,#2)}
\newcommand{\ball}[2]{\calB_{#1}(#2)}
\newcommand{\ccf}[1]{\mathsf{C}_{\mathsf{c}}(#1)}
\newcommand{\spt}[1]{\text{spt}.(#1)}
\newcommand{\mattr}[1]{\text{Tr.}(#1)}

\let\oldFootnote\footnote
\newcommand\nextToken\relax

\renewcommand\footnote[1]{%
    \oldFootnote{#1}\futurelet\nextToken\isFootnote}

\newcommand\isFootnote{%
    \ifx\footnote\nextToken\textsuperscript{,}\fi}
    
\newcommand{\dbltilde}[1]{\accentset{\approx}{#1}}

%% file: SuppMat.tex
\onecolumngrid
\begin{center}
    \textbf{\MakeUppercase{Supplementary Materials}}
    \vspace{20pt}
\end{center}
\twocolumngrid 

\section{Error Metrics}
\label{sm:metrics}
The normalized root mean square error (NRMSE) of the ML predicted electron density is defined as
\begin{equation}
NRMSE = \frac{\sqrt{\frac{1}{N} \sum_{i=1}^{N} (\rho^{i}_{DFT} - \rho^{i}_{ML})^2})}{\rho_{DFT}^{max} - \rho_{DFT}^{min}}
\end{equation}
where N is the number of grid-points for each snapshot. At $i$-th grid point, $\rho^{i}_{DFT}$ represents the electron density calculated using KS-DFT while $\rho^{i}_{ML}$ represents the electron density predicted by the ML models. $\rho_{DFT}^{max}$  and $\rho_{DFT}^{min}$ are the maximum and minimum values of electron density across all grid-points of that snapshot. \\

Let $u$ be the error field of the ML-predicted electron density, given by $u(x) = \rho_{DFT}(x) - \rho_{ML}(x)$, with $x\in\Omega$. The $H^1$ seminorm of the scalar error field $u(x)$ over a domain $\Omega$ is defined as $|u|_{H^1(\Omega)} = \left( \int_\Omega |\nabla u(x)|^2 \, dx \right)^{1/2}$. The $H^1$ norm is defined as $\|u\|_{H^1(\Omega)} = \left( \|u\|_{L^2(\Omega)}^2 + |u|_{H^1(\Omega)}^2 \right)^{1/2}$. These quantities are well defined whenever, $u$ lies in the Sobolev space $H^1(\Omega)$. \\

In the paper, the energy prediction errors are defined as the absolute differences (in Hartree per atom) between KS-DFT energies and the energies obtained by postprocessing the ML-predicted electron density.

 {
The relative $L_\infty$ norm is defined as:
\begin{equation}
    L_\infty = \frac{\max_{i=1}^N |\rho^{\mathrm{ML}}_i - \rho^{\mathrm{DFT}}_i|}{\max_i |\rho^{\mathrm{DFT}}_i|}
\end{equation}
where $\rho^{\mathrm{ML}}_i$ and $\rho^{\mathrm{DFT}}_i$ are the ML-predicted and KS-DFT electron densities at grid point $i$, respectively, and N is the number of grid-points for each snapshot.
}

\section{Postprocessing of ML predicted electron density}\label{sm:postprocessing}
For application purposes, it is of interest to compute downstream properties from the ML-predicted electron density. Since the Kohn-Sham total ground-state energy is a key property that can be computed from the electron density, it can serve as a useful test for the downstream accuracy of the ML model. We postprocess the density predictions to obtain the corresponding total ground-state energies as follows. The predicted electron density is first rescaled by the total number of electrons; this serves to ensure that the total electronic charge is consistent \cite{pathrudkar2024electronic, briling2021impact, alred2018machine, pathrudkar2022machine}. The scaled electron density $\rho^{\text{scaled}}$ at grid point $\textbf{r}$ is obtained from the predicted electron density $\rho^{\text{ML}}$ via:

\begin{align}
    \rho^{\text{scaled}}\left(\textbf{r}\right) = \rho^{\text{ML}}(\textbf{r})\frac{N_{\text{e}}}{\displaystyle\int_{\Omega}\rho^{\text{ML}}(\textbf{r})d\textbf{r}}\,.
    \label{eq:scaled_density}
\end{align}
Here $\Omega$ is the periodic supercell used in the calculations, and $N_{\text{e}}$ is the number of electrons in the system. The rescaled electron density $\rho^{\text{scaled}}$ is then used to set up the Kohn-Sham Hamiltonian and a single diagonalization step is performed by using the same electronic structure calculation framework, as was used in the data generation process \cite{xu2021sparc,xu2020m, ghosh2017sparc}. Thereafter, the energy of the system is computed using the Harris-Foulkes formula \citep{harris1985simplified, foulkes1989tight}. The difference between the energy obtained from the ML-predicted electron density and that obtained from the original DFT density serves as the energy error reported in this work.

\section{Smoothness of the electron density prediction}
To demonstrate that the ML models not only capture the electron density fields but also smooth spatial variations in them, we plot the $H^1$ seminorm and $H^1$ norm of the error in electron density fields in Figure~\ref{fig:consolidated-H1}. The ML models reproduce both the electron density fields and their smooth spatial variations, with smoothness errors following trends similar to those observed in the electron density itself. For all four material systems, the $90\%$ CCS-pruned  dataset achieves a similar degree of spatial smoothness as the original dataset (i.e., similar $H^1$ seminorm and $H^1$ norm values), whereas the $99\%$ CCS-pruned dataset shows a slight increase. The CCS method achieves better smoothness than random for $90\%$ and $99\%$ pruning, for all four systems. 
\label{sm:H1}
\begin{figure}[htbp]
    \centering    
    \includegraphics[width=\linewidth]{SI_FIGURES/Consolidated_H1.pdf} 
    \caption{\textbf{Error in smoothness of the predicted electron density:}(a) $H^1$ seminorm and (b) $H^1$ norm of the error for the ML-predicted electron density obtained from the original dataset, $90\%$, and $99\%$ randomly and CCS based pruned dataset.}
    \label{fig:consolidated-H1}
\end{figure}

\section{Robustness of Pruning}
\label{sm:robustness}
\begin{figure}
    \centering
    \includegraphics[width=\linewidth]{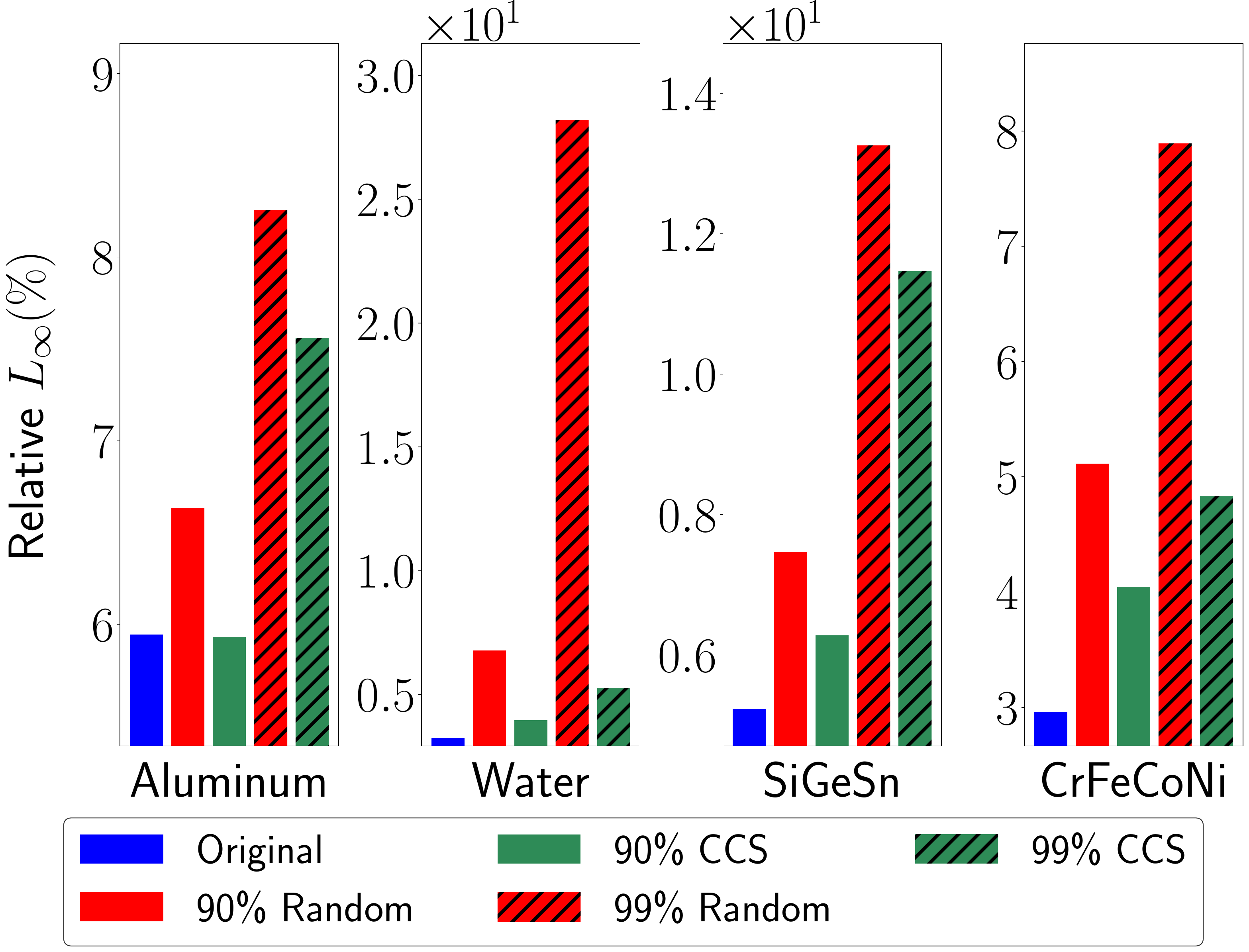}
    \caption{ {Relative $L_\infty$ error in the ML-predicted electron density for the original dataset, 90\% and 99\% randomly, and CCS-based pruned datasets.}}
    \label{fig:error-linf}
\end{figure}
\begin{figure*}[tbh!]
    \centering
    \includegraphics[width=0.8\linewidth]{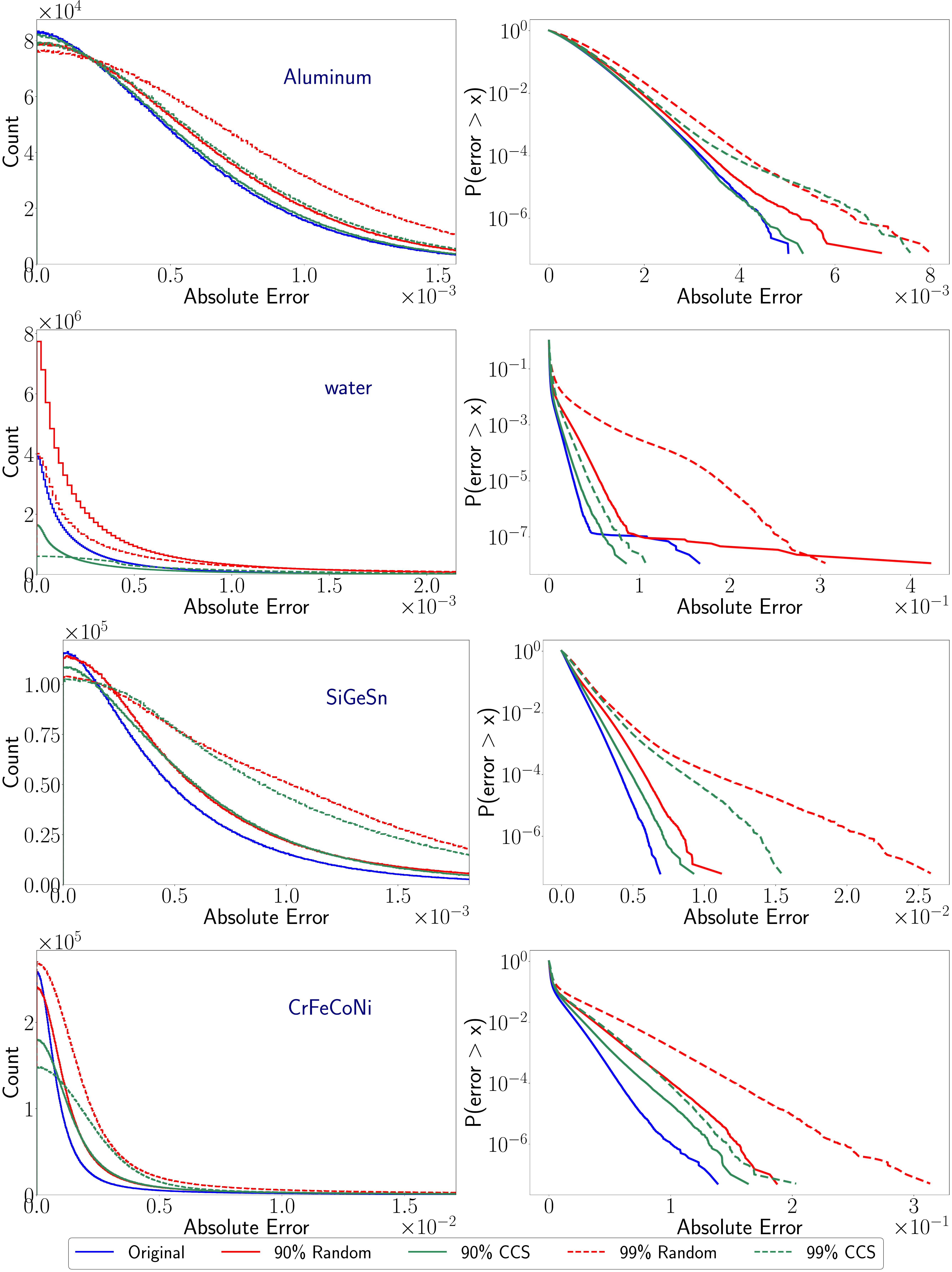}
    \caption{ {Comparison among the error distribution from the ML-predicted density obtained from the original dataset, $90\%$, and $99\%$ randomly and CCS based dataset. (left) Absolute Error Distribution for four systems. 
    (right) Probability of a pointwise error exceeding a given threshold.
    }}
    \label{fig:error-comparison}
\end{figure*}
To assess the maximum error upon pruning, we evaluate the $L_\infty$ norm of the error and the absolute error distribution of the ML prediction. Figure~\ref{fig:error-linf} shows that after 90\% pruning, there is a marginal increase in errors. At 99\% pruning, the error increase is larger. However, CCS consistently outperforms random pruning across all systems. Figure~\ref{fig:error-comparison} shows the absolute error count distribution (left panels) and the probability of the pointwise error exceeding a given threshold (right panels) for all grid points across all test snapshots, for all four systems. 
For Aluminum and SiGeSn, the 90\% CCS pruning's error distributions nearly overlap with the original, indicating negligible degradation at this pruning level. For CrFeCoNi, 90\% CCS tracks the original closely. Notably, for water, the tail of the 90\% CCS model is comparable to or better than the original. Across all systems, random pruning consistently produces the heaviest tail, while 99\% CCS remains between the original and 99\% Random, confirming that CCS-based pruning introduces significantly fewer large local errors than random pruning at high pruning factors.

\color{black}
\section{Experimental Details} 
\label{sm:experimental-details}
 {\underline{Hardware Details}: All experiments were run on NVIDIA A100 (40GB) GPUs available through NERSC Perlmutter. For the CrFeCoNi experiments, we have used 4 GPUs but for other experiments, we have used single GPU.}\\ 
\underline{Dataset details}: 
We use sufficiently large datasets to achieve high accuracy in ML for all four material systems. The  KS-DFT dataset for the aluminum system consists of 200 training snapshots (216,000 grid points each), totaling $43.2 \times 10^6$ grid points, obtained from a 32-atom simulation cell; the test set includes 60 snapshots from six temperatures  {($298 -1800$~K)}. The dataset for water includes 112 training snapshots (456,533 grid points each, totaling $51.1 \times 10^6$ grid points), and 193 test snapshots from four temperatures, obtained from a 32-atom simulation cell. The dataset for SiGeSn system, taken from 45 compositions, consists of 216 training snapshots (totaling $38.6 \times 10^6$ grid points) and test set includes 90 test snapshots from 45 compositions, obtained from a 64-atom simulation cell.  {The dataset for CrFeCoNi system, taken from 69 compositions, consists of  {552} training snapshots (totaling $161.4 \times 10^6$ grid points) and test set includes 207 test snapshots across four different temperatures and 69 compositions, also from a 32-atom simulation cell. }

 {\underline{Descriptor Size:} We have chosen descriptor size so that the original model meets the chemical accuracy level of predictions, across all systems. We have kept the all descriptor fixed throughout the pruning experiments. }

\underline{Neural Network}: 
We employ a deep neural network with 12 hidden layers, each containing 900 nodes, using GELU activation~\citep{hendrycks2016gaussian} in the hidden layers and a ReLU activation in the output layer. A  {starting} learning rate of 0.001 is used across all cases, with the AdamW optimizer~\citep{kingma2014adam, loshchilov2017decoupled}.  {ReduceLROnPlateau type learning rate scheduler was used with a minimum learning rate of $10^{-5}$. Training was not terminated at a fixed number of epochs; instead, a convergence criterion was used; training was stopped when the validation loss did not improve over a window of 35 epochs. We also want to note that all models trained on pruned datasets are trained from scratch with independent random initialization; no weights are transferred from the original model.}

\begin{figure}
\centering
\includegraphics[width=0.9\linewidth]{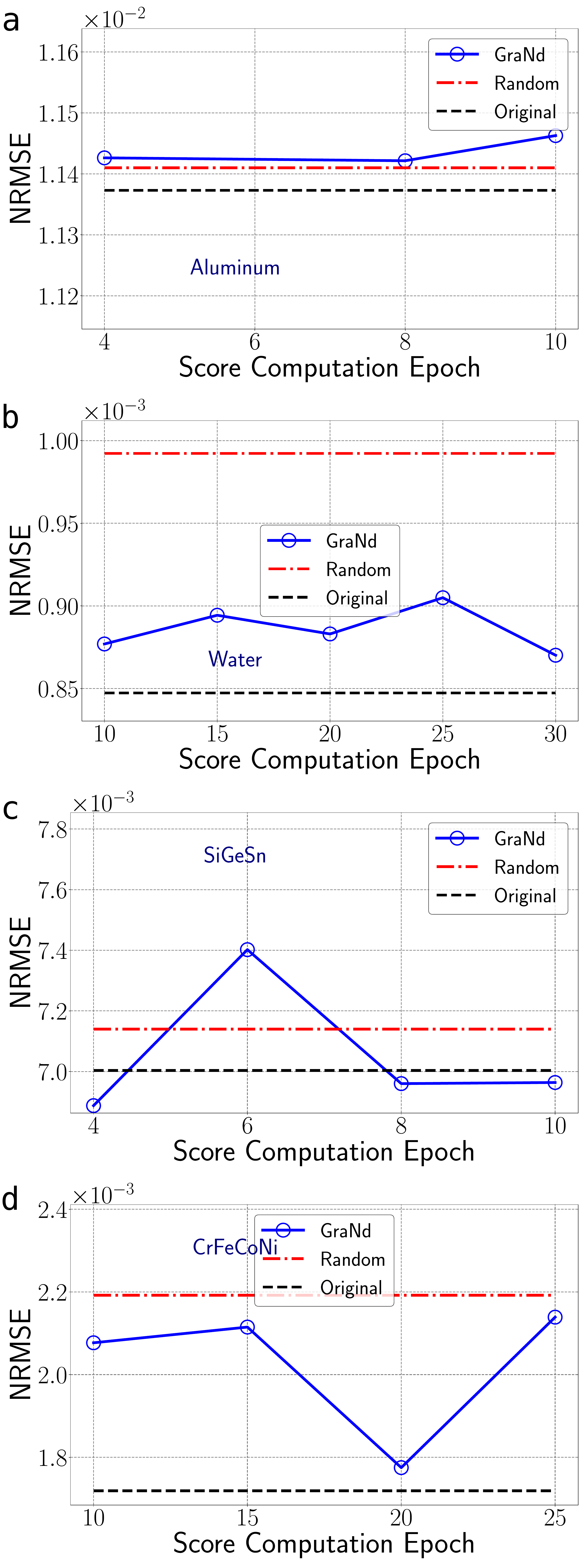}
        \caption{\textbf{Choice of score computation epoch:} The error in electron density prediction for (a) Aluminum(20\% pruning), (b) Water(40\% pruning) (c) SiGeSn(20\% pruning) and (d) CrFeCoNi(40\% pruning) after computing the GraNd score at different times in training.}
    \label{fig:sce-justification}
\end{figure}

\underline{GraNd score computation}: The epoch 10, 30, 10 and  {20} were chosen for GraNd score computation for Aluminum, water, SiGeSn and CrFeCoNi respectively. Figure~\ref{fig:sce-justification} shows the errors on the test dataset for Aluminum, water, SiGeSn, and CrFeCoNi for pruned subsets, for scores computed at different times in training. The Figure~\ref{fig:sce-justification} shows that computing the scores early in training is sufficient to identify important examples, aligning with the findings of~\citep{paul2021deep}. {For GraNd score computation, we use randomly initialized score-computation models that are separate from the models used for final training on the original dataset. The score-computation models are trained only to early checkpoints rather than to full convergence ~\citep{paul2021deep}. The computational cost of these score -computation runs is included in the full-pipeline timing comparison reported in Section \ref{sm:TimingComparison} and Supplementary Figure \ref{fig:consolidated-time}.}

\begin{figure}[h!]
    \centering
    \includegraphics[width=0.95\linewidth]{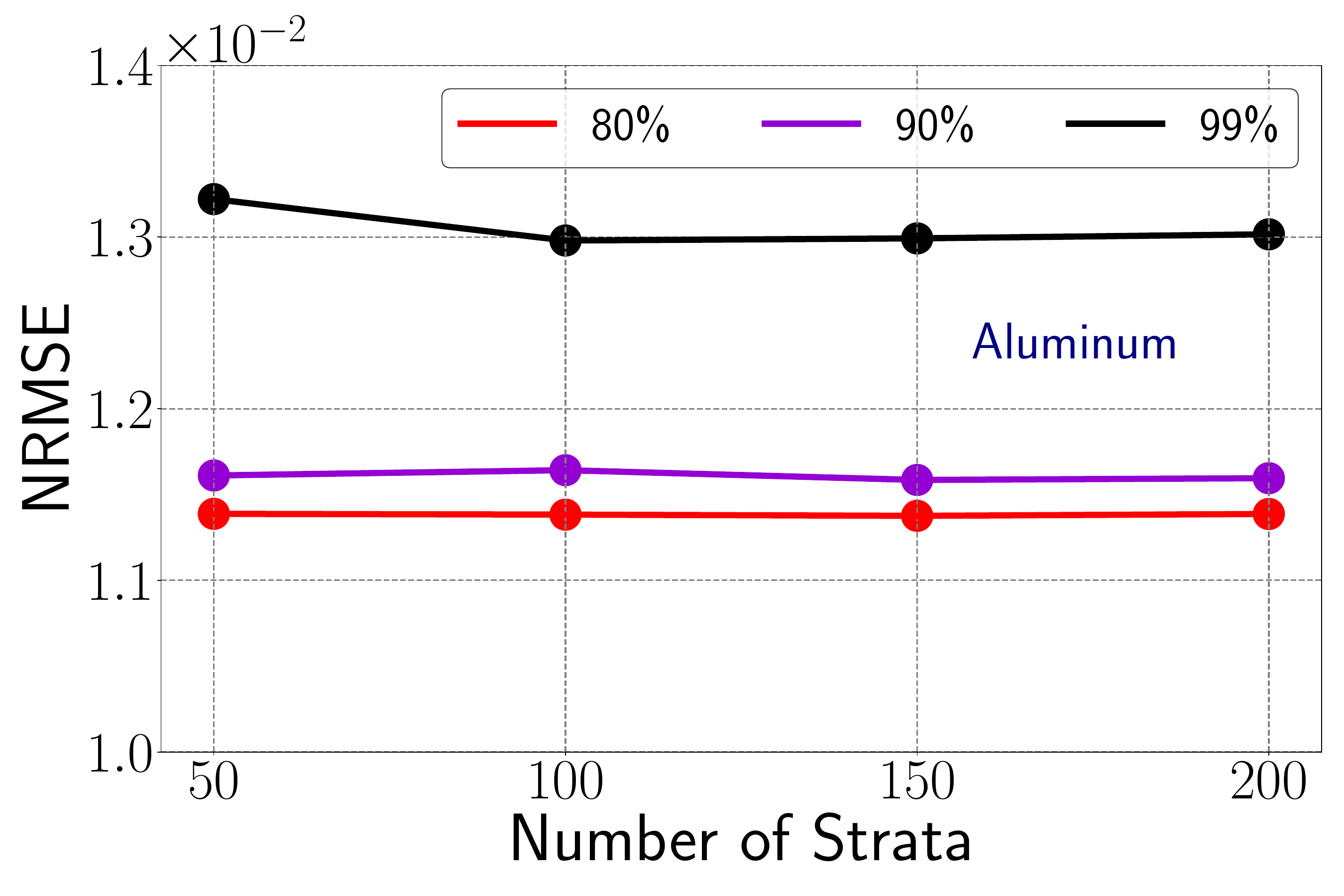}
    \caption{\textbf{Choice of number of strata($k$):}The error in electron density prediction with various numbers of strata for Aluminum system for various pruning factors for CCS method.}
    \label{fig:al-strata}
\end{figure}

\begin{figure}
\centering
    \includegraphics[width=0.9\linewidth]{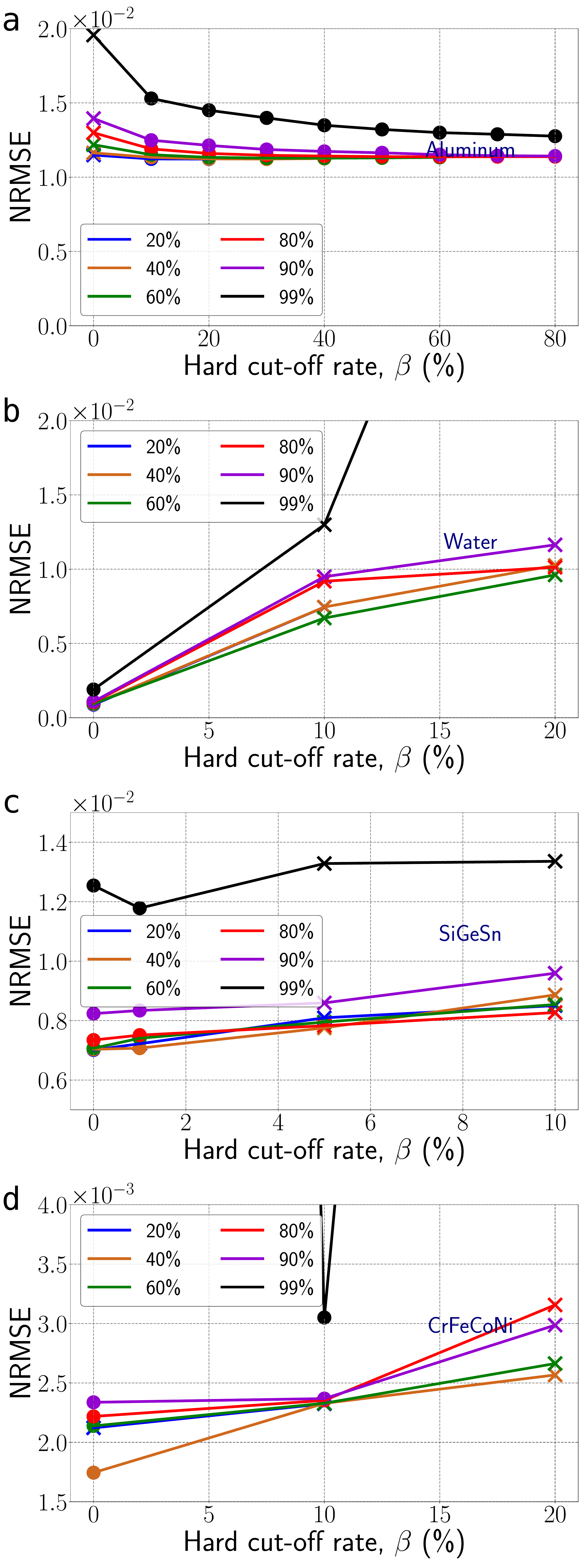}
    \caption{\textbf{Choice of hard cut-off rate($\beta$):} The error in electron density prediction with varying $\beta$ for (a) Aluminum, (b) water (c) SiGeSn (d) CrFeCoNi for the CCS based pruning. It is observed that the obtained NRMSE with pruned dataset varies with the choice of $\beta$. For higher pruning rate, a higher $\beta$ may be optimal. The points with $\times$ markers failed to beat random pruning.}
    \label{fig:beta-justify}
\end{figure}

\underline{Hyper-parameters for CCS}: There are two hyper-parameters for the CCS method. a) the hard cut-off rate ($\beta$) and b) the number of strata ($k$). The optimal $\beta$ generally increases with the pruning rate. The performance of various CCS-based coresets constructed from varying $\beta$ is shown in Figure~\ref{fig:beta-justify}. The error in charge density prediction for various numbers of strata, $k$ is  shown in Figure~\ref{fig:al-strata}.
The error is not very sensitive to the the number of strata ($k$), which is consistent with the findings of Zheng et al.~\citep{zheng2023_CCS} for image classification datasets. Hence, in this work, the number of strata ($k$) being used is 100 for all four material systems for all pruning factors. 
\clearpage


\section{Investigation into temporal correlation}
\label{sm:temporal}
\begin{figure*}[t!]
    \centering
    \includegraphics[width=0.95\linewidth]{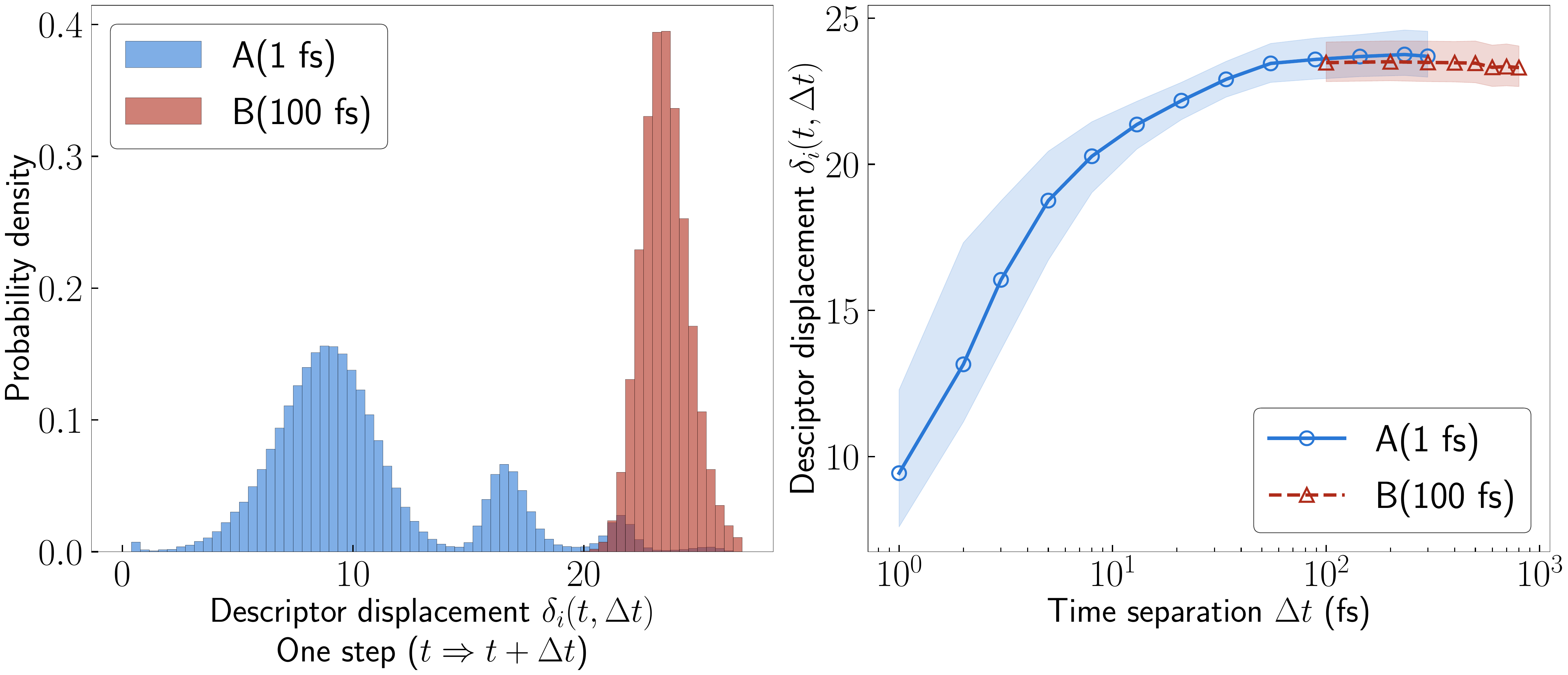}
    \caption{ {\textbf{Temporal correlation analysis for a representative system (equiatomic Cr$_{25}$Fe$_{25}$Co$_{25}$Ni$_{25}$, 32 atoms, 5000\,K)}: Dataset A contains snapshots at every timestep (1\,fs spacing) whereas dataset B contains snapshots taken at 100-timestep(100\,fs spacing). (Left) Distribution of the descriptor displacement over one sampling step, i.e. $\Delta t$ = 1\,fs (dataset A) and 100\,fs (dataset B). (Right) Descriptor displacement as a function of time separation $\Delta t$ averaged over all $t$ and grid points. }}    \label{fig:temporal}
    \end{figure*}
There will be high structural similarity (or strong correlation) between adjacent snapshots taken from an AIMD/MD trajectory. 
We have generated data so that it is (nearly) free from any structural similarity or temporal correlations present in AIMD/MD trajectory of a single simulation. To ensure that the data is structurally dissimilar (or uncorrelated) we have adopted the following three strategies in data generation:  
\begin{itemize}
    \item \underline{Large time interval between snapshots}: the training snapshots are extracted at intervals of 100 timesteps (1\, femtosecond (fs) timestep, i.e., 100\,fs spacing) from AIMD/MD trajectories.  This interval is sufficiently large compared to the typical vibrational periods in solids and liquids that fall in the range of $\sim$10--100\,fs. Such spacing between snapshots is commonly used for obtaining uncorrelated configurations from MD trajectories~\cite{ScoreDynamics2024JCTC,galindo2015convergence}.
    \item \underline{Simulations at different temperatures:} data at different temperatures are generated through entirely independent simulations, with a diverse set of initial configurations. Snapshots at different temperatures are therefore not points along a single MD trajectory and hence, carry little to no temporal correlation with one another.  Higher-temperature simulations might give a better sampling of the configuration space~\cite{pathrudkar2024electronic} and hence, various high-temperature simulations are included.
    \item \underline{Simulations for several alloy compositions}: for the alloy systems (SiGeSn and CrFeCoNi), each composition is simulated independently --- 69 compositions for CrFeCoNi and 45 compositions for SiGeSn, at various temperatures. Snapshots from different compositions at different temperatures thus originate from entirely distinct initial conditions and follow distinct trajectories, hence, are structurally dissimilar or uncorrelated. 
\end{itemize}
\textbf{Temporal Correlation Analysis:} In the following, 
we analyze the temporal correlation between snapshots taken from AIMD/MD trajectories. AIMD/MD simulations are used to generate the trajectory of snapshots from a 32-atom cell of  \text{Cr$_{25}$Fe$_{25}$Co$_{25}$Ni$_{25}$} system at 5000\,K. We consider two datasets extracted from this simulation:
\begin{itemize}
\item Dataset A: Snapshots collected at every timestep (1 fs intervals).
\item Dataset B: Snapshots collected every 100 timesteps (100 fs intervals).
\end{itemize}
In the following, we quantify the temporal correlation between snapshots directly in the descriptor space, by measuring how much the descriptor at a fixed grid point changes between two snapshots separated by a time lag $\Delta t$. Small changes indicate that the two snapshots supply nearly duplicate inputs; large changes indicate decorrelated snapshots. 
Let $\mathbf{D}_i(t) \in \mathbb{R}^{5M}$ denote the atomic-neighborhood descriptor (Eq.(1) of the main document) evaluated at the $i$-th grid point, at time $t$. Each descriptor component is standardized using its mean and standard deviation. The standardized descriptor is denoted as  $\tilde{\mathbf{D}}_i(t)$.
The \emph{descriptor displacement} over a time separation $\Delta t$ is then defined as the Euclidean distance between the standardized descriptors of the \emph{same} grid point evaluated at two snapshots separated by $\Delta t$ as
\begin{equation}
\delta_i(t, \Delta t) = \left\lVert \tilde{\mathbf{D}}_i(t + \Delta t) - \tilde{\mathbf{D}}_i(t) \right\rVert_2 .
\label{eq:fpd-2}
\end{equation}
Since the grid is fixed in the simulation cell while the atoms evolve, $\delta_i$ measures how much the local atomic environment of a given spatial location changes over $\Delta t$. The smaller the value of the \emph{descriptor displacement}, the greater the correlation between the snapshots.  Figure~\ref{fig:temporal}(left) shows the probability density of $\delta_i(t, \Delta\,t)$ for $\Delta\,t = 1$\,fs (dataset~A) and $\Delta t = 100$\,fs (dataset~B).
These distributions clearly show that the grid points in dataset A change only slightly between consecutive snapshots, whereas those in dataset B exhibit substantially larger changes. This observation further confirms that consecutive configurations in dataset A are temporally correlated, while those in dataset B are effectively uncorrelated. Figure~\ref{fig:temporal} (right) shows the descriptor distance, $\delta_i(t, \Delta t)$, as a function of the time separation, $\Delta t$. For dataset A, the descriptor distance increases with $\Delta t$ and converges after approximately 90\,fs indicating that the descriptors have become temporally uncorrelated. In contrast, for dataset B, the descriptor distance reaches the same converged value after a single sampling step, demonstrating that the selected data are already temporally uncorrelated.  \\ 
Furthermore, uncorrelated datasets (such as dataset B shown above) are constructed from independent simulations performed at different temperatures and compositions, further reducing temporal correlations within the dataset. For all systems considered in the present work, snapshots are separated by a minimum interval of $100\,\mathrm{fs}$. Therefore, our analysis indicates that the high redundancy in the data does not arise from temporal correlations.\\ 
Temporal correlations may exist within a dataset, and the degree of redundancy may depend on the extent of these correlations. However, quantifying redundancy as a function of dataset correlations is beyond the scope of the present work.

\color{black}
\section{Quantifying the coverage of a specific distribution by a subset } \label{sec:AUC_pr}
The GraNd score and other state-of-the-art importance-based pruning methods often perform poorly at higher pruning rates because of poor data coverage. 

We have evaluated the $\text{AUC}_{pr}$ by computing the expected minimum distance from the data in the validation dataset to the data in coreset.

\begin{table}[h]
    \centering
    \begin{tabular}{c|cc|cc}
        \hline
        \multicolumn{5}{c}{Aluminum} \\ \hline
        Pruning &\multicolumn{2}{c|}{$\text{AUC}_{pr}$} & \multicolumn{2}{c}{NRMSE} \\
        Rate & GraNd & CCS & GraNd & CCS\\
        \hline
        90\% & 8.396 & \textbf{8.30} & $1.58 \times 10^{-2}$ & $1.15 \times 10^{-2}$\\
        99\% & 8.727 & \textbf{8.584} & $2.55 \times 10^{-2}$ & $1.3 \times 10^{-2}$ \\
        \hline \hline
        \multicolumn{5}{c}{Water} \\ \hline
        Pruning&\multicolumn{2}{c|}{$\text{AUC}_{pr}$} & \multicolumn{2}{c}{NRMSE} \\
        Rate & GraNd & CCS & GraNd & CCS \\
        \hline
        90\% & 23.18 & \textbf{22.93} & $3.1 \times 10^{-2}$ & $1.09 \times 10^{-3}$\\
        99\% & 25.29 & \textbf{24.89} & $1.11 \times 10^{-1}$ & $1.9 \times 10^{-3}$ \\
        \hline \hline
        \multicolumn{5}{c}{SiGeSn} \\ \hline
        Pruning&\multicolumn{2}{c|}{$\text{AUC}_{pr}$} & \multicolumn{2}{c}{NRMSE} \\
        Rate & GraNd & CCS & GraNd & CCS \\
        \hline
        90\% & 76.57 & \textbf{76.30} & $2.55 \times 10^{-2}$ & $8.27 \times 10^{-3}$\\
        99\% & 81.91 & \textbf{80.537} & $8.27 \times 10^{-2}$ & $1.17 \times 10^{-2}$ \\
        \hline
        \multicolumn{5}{c}{ {CrFeCoNi}} \\ \hline
        Pruning&\multicolumn{2}{c|}{$\text{AUC}_{pr}$} & \multicolumn{2}{c}{NRMSE} \\
        Rate & GraNd & CCS & GraNd & CCS \\
        \hline
        90\% &  {13.074} & \textbf{ {13.039}} &  {$1.7 \times 10^{-2}$} &  {$2.3 \times 10^{-3}$}\\
        99\% &  {13.416} & \textbf{ {13.335}} &  {$7.45 \times 10^{-2}$} &  {$3 \times 10^{-3}$} \\
        \hline
    
    \hline
    \end{tabular}
    \caption{Comparison of $\text{AUC}_{pr}$ and NRMSE between GraNd and CCS method for different pruning percentages }
    
    \label{tab:auc_pr-nrmse}
\end{table}

From the Table \ref{tab:auc_pr-nrmse}, we can see that, for 90\% and 99\% pruning percentage, the CCS has lower $\text{AUC}_{pr}$ than the GraNd method. It confirms that the CCS method based coreset indeed selects a subset that ensures a better data coverage, as evidenced by its lower errors (NRMSE). 
\section{Generalizability}\label{sm:generalization}
In this section, we provide additional results that demonstrate the generalizability of ML models trained on pruned datasets. \\
\subsection{ {Structural Defects and Ordered Configurations}}
\begin{figure}[!h]
    \centering
    \includegraphics[width=\linewidth]{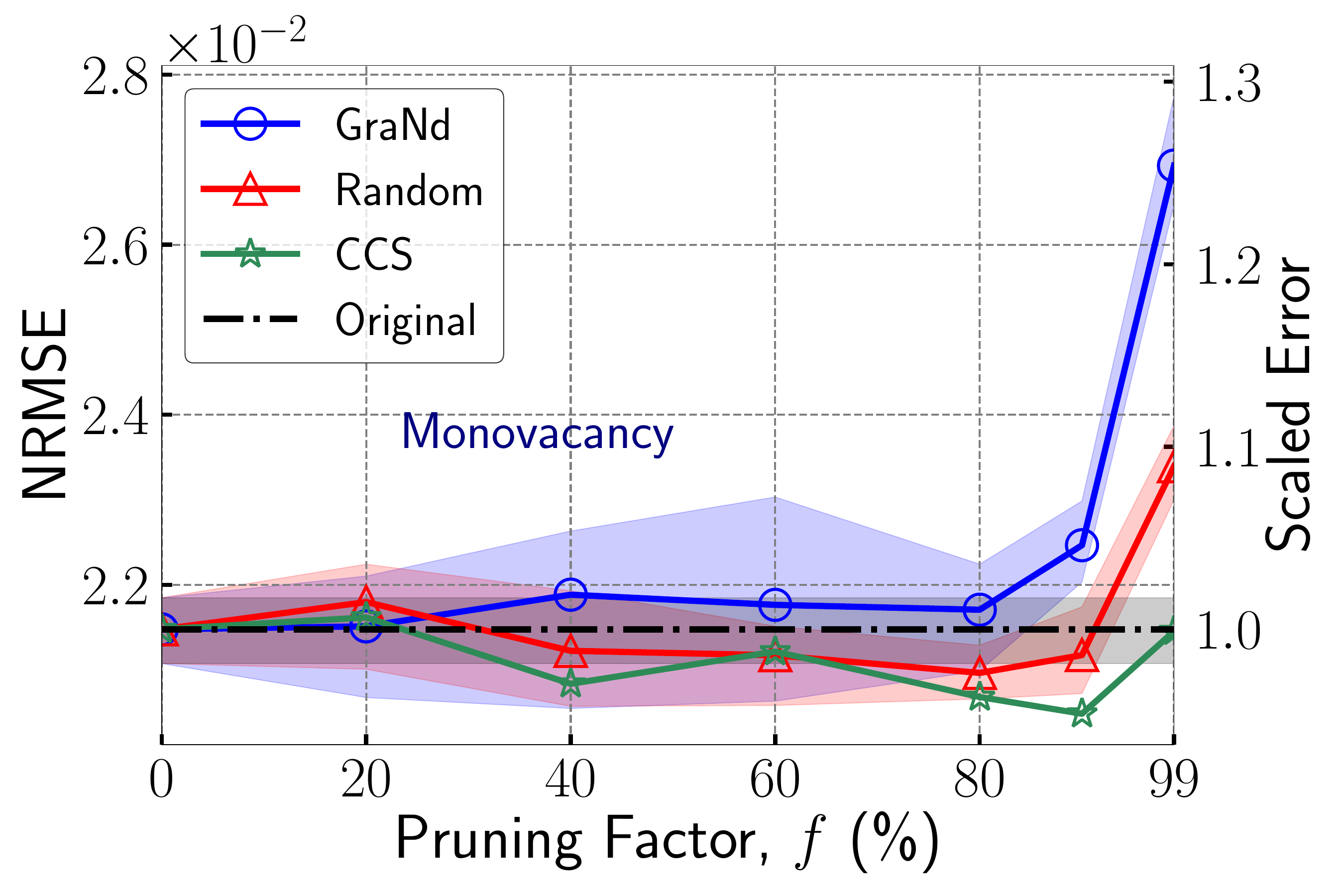}
    \caption{\textbf{ {Structural} Generalizability tests for Aluminum system:} The error in the electron density prediction for various pruning factors for monovacancies for 32-atoms Aluminum  system. The shaded region represents the range (maximum to minimum) of three ML models around the mean, shown by a solid line.}
    \label{fig:al-defects}
\end{figure}
\begin{figure}[!htbp]
    \centering
    \includegraphics[width=\linewidth]{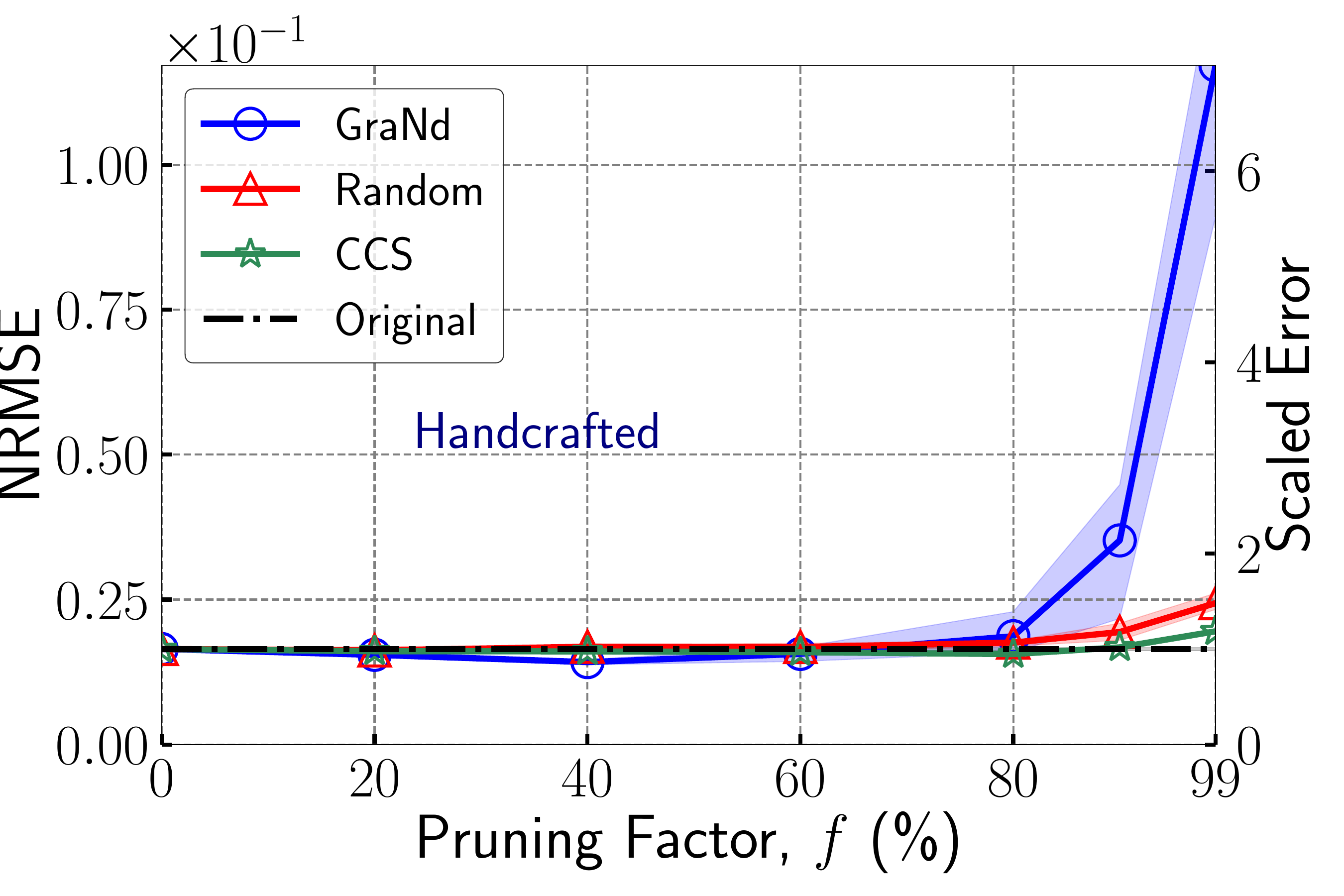}
    \caption{\textbf{ {Structural} Generalizability tests for SiGeSn system:} The error in the electron density prediction for various pruning factors for SiGeSn handcrafted checkerboard systems. The shaded region represents the range (maximum to minimum) of three ML models around the mean, shown by a solid line.}
    \label{fig:sigesn-handcrafted}
\end{figure}

\begin{figure*}[!htbp]
    \centering
    \includegraphics[width=\linewidth]{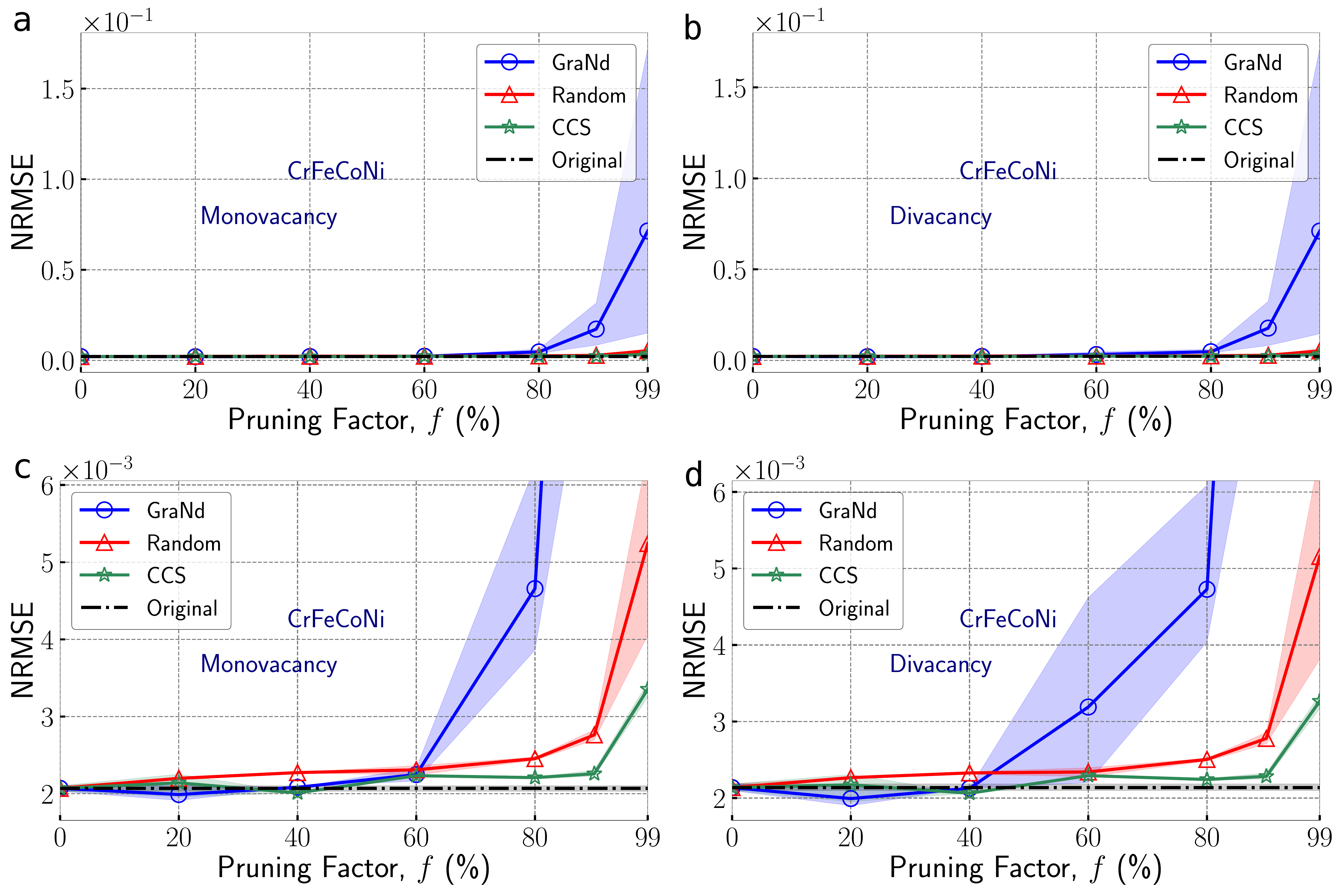}
    \caption{\textbf{ {Structural} Generalizability tests for CrFeCoNi system:} The error in the electron density prediction for various pruning factors for (a) mono- (b) di-vacancy defects for 32-atoms CrFeCoNi  system. The shaded region represents the range (maximum to minimum) of three ML models around the mean, shown by a solid line. Figures (c) and (d) show the magnified views of Figures (a) and (b).}
    \label{fig:crfeconi-defects}
\end{figure*}
The error in electron density prediction for defects in the 32-atoms aluminum system is shown in Figure~\ref{fig:al-defects}. The generalization error of the GraNd method increases rapidly beyond $80$\% pruning, whereas random pruning maintains high accuracy up to $90$\% pruning. The CCS method supports a remarkably high pruning factor ($99$\%) without any significant loss in generalization performance. 

The errors in electron density prediction for handcrafted checkerboard SiGeSn alloy systems is shown in Figure~\ref{fig:sigesn-handcrafted}. Checkerboard systems of 216 atoms and 64 atoms were considered. The generalization performance for GraNd worsens rapidly beyond $60$\%, whereas random pruning maintains high accuracy up to $80$\% pruning factor. However, CCS maintains near-original performance until $99$\%  pruning.

\begin{figure}[!h]
    \centering
\includegraphics[width=0.5\linewidth]{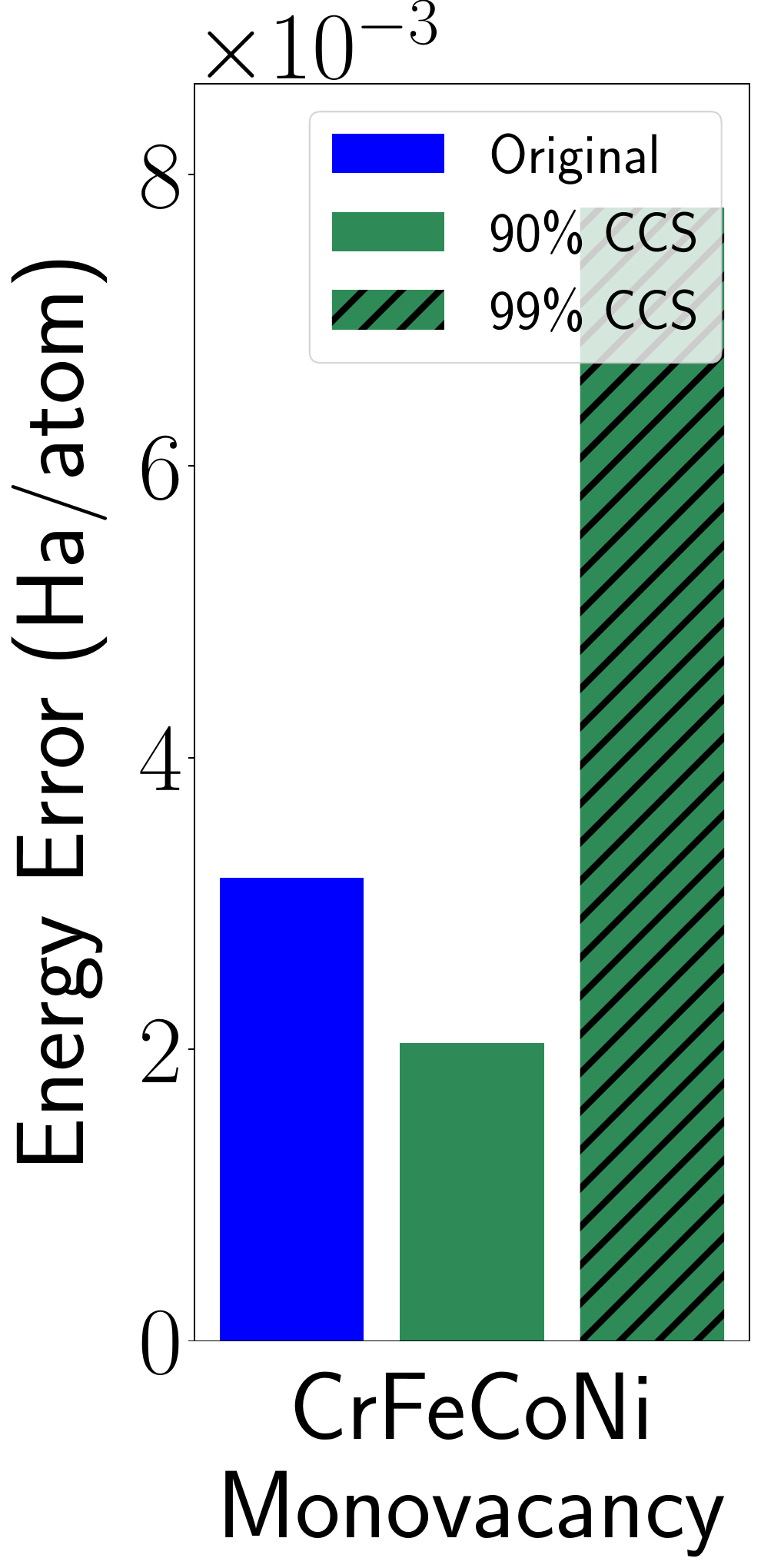}
\caption{\textbf{ {Structural} Generalizability tests for CrFeCoNi Monovacancy:} The error in the energy from the ML-predicted electron density for the original dataset, 90\% and 99\% CCS-based dataset.
    }
    \label{fig:generalizablity-crfeconi-mono-energy}
\end{figure}

 {For CrFeCoNi system with defects, the errors in electron density prediction for defects is shown in Figure~\ref{fig:crfeconi-defects}. The generalization performance for GraNd worsen rapidly beyond $60$\%, whereas random pruning maintains high accuracy up to $80$\% pruning factors. However, CCS maintains near-original performance until $90$\% pruning and show very little degradation at 99\% pruning compared to the other methods. }
We also compute the total energy of the system from the ML-predicted electron density, via postprocessing, for these generalizability tests. The total energy errors (Hartree/atom) relative to the KS-DFT calculations, are shown in Figure~\ref{fig:generalizablity-crfeconi-mono-energy},  {which shows that the 90\% pruning through CCS performs in the same level as the original dataset, achieving close to the chemical accuracy. However, 99\% pruning through CCS exceeds chemically accurate threshold.}
\subsection{ {Generalization across compositions}}
\begin{figure}[!h]
  \begin{minipage}{0.45\textwidth}
    \centering
    \includegraphics[width=0.95\linewidth]{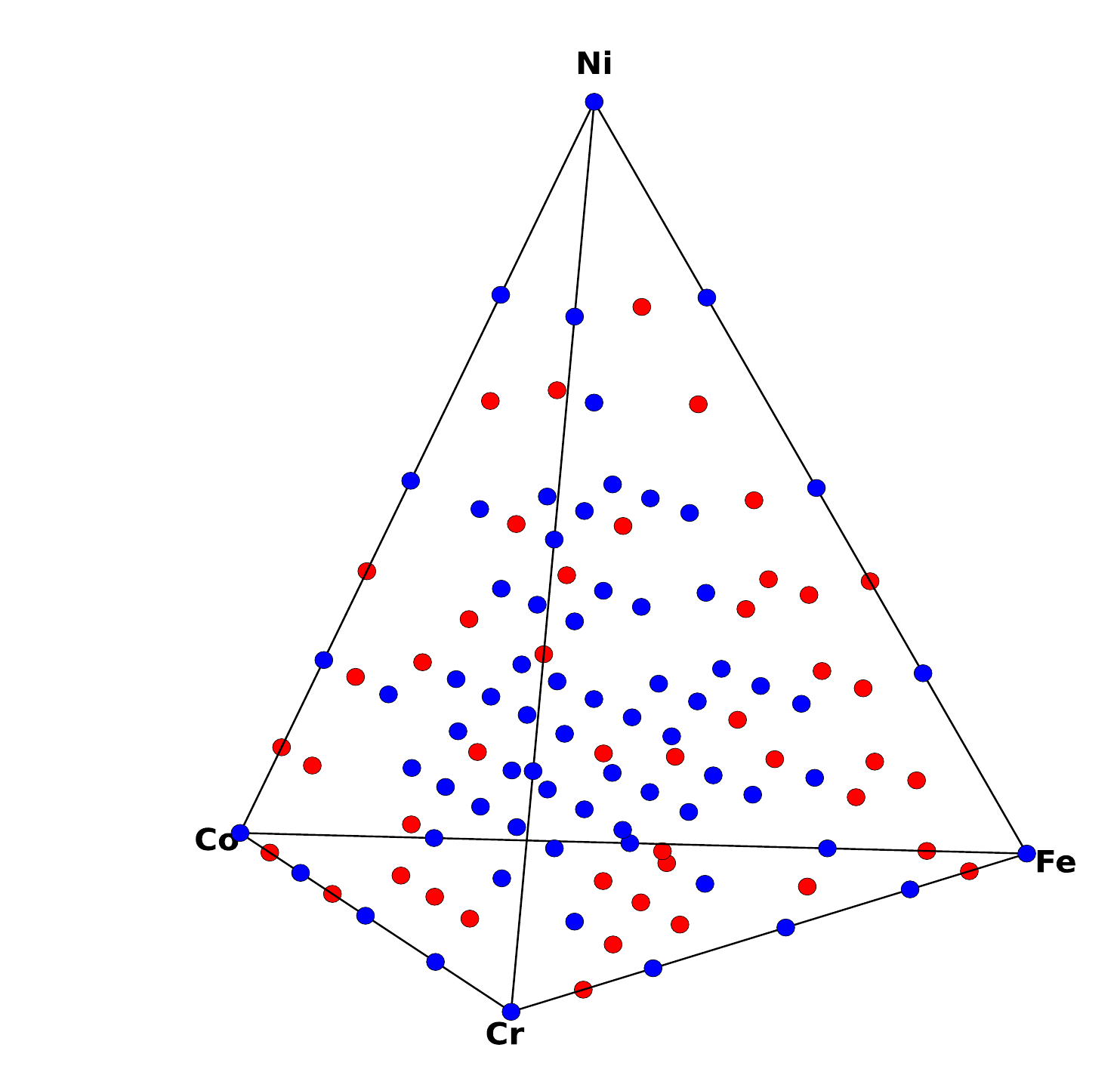}
   \caption{ {Composition space of the CrFeCoNi alloy systems. We have used 69 compositions (blue) during training; and tested on 36 unseen compositions(red) for generalization.}}
    \label{fig:comps-space}
  \end{minipage}
  \hspace{25mm}
  \begin{minipage}{0.45\textwidth}
    \centering
    \includegraphics[width=\linewidth]{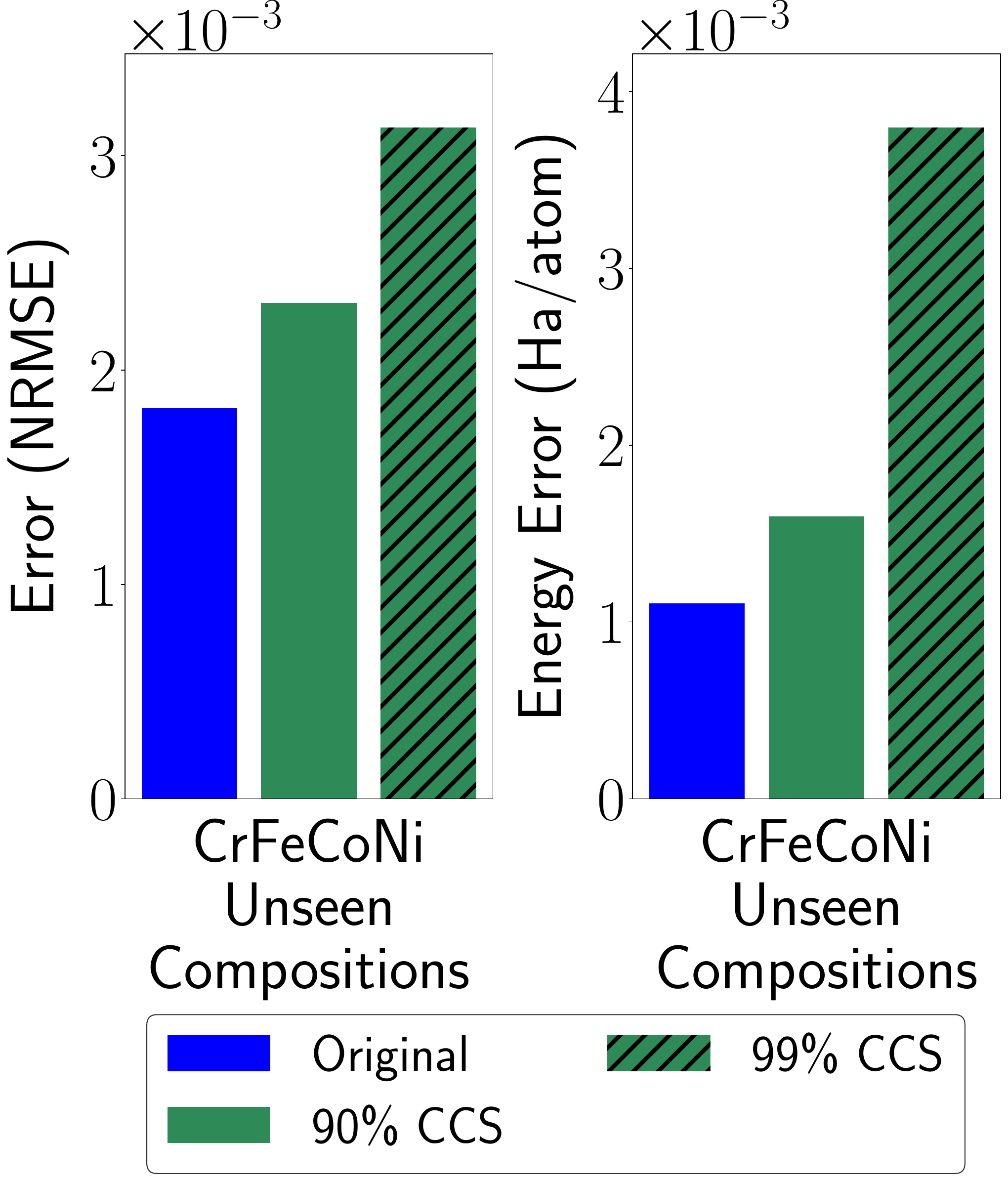}
   \caption{ {\textbf{Generalization tests for unseen compositions of CrFeCoNi}: (left) The error in the electron density prediction and (right) the energy error, for models trained on the original, 90\% CCS, and 99\% CCS pruned datasets.}}
    \label{fig:crfeconi-unseen}
  \end{minipage}
\end{figure}

 {To test extrapolation across composition space, we evaluate the ML models for CrFeCoNi on $36$ unseen compositions. The training and test compositions are shown in the composition space of CrFeCoNi in the Figure.~\ref{fig:comps-space}. This constitutes a stringent test for generalizability. The results of pruning are reported in Figure~\ref{fig:crfeconi-unseen}. The $90\%$ CCS-pruned model shows a slight increase in error relative to the original model, while remaining chemically accurate. The performance of the $99\%$ CCS-pruned model degrades further, exceeding the chemical-accuracy threshold. Such increase in error for high factors of pruning is the expected for unseen compositions. }

\subsection{ {Temperature Extrapolation}}
\begin{figure}[tbh!]
    \centering
    \includegraphics[width=0.95\linewidth]{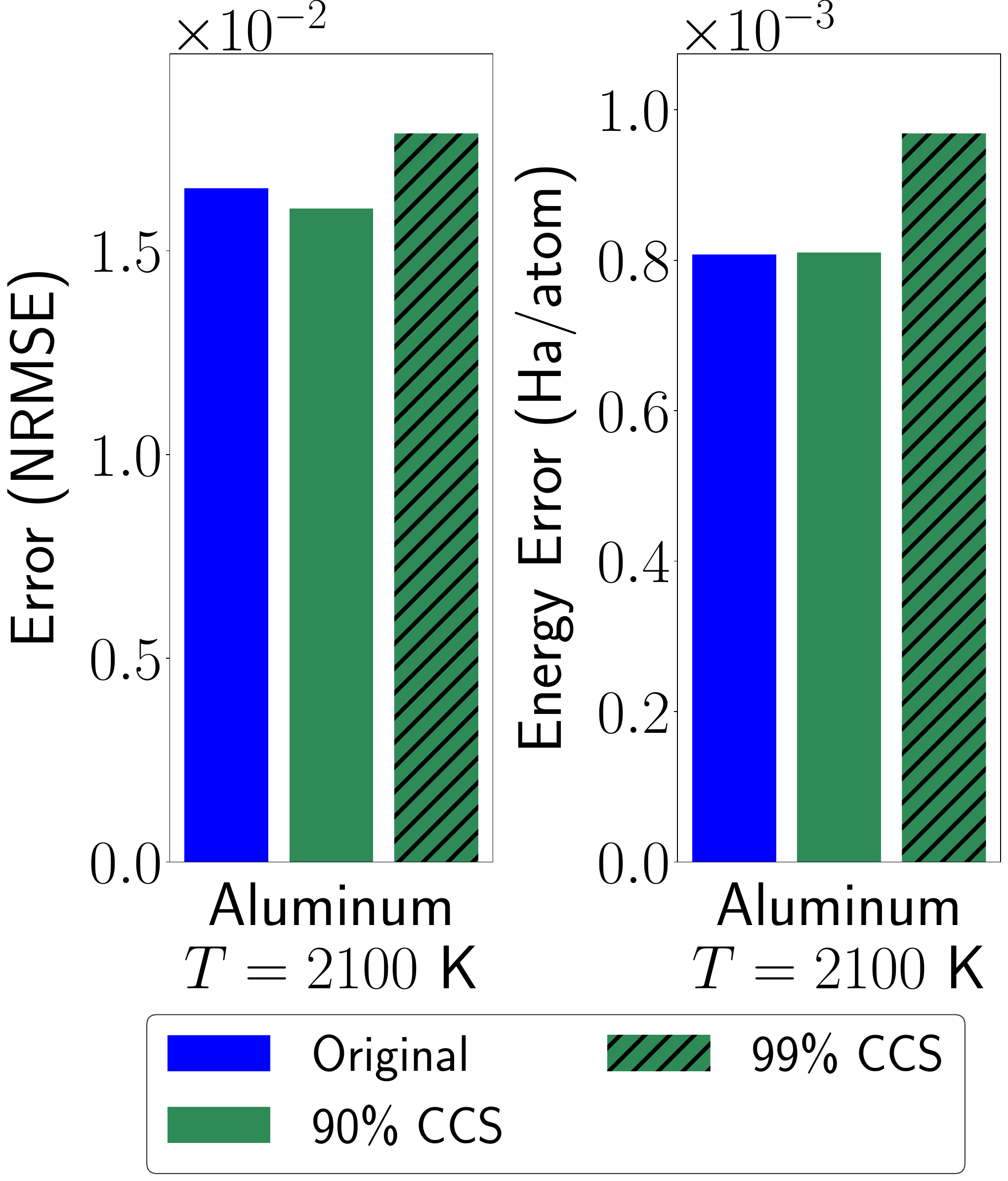}
   \caption{ {\textbf{Generalization tests for temperature extrapolation for aluminum:} (left) The error in electron density prediction and (right) the energy error for Aluminum 32-atoms system at 2100K, for models trained on the original, 90\% CCS, and 99\% CCS pruned datasets.}}
    \label{fig:aluminum-2100k}
\end{figure}
 {Generalizability beyond the temperature range of training data is tested. The aluminum models are trained on configurations sampled at six simulation temperatures ($298$, $600$, $900$, $1200$, $1500$, and $1800$~K) and tested on unseen dataset generated at $2100$K. As shown in Figure~\ref{fig:aluminum-2100k}, the errors in the electron density prediction and energy are very close for the original, $90\%$ CCS, and $99\%$ CCS datasets, and all three energy errors remain well within chemical accuracy. The pruned-data models therefore reproduce the temperature extrapolation behavior of the full-dataset model with minimal degradation even at $99\%$ pruning. }

\subsection{ {Size Extrapolation}}
\begin{figure}[tbh!]
    \centering
    \includegraphics[width=0.95\linewidth]{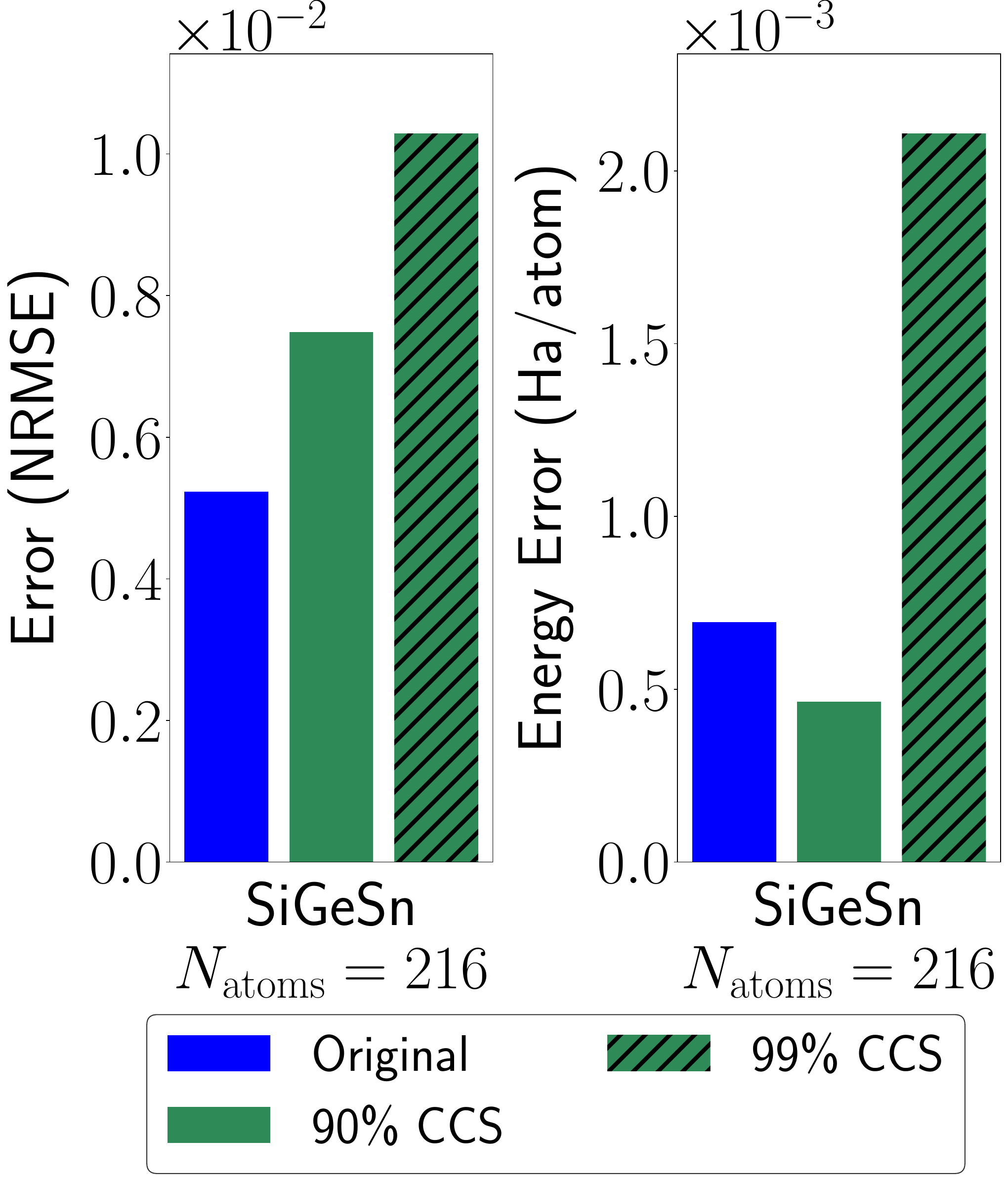}
   \caption{ {\textbf{Generalization tests for size extrapolation for SiGeSn:} (left) The error in electron density prediction and (right) the energy error for SiGeSn 216-atoms system at 2400K, for models trained on the original, 90\% CCS, and 99\% CCS pruned datasets.}}
    \label{fig:sigesn-216atoms}
\end{figure}
 {Generalizability beyond the size  of training system is tested. We evaluate the SiGeSn models on 216-atom systems whereas only 64-atom systems are used in training. We have used three compositions of SiGeSn at 2400~K. The density prediction error increases with pruning fraction slightly; however, the 90\% CCS model maintains chemical accuracy in energy, and the 99\% CCS model remains close to the chemical-accuracy threshold, as shown in Figure~\ref{fig:sigesn-216atoms}.}
\begin{figure*}[tbh!]
    \centering
    \includegraphics[width=\textwidth]{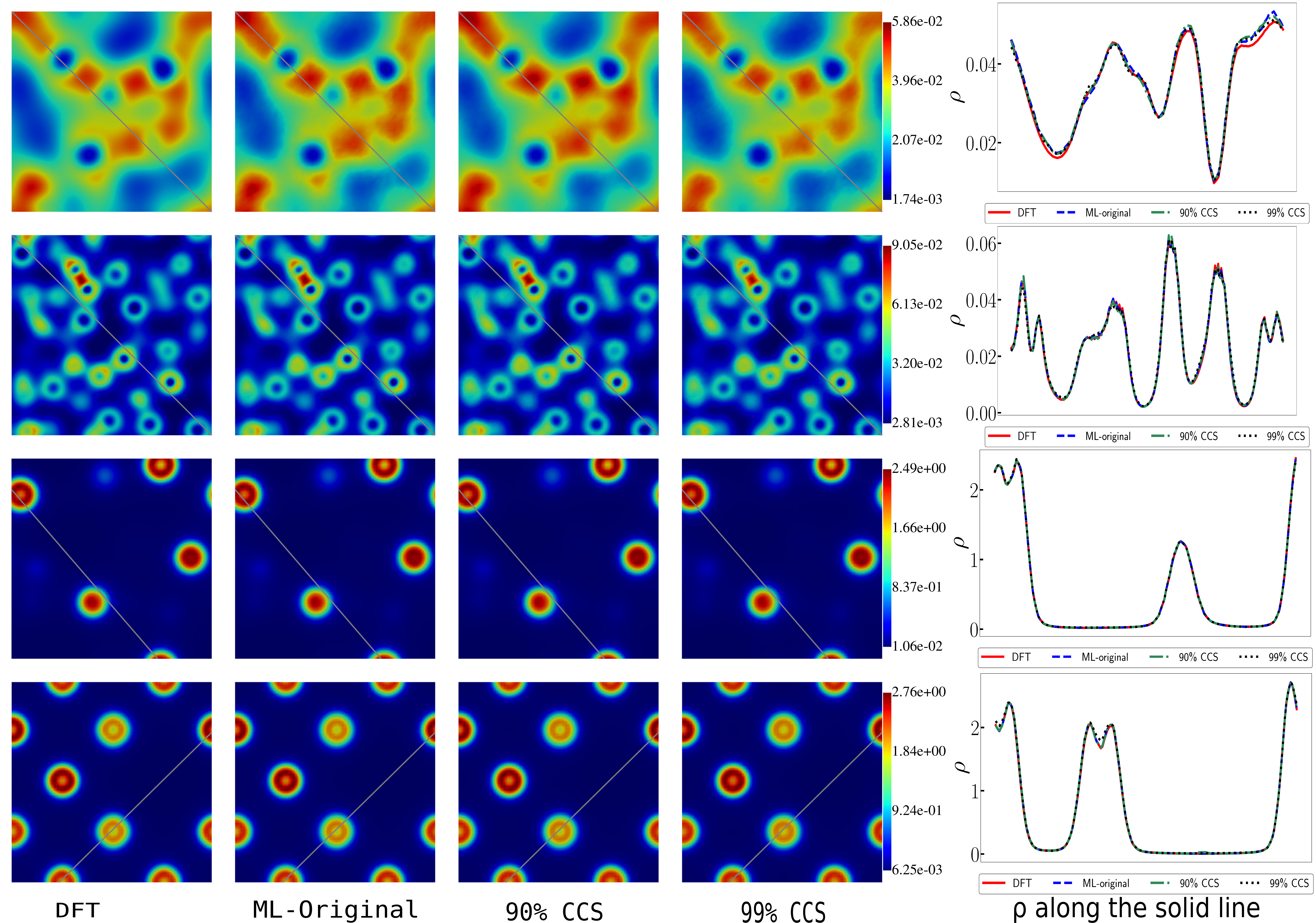}
   \caption{ {\textbf{Model generalizability to unseen configurations:} 
    2D slices of electron density obtained by KS-DFT, ML model trained on the  original dataset, 90\% CCS, and 99\% CCS based pruned datasets. The electron density($\rho$) along the solid line is also compared with KS-DFT, showing remarkable agreement. \textit{(First row) }Temperature extrapolation test for 32-atoms aluminum systems. \textit{(Second row)} System size extrapolation test for SiGeSn system. \textit{(Third row)} Unseen compositions test for the CrFeCoNi system \textit{(Fourth row)} Mono-vacancy defect for 32-atoms CrFeCoNi systems. The unit of electron density is $\text{Bohr}^{-3}$ in atomic units.}}
    \label{fig:generalization-extra}
\end{figure*}
\clearpage

\section{Time Comparison}\label{sm:TimingComparison}
\begin{figure}[h!]
    \centering
    \includegraphics[width=1\linewidth]{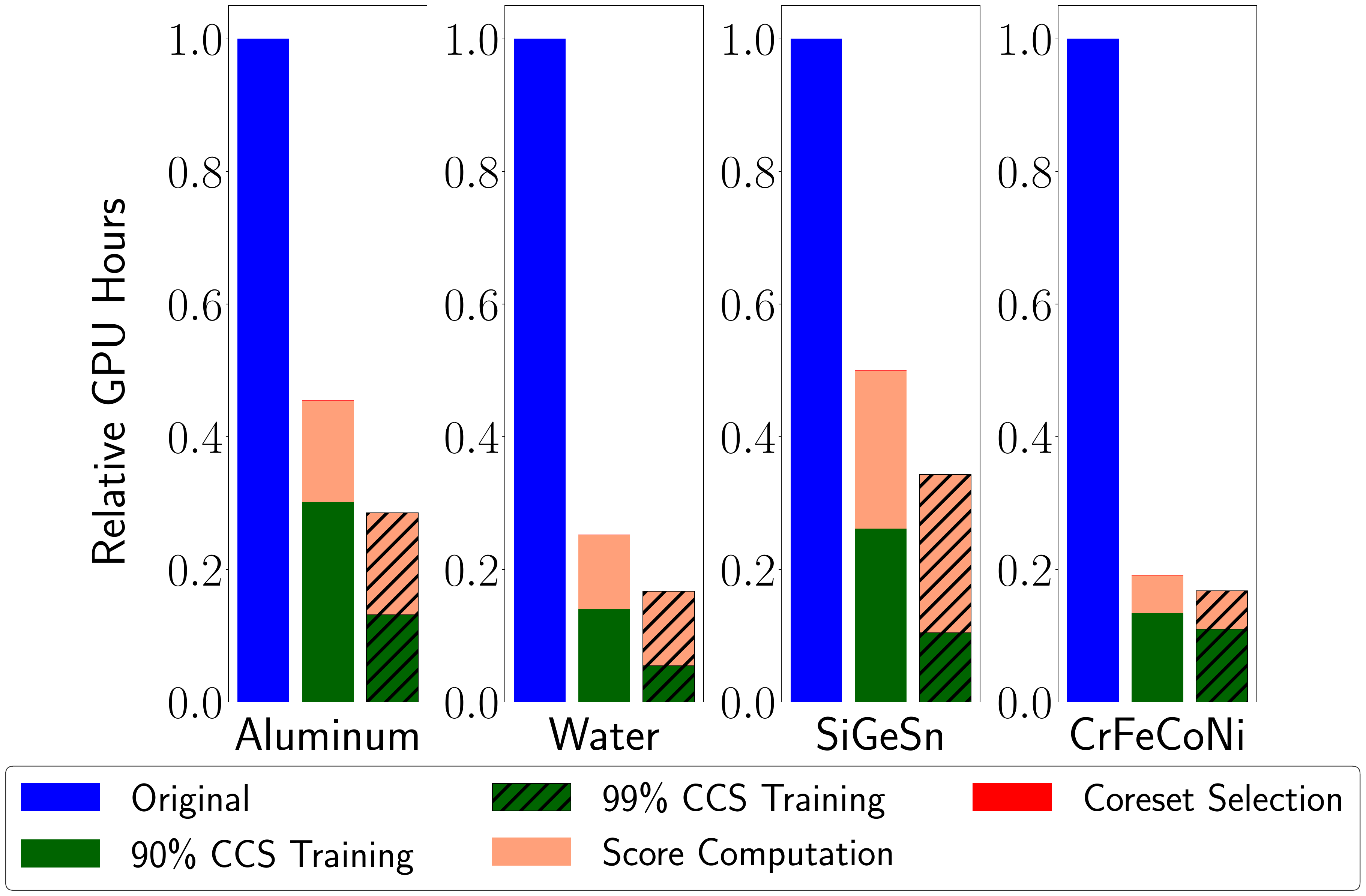}
    \caption{ {Time comparison between the ML model trained on the original dataset, 90\% CCS and 99\% CCS based pruned dataset. The coreset selection time is small and hence is not visible.}}
    \label{fig:consolidated-time}
\end{figure}
The full pipeline cost, including GraNd score computation, coreset selection, and training on the pruned dataset, relative to training on the original dataset are shown in Figure~\ref{fig:consolidated-time}. The cost of coreset selection step through CCS is negligible across all systems, taking 30 to 60 seconds. The GraNd score computation costs significant fraction of computational overhead. Even accounting for this cost, the total computational cost is substantially lower than training on the original dataset for all systems: at 99\% CCS pruning, corresponding to speedups of roughly $3\times$ to $6\times$ over the original training time. At 90\% CCS pruning, the speedups range from approximately $2\times$ to $5\times$.
\clearpage
\color{black}
\section{Visualization of Redundancy}
\label{sm:viz-pruning}
\begin{figure*}
    \centering
    \includegraphics[width=0.95\linewidth]{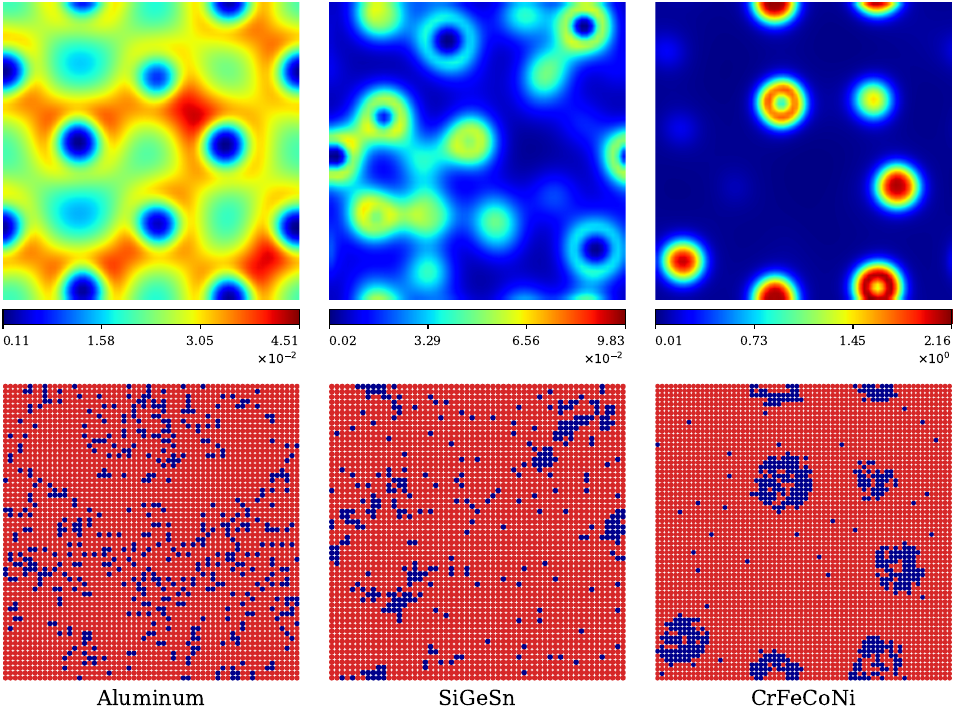}
    \caption{Spatial distribution of pruned and retained grid points for Aluminum, SiGeSn, and CrFeCoNi at 90\% CCS pruning. For each material, the top figure shows the KS-DFT electron density on a representative 2D slice, and the bottom figure shows the corresponding spatial map of pruned (red) and retained (blue) grid points. The unit of electron density is $\text{Bohr}^{-3}$ in atomic units.}
    \label{fig:pruned-slices}
\end{figure*}
 {To visualize which grid-points were kept or pruned, we have plotted the spatial distribution of retained and pruned grid points for Aluminum, CrFeCoNi and SiGeSn at 90\% CCS pruning, as shown in Figure~\ref{fig:pruned-slices}. The figures reveal that CCS retains grid points in regions of high and rapidly varying electron density --- e.g. near atoms and bonding regions --- while predominantly pruning the low-density interstitial regions. For aluminum, whose density variation is comparatively less rapid, the retained points are correspondingly more uniformly distributed. In all cases, a small number of scattered interstitial points are also retained, consistent with CCS's stratified sampling across the importance scores. The CrFeCoNi system does show strong clustering of retained data points near the atomic nuclei, very likely due to the strongly localized pseudocharges associated with the elements in this system, arising from the inclusion of semi-core states in the corresponding pseudopotentials. Notably, the spatial pattern of data retention reflects the underlying electronic structure of the material. Importantly, this physically meaningful spatial selection pattern emerges naturally from the ML-based importance scores, without any explicit physics-informed assumption about which spatial regions are informative.}

\section{Validation of Forces}
\label{sm:forces}
\begin{figure}
    \centering
    \includegraphics[width=0.85\linewidth]{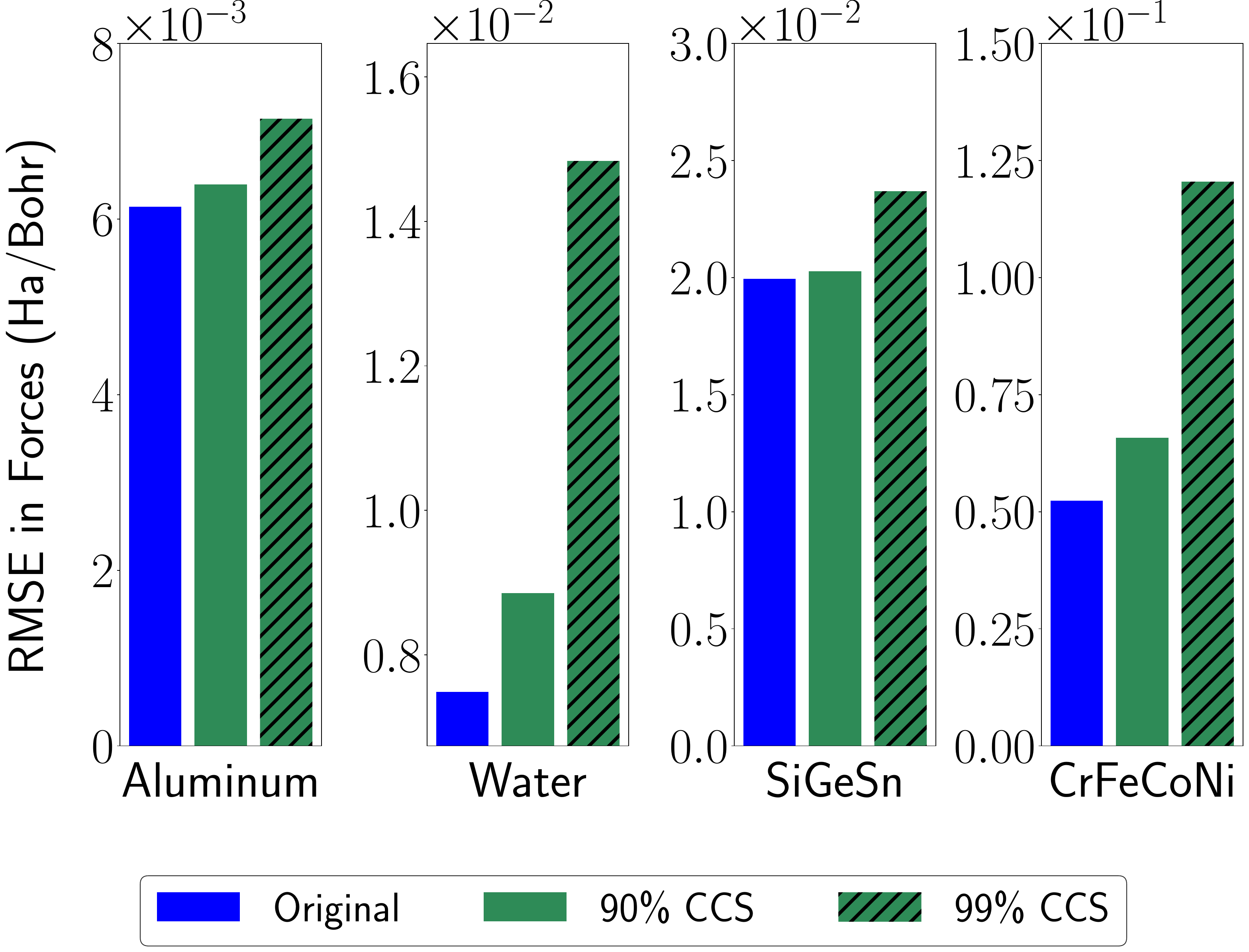}
    \caption{ {Error(RMSE) in atomic forces, as obtained from the ML-predicted electron density for the original dataset, 90\% CCS, and  99\% CCS-based pruned datasets.}}
    \label{fig:forces}
\end{figure}

 {It should be noted that the ML model is not trained using force data. Consequently, the forces, which are computed solely from the ML predicted electron densities, may be predicted less accurately than the electron densities themselves. We have carried out this analysis of force calculation, and our key finding is that, like the energies, the force errors do not necessarily get much worse, depending on the level of pruning, i.e., the pruned ML models have similar force errors (with respect to KS-DFT references), as do the forces computed from the original ML model. Specifically, from  Figure~\ref{fig:forces} above, we see that the errors in the forces of the ML models, with respect to reference KS-DFT results (force errors with respect to KS-DFT results for each snapshot were computed as the ${L}^2$ norm of the difference, over all atoms and all force components). Although the errors in the forces remain low for the $90\%$ pruning case, they increase significantly for the $99\%$ pruning case for some cases. 
We find it pertinent to point out, that in general, the errors in the forces of the ML models are somewhat worse than errors in the energies. However, this is to be expected, since force errors are known to be far more sensitive to self-consistency errors in the electron density \cite{kim2014ions}, as compared to the energy, which benefits from a variational structure and the quadratically correct Harris-Foulkes formula \cite{harris1985simplified}.  Notably, in terms of force errors, the performance of the original ML model is the worst for the \ce{CrFeCoNi} system, due to the hard pseudopotentials associated with the elements present. Therefore, the error in forces does not  increase significantly up to $90\%$ pruning.}

\section{Pruning using the SOAP descriptor}
\label{sm:soap}
\begin{figure}
    \centering
    \includegraphics[width=0.95\linewidth]{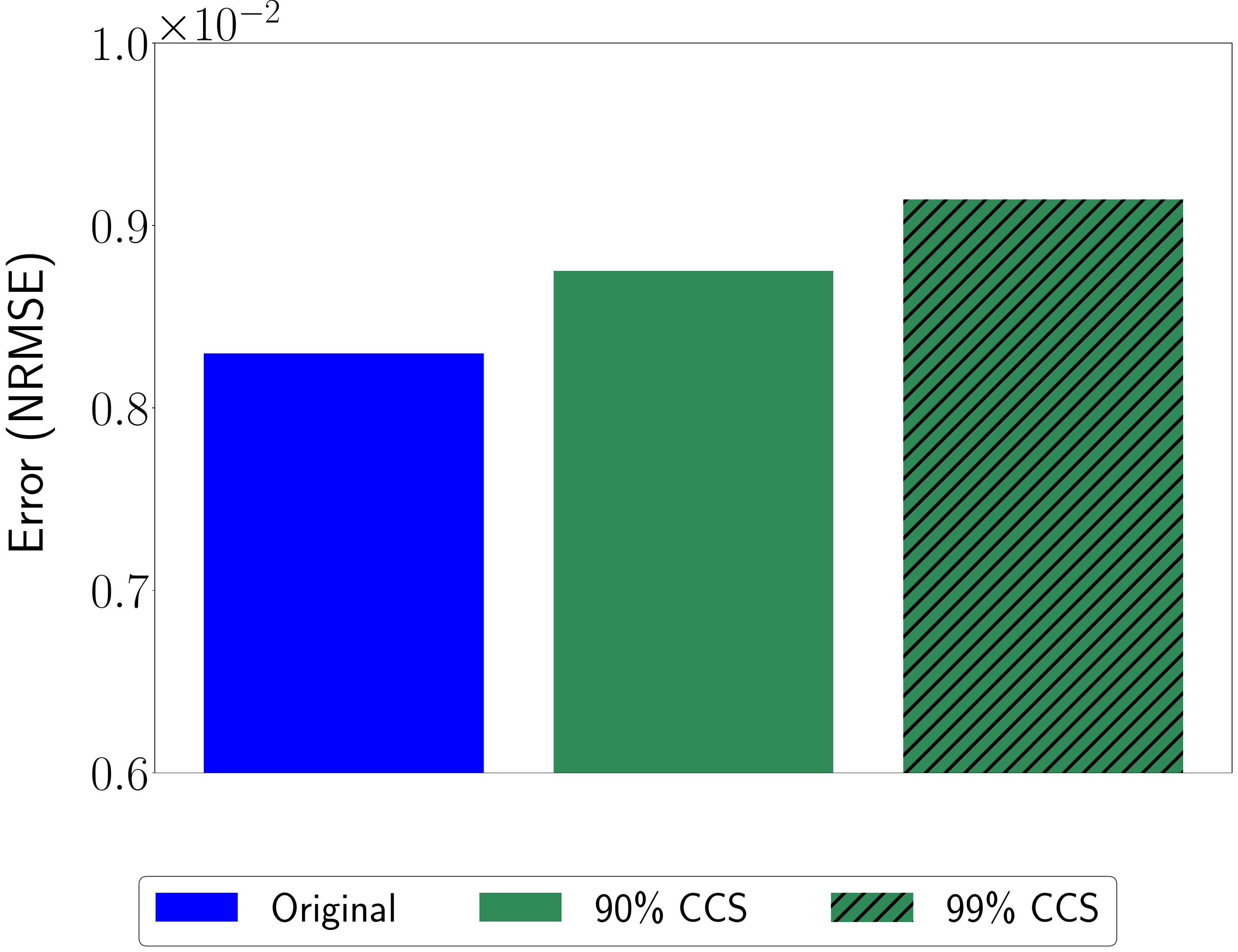}
    \caption{\textbf{Redundancy using the SOAP descriptor for the SiGeSn system}: Error in electron density prediction for the ML models trained on the original, 90\% and 99\% CCS pruned datasets. }
    \label{fig:soap}
\end{figure}
 {We note that the pruning approaches themselves (GraNd and CCS) are descriptor-agnostic by construction, as they operate on training dynamics and stratified sampling across importance scores. To test whether the observed redundancy is an artifact of our descriptor choice, we have repeated the pruning experiments for a representative system (SiGeSn) using the Smooth Overlap of Atomic Positions (SOAP) descriptors~\cite{bartok2013representing} in place of the currently used descriptors, keeping everything else the same. The results are shown in Figure~\ref{fig:soap}. The redundancy findings persist under the SOAP descriptor as well. The models trained on 90\% CCS-pruned data achieve near-original performance, and even at 99\% pruning, the error increase is very small. Such findings imply that the observed redundancies are generalizable to other descriptors. }\\
\section{Estimating Intrinsic Dimensions}
\label{sm:id}
\begin{table*}[b!] 
\centering
\begin{tabular}{lccccccc} 
\toprule
System & MLE & IID & GRIDE & MiND\textsubscript{ML} & MID & PCA (99\%) & Ambient \\
\midrule
\midrule
Aluminum  & 60 - 65 & 72 & 55 - 60 & 68 - 70 & 131 & 206 & 236 \\
Water     & 83 - 92 & 95 & 75 - 83 & 90 - 93 & 200 & 247 & 295 \\
SiGeSn    & 78 - 89 & 97 & 69 - 78 & 89 - 94 & 196 & 243 & 295 \\
CrFeCoNi  & 78 - 88 & 96 & 68 - 80 & 89 - 93 & 201 & 259 & 295 \\
\bottomrule
\end{tabular}
\caption{ {Intrinsic dimension (ID) estimates for the four material systems using several estimators, compared with the number of principal components required to capture 99\% of the variance (PCA) and the ambient descriptor dimension. The neighbourhood-based estimators (MLE, GRIDE, MiND\textsubscript{ML}) are reported as the range obtained by considering the following number of  neighbors: $100, 250, 500, 750, 1000$.  The IID as well as the MID sweep across 100 -- 1000 nearest neighbors for their estimation.}}
\label{tab:id}
\end{table*}

 {We estimated the intrinsic dimension (ID) using several additional estimators, namely Gride~\cite{denti2022gride}, MiND\textsubscript{ML}~\cite{rozza2012mind}, iterative-based intrinsic Dimension(IID), and the modified intrinsic dimension (MID) estimator~\cite{qiu2023underestimation}. The results are summarized in Table~\ref{tab:id}. The Gride, MiND\textsubscript{ML}, and IID estimators produce ID values that corroborate closely with the maximum likelihood estimator (MLE) values reported in the manuscript for each system. Although there is no universally accepted method for accurately estimating the intrinsic dimension of high-dimensional nonlinear manifolds for real datasets, these independent estimators shows close agreements. However, the manifold nonlinearity may lead to  underestimation of the intrinsic dimension depending on the dataset's high dimensionality. To account for this effect, we also employed the modified intrinsic dimension (MID) estimator, which yields the most conservative (largest) estimates among the methods considered. The MID estimator yields systematically larger IDs: 131 for aluminum, 196 for SiGeSn, 200 for water, and 201 for CrFeCoNi. Importantly, even the largest estimates remain below the corresponding PCA dimensions required to explain $99\%$ of the variance (206--259) and are substantially lower than the ambient descriptor dimensions (236--295) for all systems considered. Therefore, our conclusions are robust with respect to the choice of intrinsic-dimension estimator. Regardless of the estimation method, the electronic-structure descriptors occupy a nonlinear manifold whose intrinsic dimension is significantly smaller than the ambient descriptor space. At the same time, the variation among the estimators highlights the inherent uncertainty in assigning a single numerical value to the intrinsic dimension of complex, high-dimensional nonlinear manifolds and suggests that the reported IDs should be interpreted as method-dependent estimates rather than exact quantities.}\\

\section{Additional Results}\label{sm:additional-results}
\begin{figure}[tbhp]
    \centering
    \includegraphics[width=\linewidth]{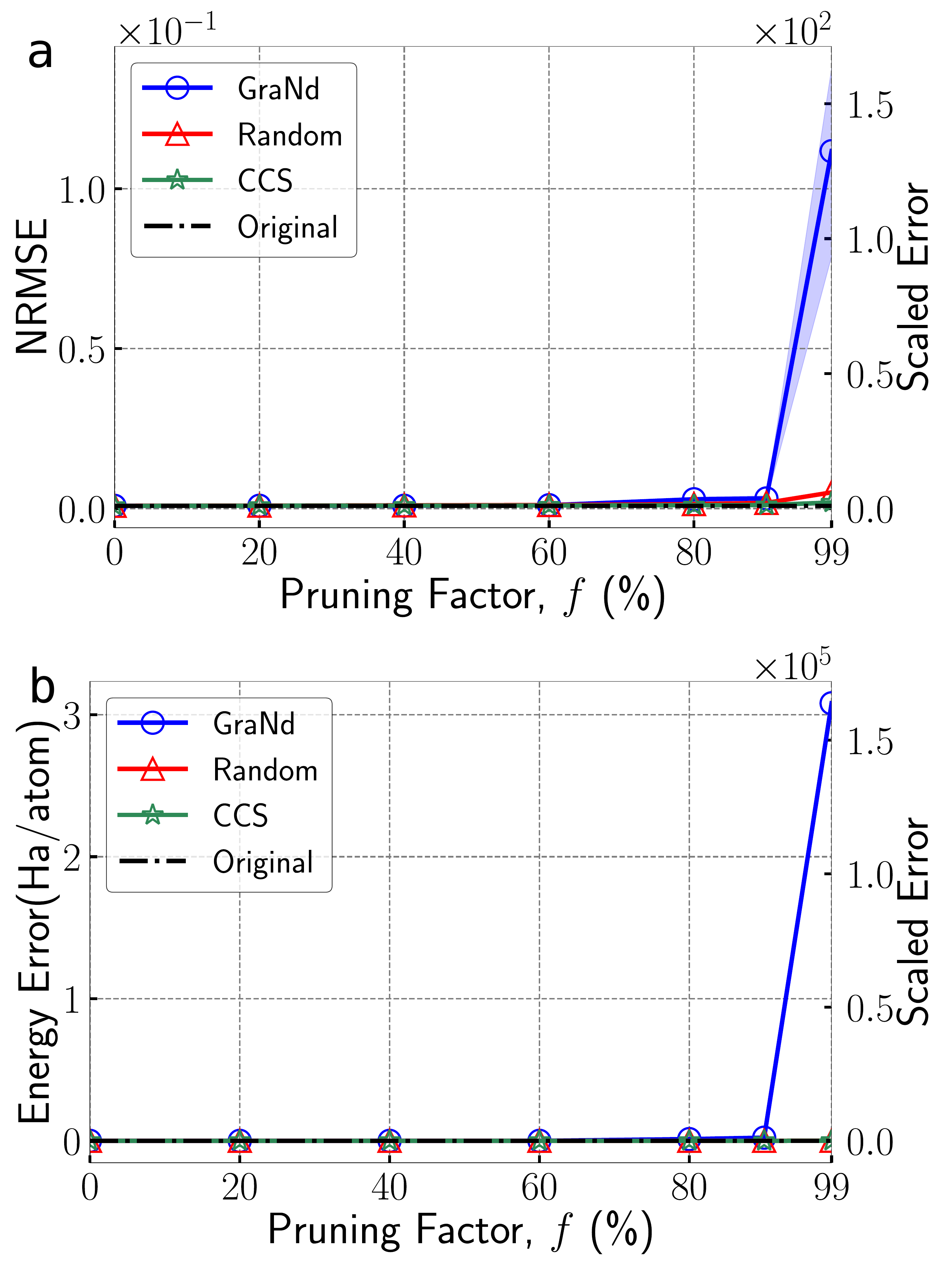}
        \caption{\textbf{Water system:} (a)  Error in the predicted electron density for various pruning factors. (b)  Error in energy obtained from the predicted electron density for various pruning factors. 
        }
    \label{fig:water-consolidated-full}
\end{figure}

\begin{figure}[tbhp]
    \centering
    \includegraphics[width=\linewidth]{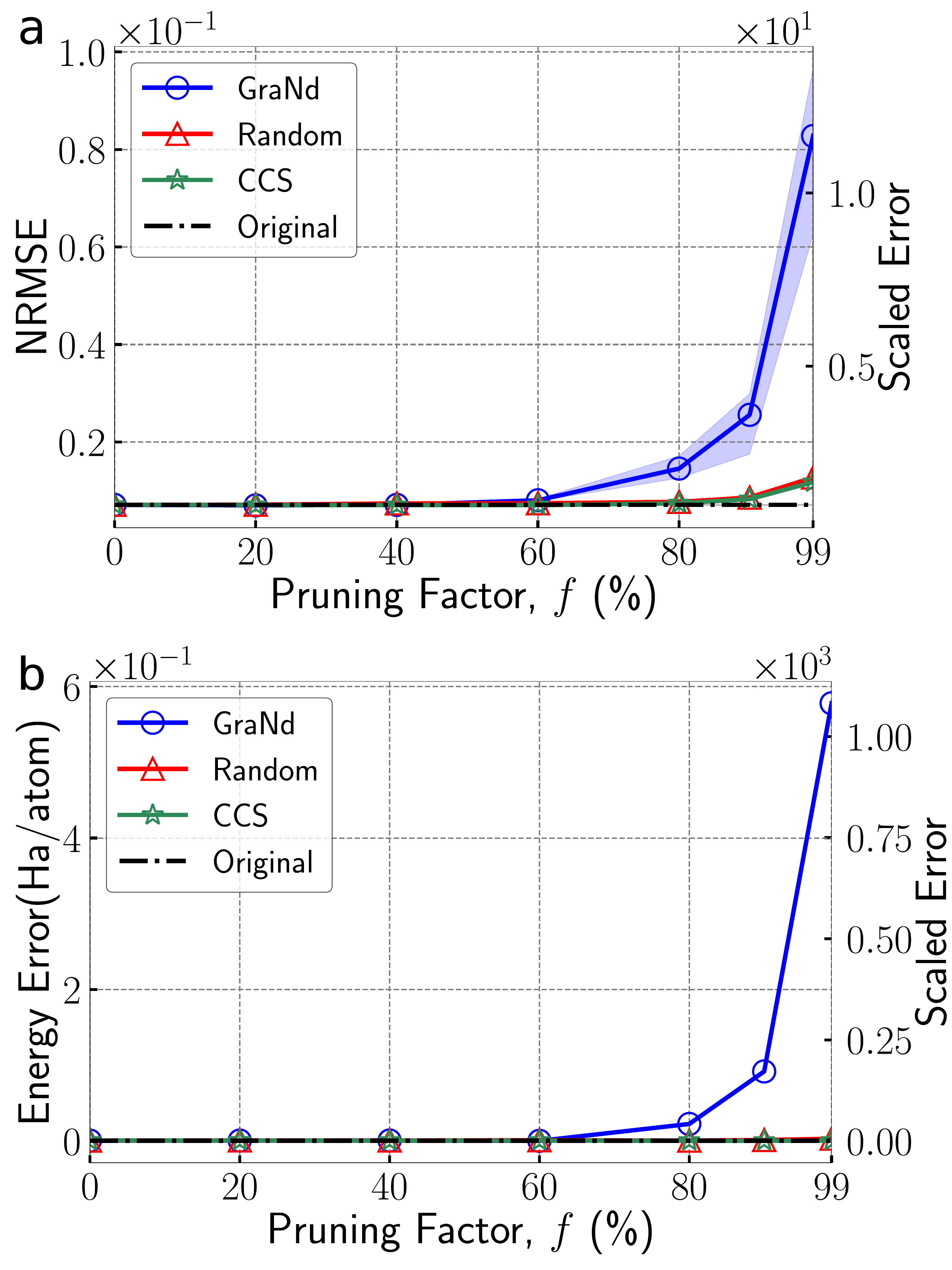}
        \caption{\textbf{SiGeSn system:} (a)  Error in the predicted electron density for various pruning factors. (b)  Error in energy obtained from the predicted electron density for various pruning factors. 
        }
    \label{fig:sigesn-consolidated-full}
\end{figure}

\begin{figure}[!t]
    \centering
    \includegraphics[width=\linewidth]{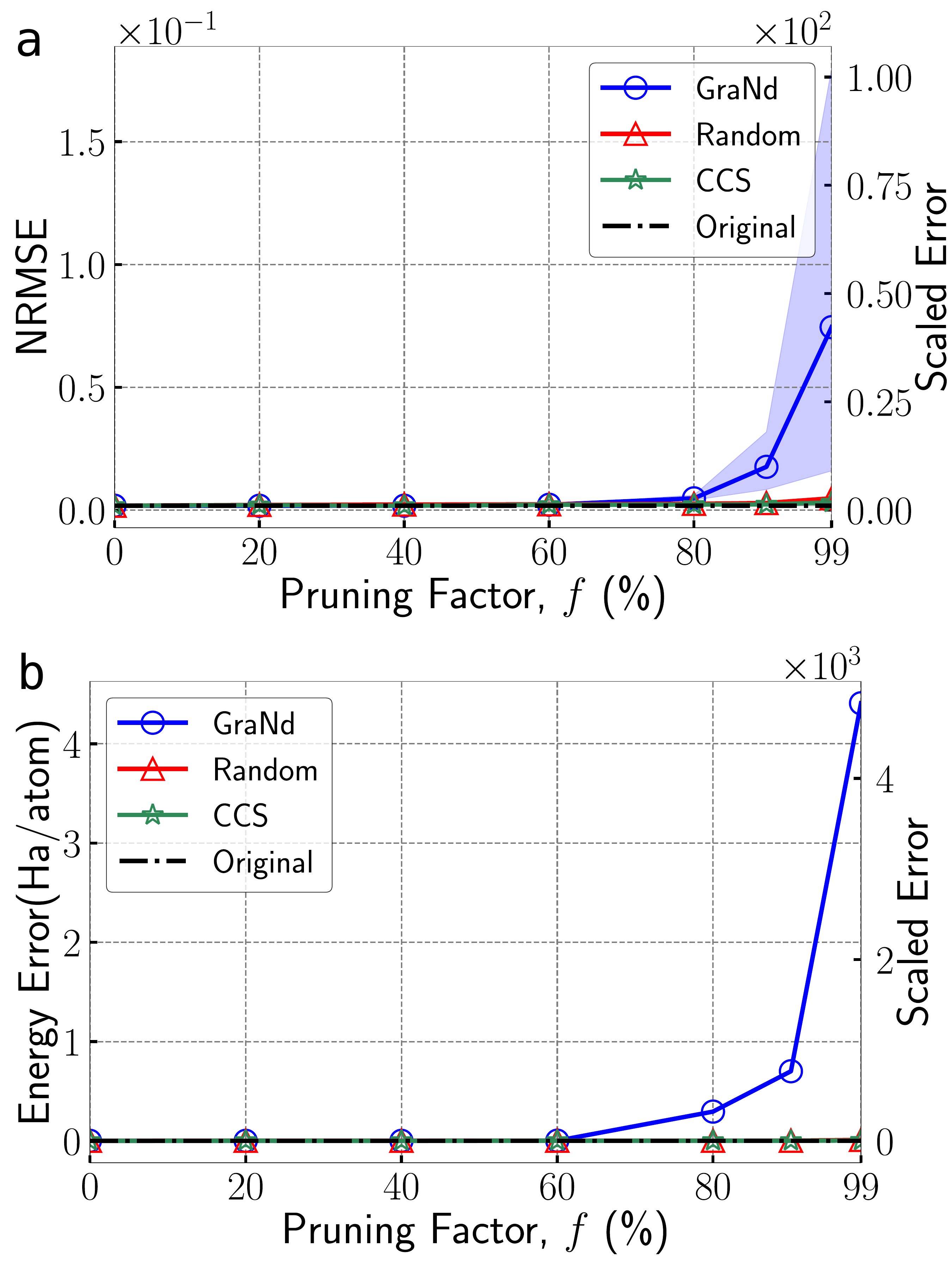}
        \caption{\textbf{CrFeCoNi system:} (a)  Error in the predicted electron density for various pruning factors. (b)  Error in energy obtained from the predicted electron density for various pruning factors. 
        }
    \label{fig:crfeconi-consolidated-full}
\end{figure}

A complete and unscaled view of the error versus pruning factor for water, SiGeSn, and CrFeCoNi, (corresponding to Figure~\ref{fig:consolidated-all-nrmse-energy} of the main text) is provided in Figure~\ref{fig:water-consolidated-full}, Figure~\ref{fig:sigesn-consolidated-full}, and Figure~\ref{fig:crfeconi-consolidated-full} for completeness. It shows that the GraNd method becomes too erroneous beyond 60\% pruning for water, SiGeSn and CrFeCoNi. However, the random pruning provides high accuracy up to 90\%. 

\clearpage

\subsection{Isosurface Plots}\label{sm:isosurface}
\begin{figure*}[p]
    \centering
\includegraphics[width=\textwidth]{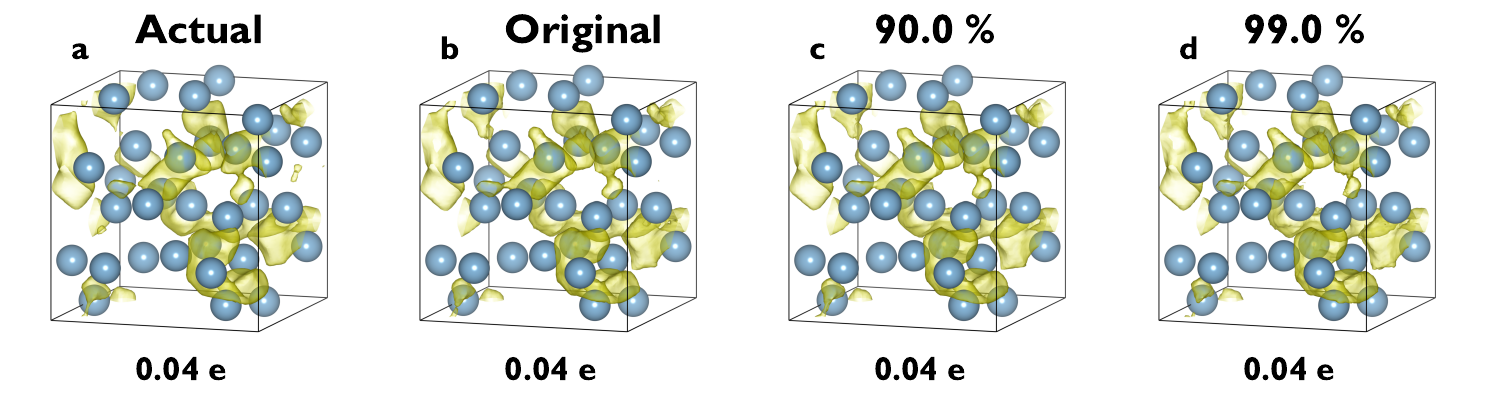}
    \caption{\textbf{Aluminum system at 1500 K.} Iso-surfaces of electron density  obtained by KS-DFT (a) and ML model for original (unpruned) (b), $90$\% CCS (c), and $99$\% CCS (d).}  
    \label{fig:al-iso-rho}
\end{figure*}

\begin{figure*}
    \centering
    \includegraphics[width=\textwidth]{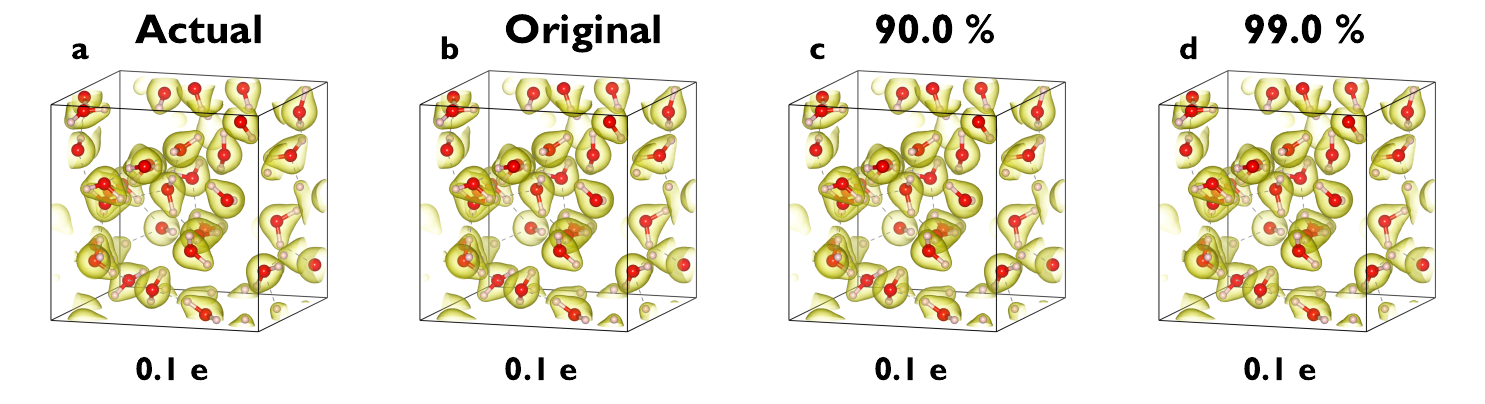}
        \caption{\textbf{Water system at 600K}: Iso-surfaces of electron density (at 0.1 e) obtained by DFT (a) and ML model for original (b), 90\% CCS (c), and 99\% CCS (d). The red and pink spheres represent the Oxygen and Hydrogen atoms.}
    \label{fig:water-iso-rho}
\end{figure*}

\begin{figure*}
    \centering    
    \includegraphics[width=\textwidth]{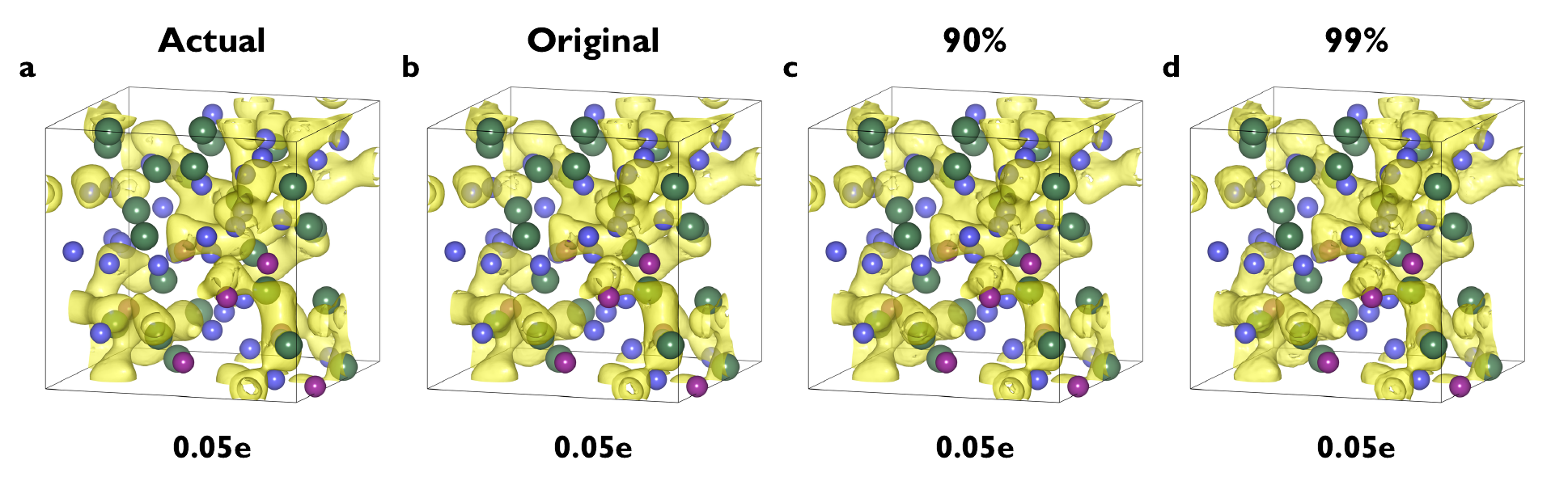}\caption{\textbf{$\text{Si}_{50}\text{Ge}_{12.5}\text{Sn}_{37.5}$ system at 2400K}: Iso-surfaces of electron density obtained by DFT (a) and ML model for original (unpruned) (b), 90\% CCS (c), and 99\% CCS (d). The blue, purple, and green spheres represent the Si, Ge, and Sn atoms.}
    \label{fig:sigesn-iso-rho}
\end{figure*}

\begin{figure*}
    \centering    
    \includegraphics[width=\textwidth]{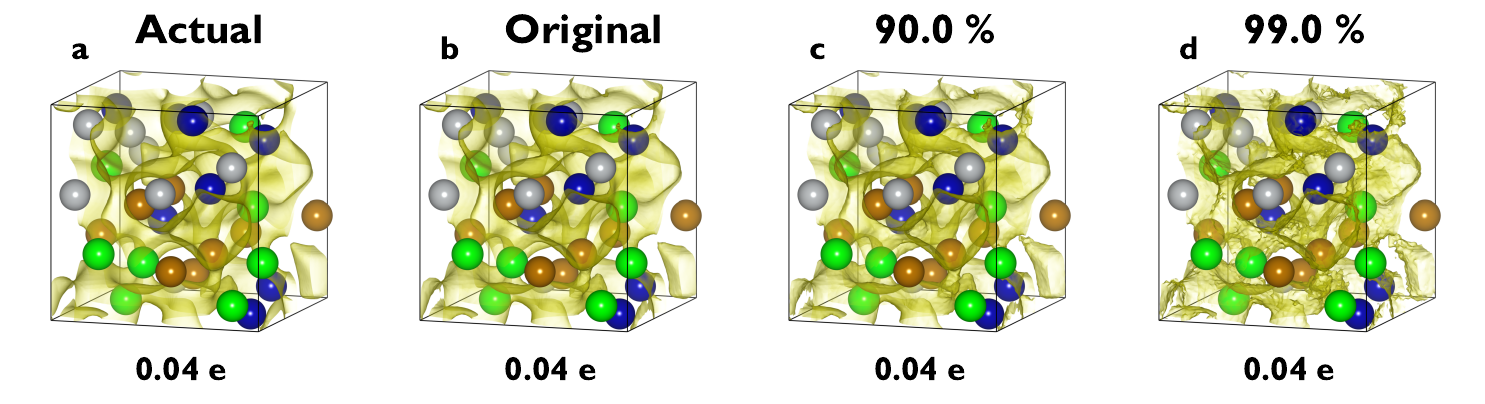}\caption{\textbf{$\text{Cr}_{25}\text{Fe}_{25}\text{Co}_{25}\text{Ni}_{25}$ system at 5000K}: Iso-surfaces of electron density obtained by DFT (a) and ML model for original (unpruned) (b), 90\% CCS (c), and 99\% CCS (d). The green, orange, blue, and gray spheres represent the Cr, Fe, Co, and Ni atoms.}
    \label{fig:crfeconi-iso-rho}
\end{figure*}

The 3D electron density fields are shown as iso-surface plots for the four materials in Figures~\ref{fig:al-iso-rho}, \ref{fig:water-iso-rho}, \ref{fig:sigesn-iso-rho}, and \ref{fig:crfeconi-iso-rho}. The overall pattern of electron density iso-surfaces is well captured by the original dataset, 90\%, and 99\% CCS based pruned datasets. 

\begin{figure*}
    \centering
    \includegraphics[width=\textwidth]{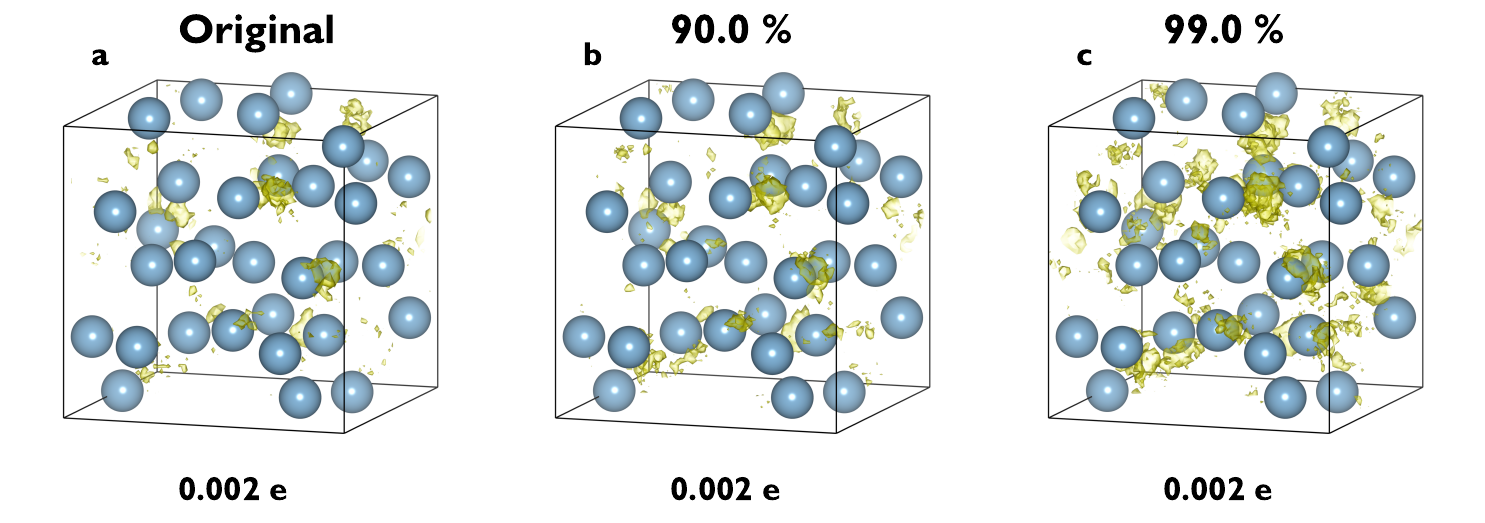}
        \caption{\textbf{Aluminum system at 1500 K.} Iso-surfaces of errors in electron density obtained by ML model for original (a), 90\% CCS (b), and 99\% CCS(c) respectively.}
    \label{fig:al-iso-error}
\end{figure*}

\begin{figure*}
    \centering
    \includegraphics[width=\textwidth]{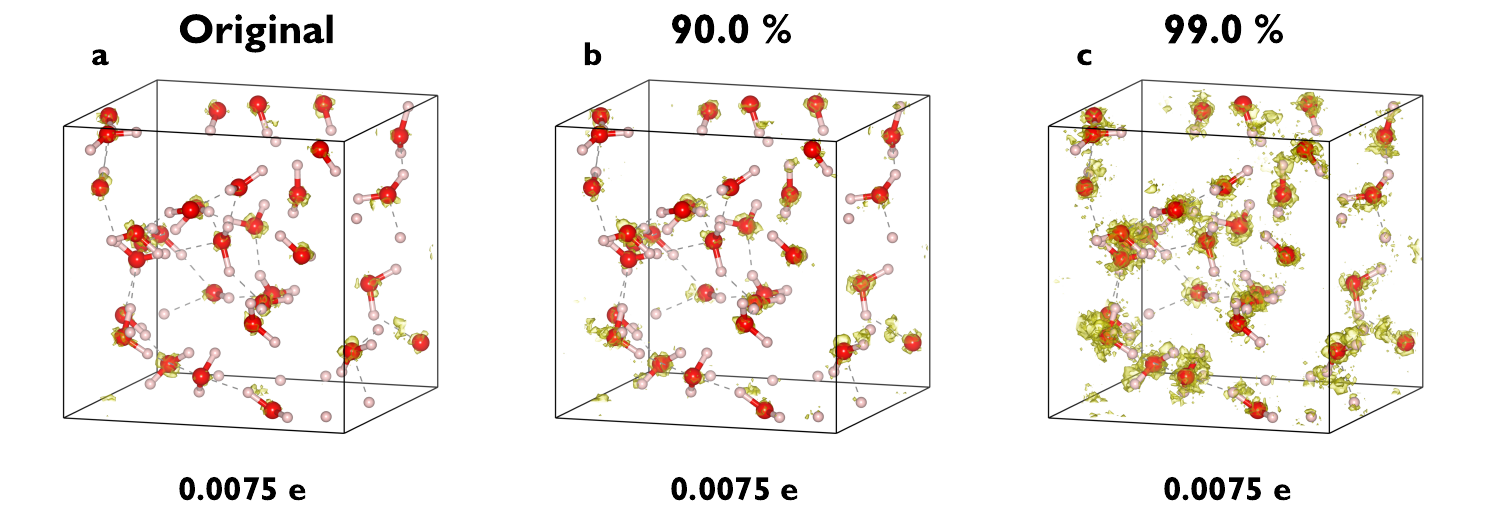}
        \caption{\textbf{Water system at 600K}: Iso-surfaces of errors in electron density (at 0.0075 e) obtained by ML model for original (unpruned) (a), 90\% CCS (b), and 99\% CCS (c) respectively. }
    \label{fig:water-iso-error}
\end{figure*}

\begin{figure*}
    \centering
    \includegraphics[width=\textwidth]{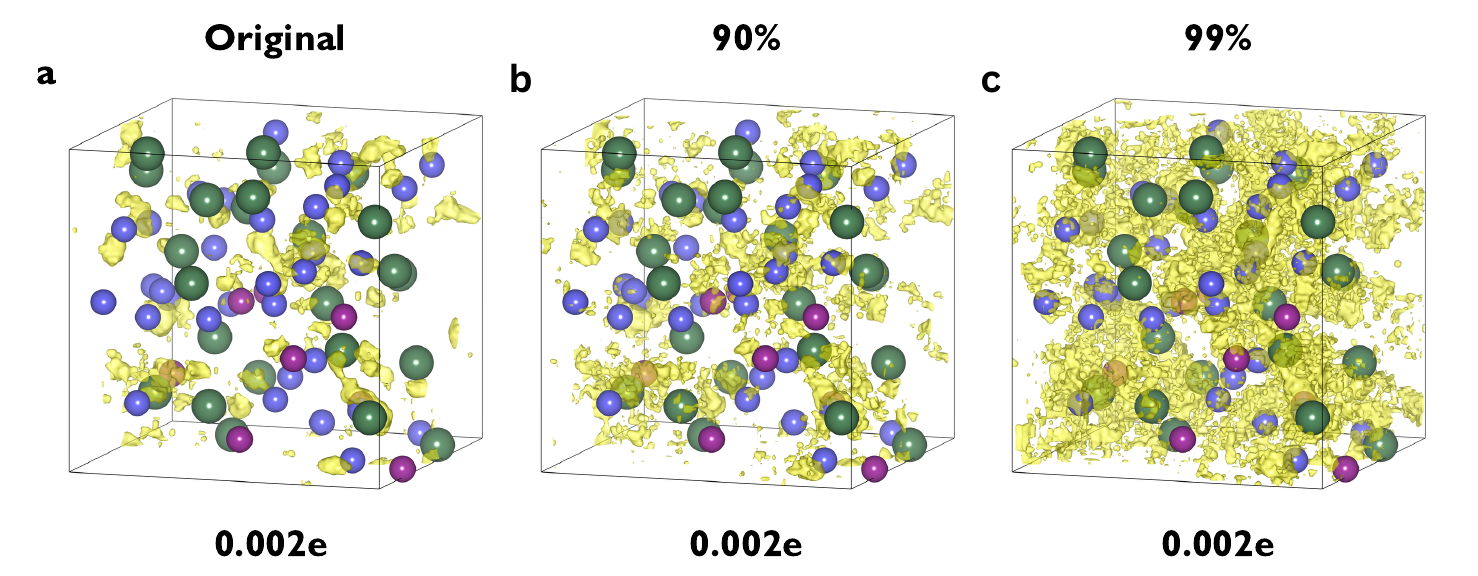}\caption{\textbf{$\text{Si}_{50}\text{Ge}_{12.5}\text{Sn}_{37.5}$ system at 2400K}: Iso-surfaces of errors in electron density obtained by ML model for original (a), 90\% CCS (b), and 99\% CCS (c) respectively.}
    \label{fig:sigesn-iso-error}
\end{figure*}

\begin{figure*}
    \centering
    \includegraphics[width=\textwidth]{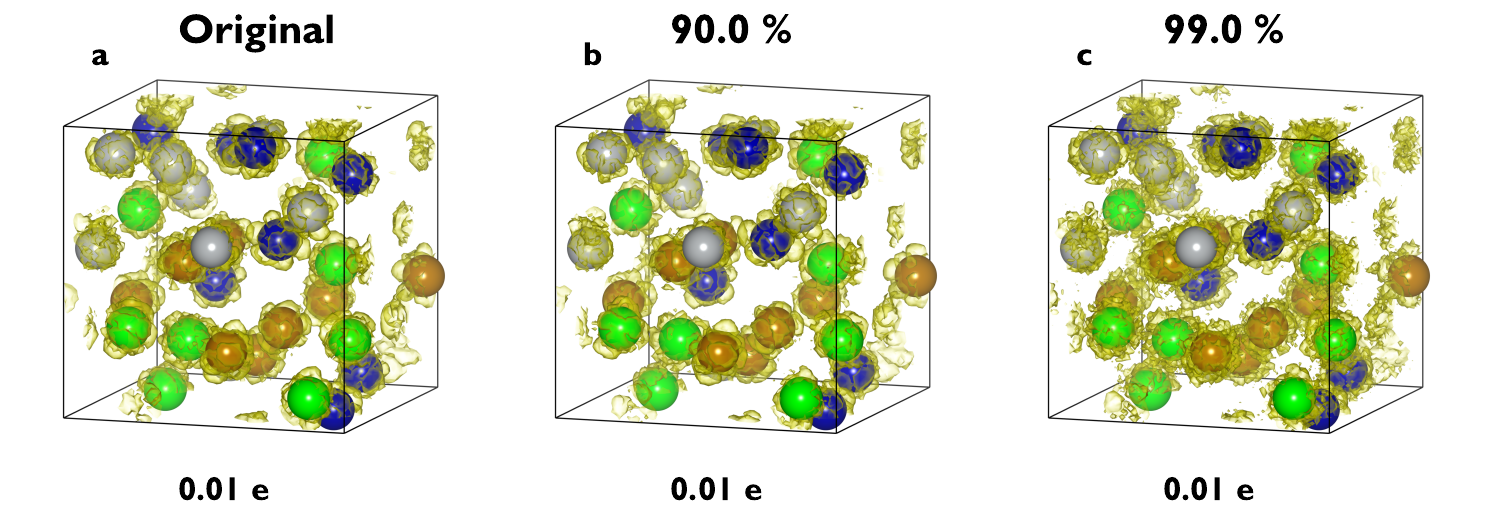}\caption{\textbf{$\text{Cr}_{25}\text{Fe}_{25}\text{Co}_{25}\text{Ni}_{25}$ system at 5000K}: Iso-surfaces of errors in electron density obtained by ML model for original (a), 90\% CCS (b), and 99\% CCS (c) respectively.}
    \label{fig:crfeconi-iso-error}
\end{figure*}

The corresponding 3D error fields are shown in Figures~\ref{fig:al-iso-error}, \ref{fig:water-iso-error}, \ref{fig:sigesn-iso-error}, and \ref{fig:crfeconi-iso-error}. Although the error iso-surfaces closely resemble those of the original datasets even after $90$\% and $99$\% pruning, a slight yet progressive increase in error is observed as pruning levels increase. These iso-surface visualizations are generated with the VESTA~\cite{momma2008vesta} software.



